%% file: physic2016_paper.tex
\documentclass[twocolumn, showpacs, preprintnumbers,prd, superscriptaddress, floatfix, nofootinbib, raggedbottom]{revtex4-2}
\pdfoutput=1

\usepackage{color}
\usepackage{cancel}
\usepackage{epsfig}
\usepackage{booktabs}
\usepackage{diagbox}
\usepackage{array,multirow}
\usepackage{amsmath, amssymb, mathrsfs}

\usepackage{graphicx}
\usepackage{epstopdf}

\usepackage{dcolumn}

\usepackage{bm}

\usepackage[utf8]{inputenc}

\usepackage{xcolor}

\usepackage{siunitx}
\DeclareSIUnit\clight{\text{\ensuremath{c}}}

\usepackage{natbib}
\usepackage{hyperref}

\usepackage{float}
\usepackage{cleveref}

\newcommand{\pos}{\mathrm{e^{+}}}
\newcommand{\ele}{\mathrm{e^{-}}}
\newcommand{\epem}{\pos\ele}

\newcommand{\aprime}{A^\prime}

\def \rarr {\rightarrow}
\def \grinp {\includegraphics}
\def \tw {\textwidth}

\def \Mlr {M\o ller }
\def \Ap {$\mathrm{A^{\prime}}$ }

\def \ee {$\mathrm{e^{+}e^{-}}$}

\begin{document}

	\title{Searching for Prompt and Long-Lived Dark Photons in 
	Electro-Produced \texorpdfstring{\ee}{} Pairs with the Heavy Photon Search
	Experiment at JLab}
    
    \input{authors}

    \date{\today}
    \begin{abstract}
        
        The Heavy Photon Search  experiment (HPS) at the Thomas Jefferson National Accelerator Facility
        searches for electro-produced dark photons.  We report  results
        from the 2016 Engineering Run consisting of \SI{10608}{nb^{-1}} of data
        for both the prompt and displaced vertex searches.  
A search for a prompt resonance in the $\epem$ invariant mass distribution between 39 and \SI{179}{MeV} showed no evidence of dark photons above the large QED background, limiting the coupling of $ \epsilon^2\gtrsim 10^{-5}$, in agreement with previous searches. The search for displaced vertices showed no evidence of excess signal over background in the masses between 60 and \SI{150}{MeV}, but had insufficient luminosity to limit canonical heavy photon production.
        This is the first displaced vertex search result published by HPS.
        HPS has taken high-luminosity data runs in 2019 and 2021 that will explore new
        dark photon phase space.  
        
	\end{abstract}
	\pacs{JLAB-PHY-23-3738}
    \maketitle
    \raggedbottom

    %Introduction
    \input{Intro.v2}
    %Theory and Experimental Landscape
    %\input{Landscape}
    %Detector Overview
    \input{detector_overview}
    \input{CEBAF}
    \input{BeamlineTarget}

    \input{SVT}

    \input{ECal}
    \input{Trigger}
    %Analysis Overview
    \input{anaOverview}
    %Data Samples
    \input{DataSamples}
    %Tracking & Vertexing (should be recon generally)
    \input{Recon}
    \input{EventSelection}
%    \input{SampleComposition}
    \input{radFrac}
    \input{massReso}
    %Resonance Search
    \input{ResonanceSearch}
    %Displaced Vertex Search
    \input{DisplacedVertexSearch}

    %Summary and Future Plans
    \input{Summary}

    %acknowledgements    
    \input{acknowledgements}
%    \nocite{*}
\bibliography{Bibl}
\end{document}

%% file: authors.tex
%
% 2016 Engineering Run Author List
%
%

\newcommand*{\SACLAY}{IRFU, CEA, Universit\'e Paris-Saclay, F-91190 Gif-sur-Yvette, France}
\newcommand*{\SACLAYindex}{1}
\newcommand*{\INFNGE}{INFN, Sezione di Genova, 16146 Genova, Italy}
\newcommand*{\INFNGEindex}{2}
\newcommand*{\INFNTUR}{INFN, Sezione di Torino, 10125 Torino, Italy}
\newcommand*{\INFNTURindex}{21}
\newcommand*{\INFNMB}{University di Milano Bicocca, 20126 Milano, Italy}
\newcommand*{\INFNMBindex}{56}
\newcommand*{\ORSAY}{Université Paris-Saclay, CNRS, IJCLab, 91405, Orsay, France}
\newcommand*{\ORSAYindex}{22}
\newcommand*{\UNH}{University of New Hampshire, Durham, New Hampshire 03824, USA}
\newcommand*{\UNHindex}{27}
\newcommand*{\NSU}{Norfolk State University, Norfolk, Virginia 23504, USA}
\newcommand*{\NSUindex}{28}
\newcommand*{\ODU}{Old Dominion University, Norfolk, Virginia 23529, USA}
\newcommand*{\ODUindex}{30}
\newcommand*{\ROMAII}{Universit\`a di Roma Tor Vergata, 00133 Rome Italy}
\newcommand*{\ROMAIIindex}{32}
\newcommand*{\JLAB}{Thomas Jefferson National Accelerator Facility, Newport News, Virginia 23606, USA}
\newcommand*{\JLABindex}{36}
\newcommand*{\GLASGOW}{University of Glasgow, Glasgow G12 8QQ, United Kingdom}
\newcommand*{\GLASGOWindex}{39}
\newcommand*{\WM}{College of William \& Mary, Williamsburg, Virginia 23187, USA}
\newcommand*{\WMindex}{42}
\newcommand*{\YEREVAN}{Yerevan Physics Institute, 375036 Yerevan, Armenia}
\newcommand*{\YEREVANindex}{43}
\newcommand*{\PERIMETER}{Perimeter Institute, Ontario, Canada N2L 2Y5}
\newcommand*{\PERIMETERindex}{44}
\newcommand*{\SASSARI}{Universit\`a di Sassari, 07100 Sassari, Italy}
\newcommand*{\SASSARIindex}{45}
\newcommand*{\FNAL}{Fermi National Accelerator Laboratory, Batavia, IL 60510, USA}
\newcommand*{\FNALindex}{46} 
\newcommand*{\CATANIA}{INFN, Sezione di Catania, 95123 Catania, Italy}
\newcommand*{\CATANIAindex}{47}
\newcommand*{\STONYBROOK}{C.~N.~Yang Institute for Theoretical Physics, 
                          Stony Brook University, Stony Brook, NY 11794, USA}
\newcommand*{\STONYBROOKindex}{48}
\newcommand*{\SLAC}{SLAC National Accelerator Laboratory, Stanford University, Stanford, CA 94309, USA}
\newcommand*{\SLACindex}{49}
\newcommand*{\UCSC}{Santa Cruz Institute for Particle Physics, University of California, Santa Cruz, CA 95064, USA}
\newcommand*{\UCSCindex}{50}
\newcommand*{\PADOVA}{Universit\`a di Padova, 35122 Padova, Italy} 
\newcommand*{\PADOVAindex}{51}
\newcommand*{\INFNPA}{INFN, Sezione di Padova, 16146 Padova, Italy}
\newcommand*{\INFNPAindex}{52}
\newcommand*{\IDAHO}{Idaho State University, Pocatello, ID, 83209, USA}
\newcommand*{\IDAHOindex}{53}
\newcommand*{\INFNROMA}{INFN, Sezione di Roma Tor Vergata, 00133 Rome, Italy}
\newcommand*{\INFNROMAindex}{54}
\newcommand*{\INFNSUD}{INFN, Laboratori Nazionali del Sud, 95123 Catania, Italy}
\newcommand*{\INFNSUDindex}{55}

\author{P.~H.~Adrian}\affiliation\SLAC
\author{N.~A.~Baltzell}\affiliation\JLAB
\author{M.~Battaglieri}\affiliation\INFNGE
\author{M.~Bond\'i}\affiliation\CATANIA
\author{S.~Boyarinov}\affiliation\JLAB
\author{C.~Bravo}\thanks{Corresponding Author: bravo@slac.stanford.edu}\affiliation\SLAC
\author{S.~Bueltmann}\affiliation\ODU
\author{P.~Butti}\affiliation\SLAC
\author{V.~D.~Burkert}\affiliation\JLAB
\author{D.~Calvo}\affiliation\INFNTUR
\author{T.~Cao}\affiliation\UNH
\author{M.~Carpinelli}\affiliation\INFNSUD\affiliation\INFNMB
\author{A.~Celentano}\affiliation\INFNGE
\author{G.~Charles}\affiliation\ORSAY
\author{L.~Colaneri}\affiliation\ROMAII\affiliation\INFNROMA
\author{W.~Cooper}\affiliation\FNAL
\author{B.~Crowe}\affiliation\UNH
\author{C.~Cuevas}\affiliation\JLAB
\author{A.~D'Angelo}\affiliation\ROMAII\affiliation\INFNROMA
\author{N.~Dashyan}\affiliation\YEREVAN
\author{M.~De~Napoli}\affiliation\CATANIA
\author{R.~De~Vita}\affiliation\INFNGE
\author{A.~Deur}\affiliation\JLAB
\author{M.~Diamond}\affiliation\SLAC
\author{R.~Dupre}\affiliation\ORSAY
\author{H.~Egiyan}\affiliation\JLAB
\author{L.~Elouadrhiri}\affiliation\JLAB
\author{R.~Essig}\affiliation\STONYBROOK
\author{V.~Fadeyev}\affiliation\UCSC
\author{C.~Field}\affiliation\SLAC
\author{A.~Filippi}\affiliation\INFNTUR
\author{A.~Freyberger}\affiliation\JLAB
\author{M.~Gar\c{c}on}\affiliation\SACLAY
\author{N.~Gevorgyan}\affiliation\YEREVAN
\author{F.~X.~Girod}\affiliation\JLAB
\author{N.~Graf}\affiliation\SLAC
\author{M.~Graham}\thanks{Corresponding Author: mgraham@slac.stanford.edu}\affiliation\SLAC
\author{K.~A.~Griffioen}\affiliation\WM
\author{A.~Grillo}\affiliation\UCSC
\author{M.~Guidal}\affiliation\ORSAY
\author{R.~Herbst}\affiliation\SLAC
\author{M.~Holtrop}\affiliation\UNH
\author{J.~Jaros}\affiliation\SLAC\
\author{R.~P.~Johnson}\affiliation\UCSC
\author{G.~Kalicy}\affiliation\ODU
\author{M.~Khandaker}\affiliation\IDAHO
\author{V.~Kubarovsky}\affiliation\JLAB
\author{E.~Leonora}\affiliation\CATANIA
\author{K.~Livingston}\affiliation\GLASGOW
\author{L.~Marsicano}\affiliation\INFNGE
\author{T.~Maruyama}\affiliation\SLAC
\author{S.~McCarty}\affiliation\UNH
\author{J.~McCormick}\affiliation\SLAC
\author{B.~McKinnon}\affiliation\GLASGOW
\author{K.~Moffeit}\affiliation\GLASGOW
\author{O.~Moreno}\affiliation\SLAC\affiliation\UCSC
\author{C.~Munoz~Camacho}\affiliation\ORSAY
\author{T.~Nelson}\affiliation\SLAC
\author{S.~Niccolai}\affiliation\ORSAY
\author{A.~Odian}\affiliation\SLAC
\author{M.~Oriunno}\affiliation\SLAC
\author{M.~Osipenko}\affiliation\INFNGE
\author{R.~Paremuzyan}\affiliation\JLAB\affiliation\UNH
\author{S.~Paul}\affiliation\WM
\author{N.~Randazzo}\affiliation\CATANIA
\author{B.~Raydo}\affiliation\JLAB
\author{B.~Reese}\affiliation\SLAC
\author{A.~Rizzo}\affiliation\ROMAII\affiliation\INFNROMA
\author{P.~Schuster}\affiliation\SLAC\affiliation\PERIMETER
\author{Y.~G.~Sharabian}\affiliation\JLAB
\author{G.~Simi}\affiliation\PADOVA\affiliation\INFNPA
\author{A.~Simonyan}\affiliation\ORSAY
\author{V.~Sipala}\affiliation\SASSARI\affiliation\INFNSUD
\author{A.~Spellman}\affiliation\UCSC
\author{D.~Sokhan}\affiliation\GLASGOW
\author{M.~Solt}\affiliation\SLAC
\author{S.~Stepanyan}\affiliation\JLAB
\author{H.~Szumila-Vance}\affiliation\JLAB\affiliation\ODU
\author{N.~Toro}\affiliation\SLAC\affiliation\PERIMETER
\author{S.~Uemura}\affiliation\SLAC
\author{M.~Ungaro}\affiliation\JLAB
\author{H.~Voskanyan}\affiliation\YEREVAN
\author{L.~B.~Weinstein}\affiliation\ODU
\author{B.~Wojtsekhowski}\affiliation\JLAB

%% file: Intro.v2.tex
\section{Introduction}

Interest in searching for new, sub-GeV mediators with weak couplings to ordinary matter has grown exponentially in recent years, where such forces could play an essential role in production of sub-GeV dark matter in the early universe~\cite{Hewett:2012ns,Essig:2013lka,Alexander:2016aln,Battaglieri:2017aum}. Additionally and more generally, such experiments are a key complement to searches for new physics at high energies where new weakly coupled physics at low mass scales can be difficult to identify. Heavy photons, also known as ``hidden-sector'' or ``dark'' photons, are a benchmark example of such a mediator that also appears in many scenarios for physics beyond the Standard Model. Kinetic mixing of the heavy photon with the Standard Model photon through radiative loops of massive particles generates a weak coupling of the heavy photon to electrically charged particles~\cite{Holdom:1985ag,Galison:1983pa,Fayet:1990wx}.  As a result, heavy photons would be radiated by energetic electrons passing through a target in a process analogous to bremsstrahlung, but at parametrically lower rate, and can also decay to lepton-anti-lepton pairs~\cite{AprimeFixedTargetTheory}. While our search is focused on heavy photons, it is also sensitive to dark forces with vector, axial-vector, scalar, or pseudo-scalar couplings to matter which will have similar signatures and could also be produced in our experiment.

The Heavy Photon Search Experiment (HPS) at the Thomas Jefferson National Accelerator Facility (JLab) in Newport News, Virginia, searches for heavy photons and other new force carriers that are produced via electro-production and decay to electron-positron pairs~\cite{AprimeFixedTargetTheory}. Note that if direct decays to dark matter (or other dark-sector particles) are kinematically allowed, those decays are expected to dominate over the decay to SM particles, so HPS is only sensitive to heavy photons with less than twice the mass of the dark matter particle. Experimental signatures are either a resonance in the invariant electron-positron mass distribution or displaced decay vertices with a particular invariant mass, depending on the heavy photon mass and the heavy photon mass. Over the past decade, searches for dark photons have been conducted over large regions of the dark photon mass/coupling parameter space~\cite{Bjorken:1988as, riordan1987, bross1991, konaka1986,
             davier1989, andreas2012, Blumlein:1990ay, Blumlein:1991xh, Banerjee:2019pds, NA64:2019auh, Gninenko:2012eq, Aubert:2009cp, Babusci:2012cr, Archilli:2011zc, Aaij:2017rft, PhysRevLett.124.041801, Yamaguchi:2016mbc, Abrahamyan:2011gv, Merkel:2014avp,
            Agakishiev:2013fwl, Batley:2015lha, Aaij:2017rft, PhysRevLett.124.041801}, but much of that parameter space, including territory favored by thermal dark matter production in the early universe, remains unexplored and accessible to HPS~\cite{Battaglieri:2017aum}. Evidence for a dark force could be the first compelling evidence for a hidden sector and lead to identifying the nature of dark matter. 

For concreteness, we focus our discussion on the heavy photon, denoted $\aprime$.  The $\aprime$ is the mediator of a spontaneously broken ``hidden'' $U(1)'$ gauge symmetry. The $\aprime$ interacts with SM particles through kinetic mixing with the SM $U(1)_Y$ (hypercharge) gauge boson, resulting at low energies in the effective Lagrangian density 
\begin{equation}
\mathcal{L} \supset -\frac{\epsilon}{2} F'_{\mu\nu}F^{\mu\nu}\,,
\end{equation}
where $\epsilon$ denotes the strength of the kinetic mixing, 
$F'_{\mu\nu}=\partial_\mu \aprime_\nu - \partial_\nu \aprime_\mu$ is the $U(1)'$ field strength tensor, and similarly $F^{\mu\nu}$ denotes the field strength of the SM photon. This $\aprime$--photon mixing allows heavy photons to be produced in interactions involving electromagnetically charged particles and, if sufficiently massive, to decay into pairs of charged particles like electron-positron pairs or muon-antimuon pairs, or to hidden-sector states. The value of $\epsilon$ and the $\aprime$ mass ($m_{\aprime}$) generated in the fundamental theory naturally fall into the sensitivity range of HPS in certain model scenarios~\cite{ArkaniHamed:2008qn,ArkaniHamed:2008qp,Baumgart:2009tn,Essig:2009nc,Cheung:2009qd,Morrissey:2009ur}. 

 %%%%%%%%%%%%%%%%%%%%%%%%%%%%%%%%%%%%%%%%%%%%%%%% F I G U R E %%%%%%%%%%%%%%%%%%%%%%%%%%%%%%%%%%%%%%%%%%%%%%%%%%%%
\begin{figure}[!htb]
    \centering
    \includegraphics[width=0.45\tw]{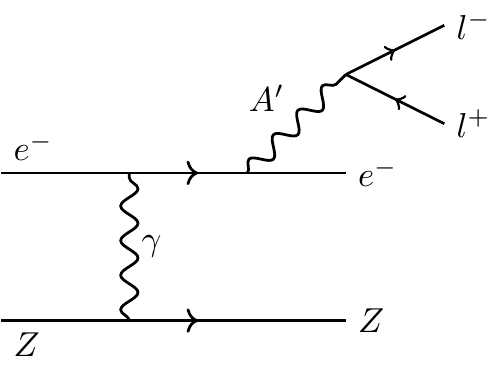}
    \caption{ Diagram of the $\aprime$ production off the tungsten target and decay to an $\epem$ pair.}
\label{fig:ap_feynman}
\end{figure}
 %%%%%%%%%%%%%%%%%%%%%%%%%%%%%%%%%%%%%%%%%%%%%%%% F I G U R E %%%%%%%%%%%%%%%%%%%%%%%%%%%%%%%%%%%%%%%%%%%%%%%%%%%%

The HPS experiment, which utilizes the Continuous Electron Beam Accelerator Facility (CEBAF) at JLab, can explore a wide range of heavy photon masses ($m_{\aprime} \sim 20-\SI{220}{MeV}$) and couplings ($\epsilon^2 \sim 10^{-10} - 10^{-6}$) by exploiting a range of beam energies and utilizing both resonance search and separated vertex search
 strategies. In this paper, results from both strategies are reported, using the data from the 2016 Engineering Run which employed an electron beam with a current of \SI{200}{nA} and an energy of $E_{\rm beam}=\SI{2.3}{GeV}$ incident on a thin ($\SI{4}{\micro\meter}$) tungsten target, and integrating a luminosity of $\SI{10608}{nb^{-1}}$. We have previously reported on the resonance search from our 2015 Engineering Run at \SI{1.03}{GeV}~\cite{Adrian:2018scb}. In HPS, the $\aprime$s would be electro-produced on the target nuclei, and would subsequently decay to electron-positron pairs, shown in \Cref{fig:ap_feynman}. A charged particle spectrometer, triggered by an electromagnetic calorimeter, measures the momenta and trajectories of the pair, from which its invariant mass and decay position can be reconstructed. The $\aprime$ decay length in the laboratory frame is given by
\begin{equation}
    \ell_0 \simeq \frac{\SI{1.8}{mm}}{N_{\rm eff}} \left(\frac{E_{\rm beam}}{\SI{2.3}{GeV}}\right)
    \left(\frac{10^{-4}}{\epsilon}\right)^2 \left(\frac{\SI{100}{MeV}}{m_{\aprime}}\right)^2\,,
\end{equation}
where $N_{\rm eff}$ is the number of decay channels kinematically accessible ($N_{\rm eff} = 1$ for HPS searches below the di-muon threshold)~\cite{AprimeFixedTargetTheory}. 
For larger couplings, the $\aprime$ is essentially prompt and would appear as a narrow resonance, with a width set by the experimental resolution, on top of a broad distribution of background events from ordinary quantum electrodynamics (QED) processes. At smaller couplings the $\aprime$ lifetime is long enough to give rise to secondary decay vertices, which can be distinguished from the prompt QED background, providing a second signature for heavy photon production.  A recent search, motivated by the same models as discussed here, looked for muon pairs with a displaced vertex was conducted by the LHCb experiment~\cite{Aaij:2017rft, PhysRevLett.124.041801}  %I don't like this last sentence or where it is....

The HPS experiment records copious QED trident production, as well as wide-angle bremsstrahlung production with subsequent conversion in the target or detector material, both of which produce the same final state. While these processes constitute physics backgrounds for the heavy photon search, they also enable important experimental checks and provide an experimental determination of our sensitivity, since the expected heavy photon production can be related to the measured trident production. 
The experimental mass resolution impacts the reach and is a critical input to the fits of the mass spectrum and to setting the width of the mass bins for the vertex search. It is calibrated directly from the data by measuring the invariant mass of M\o ller pairs, which have a unique invariant mass for any given incident electron energy. Similarly, the measured decay length distribution of the prompt trident signal provides a critical estimate of the decay length resolution.

The outline of the rest of the paper is as follows. \Cref{sec:detector} describes the beamline, target, and detector used by the HPS experiment. \Cref{sec:anaOverview} gives an overview of the common elements of the data analysis described in the paper. Sections \ref{sec:bumphunt} and \ref{sec:vtxAna} describe in detail the resonance search and displayed vertex search, respectively.  Finally, \Cref{sec:summary} gives a summary of the paper.

%% file: detector_overview.tex
\section{Detector Overview}
\label{sec:detector}

While the rejection of QED backgrounds motivates the best possible resolutions for $\epem$ mass and vertex position, the kinematic characteristics of the signal and beam backgrounds determine the overall layout of the HPS apparatus. Radiation of a mediator that is heavy compared to the incoming electron carries away most of the energy in the reaction, so $x = E_{A'}/E_\mathrm{beam}$ is peaked strongly at \num{1} \cite{AprimeFixedTargetTheory}. Since HPS operates at beam energies beyond \SI{1}{GeV}, which are large compared to $A'$ masses of interest, the $A'$ is highly boosted with its momentum closely aligned with the beam direction. The $A'$ subsequently decays to an $\epem$ pair, leaving that pair also boosted in the very forward direction and azimuthally back-to-back with respect to the beamline. Therefore, a detector with excellent forward acceptance immediately downstream of the target is required to detect the $\epem$ decay products and cleanly identify secondary vertices as close to the target and through-going beam as possible.

\begin{figure}
    \centering
    \includegraphics[width=0.45\textwidth]{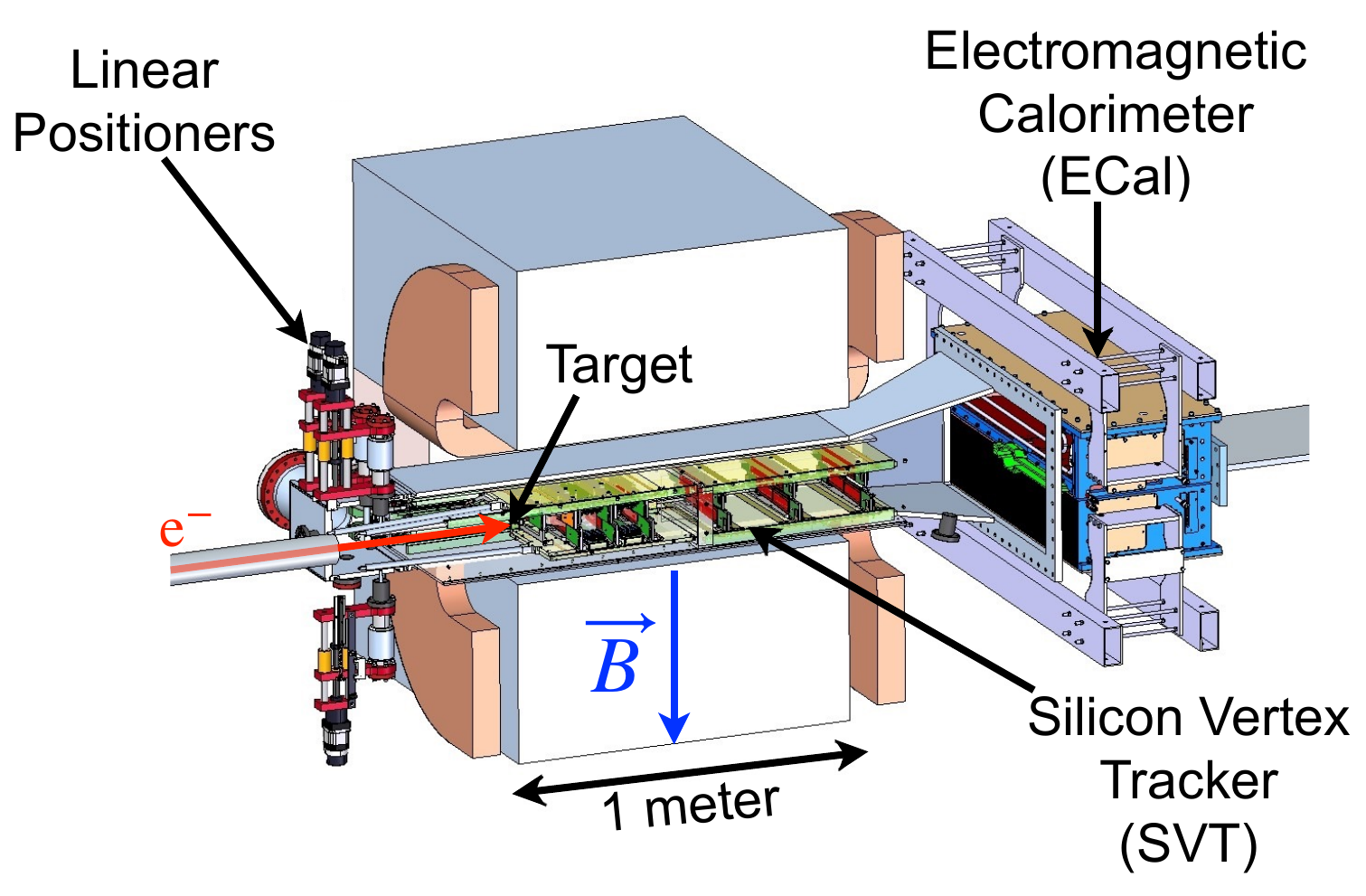}
    \caption{A cutaway view of the HPS detector showing the SVT in a vacuum chamber inside the bore of the spectrometer magnet and the ECal downstream. The positions of the target and the front portions of the SVT are controlled by a set of linear positioning motors upstream of the detector.}
    \label{fig:hps_generalView}
\end{figure}

HPS realizes this concept with a magnetic spectrometer, consisting of a multi-layer Silicon Vertex Tracker (SVT) situated within a large dipole magnet (\SI{0.24}{T} for the beam energy described in this paper), to measure the momenta and trajectories of the $\epem$ pair. The field of the dipole is vertical, dispersing most of the beam electrons that have radiated in the target, as well as other electromagnetic backgrounds, into the horizontal plane containing the beam. As a result, the SVT is split into two segments, one above and one below the beam plane, which are positioned as close to it as possible to maximize acceptance. The extent of the forward acceptance is limited by the background rate of single beam electrons that scatter in the target, which cannot mimic the signal but creates extreme occupancies ($\approx$\SI{4}{MHz/mm^2}) at the edge of the first layer of the SVT. 
To minimize occupancy and accidentals while preserving sensitivity to small signals, the high repetition rate of the CEBAF beam (499 MHz) effectively spreads the luminosity out in time, and the high-rate $\epem$ trigger selectively picks only events of interest. High rate capability in the SVT and the lead tungstate electromagnetic calorimeter (ECal) allow selection of only those hits in time with the trigger for readout and reconstruction. These key components of the HPS apparatus are shown in  \Cref{fig:hps_generalView}.

%As a consequence, the high repetition rate of the CEBAF beam (\SI{499}{MHz}), in tandem with a high-rate $\epem$ trigger with precision timing and similarly precise timing in the SVT, is required to select only in-time hits for readout and reconstruction. HPS uses a lead tungstate electromagnetic calorimeter (ECal) -- also split above and below the beam plane -- to provide this trigger and an offline estimate of the precise hit time, along with particle identification for the tracks reconstructed in the SVT. These key components of the HPS apparatus are shown in  \Cref{fig:hps_generalView}.

%%%%%%%%%%%%%%%%%%%%%%%%%%%% F I G U R E %%%%%%%%%%%%%%%%%%%%%%%%%%%%%%%%%%%%%%%%%%%%%%%
%%%%%%%%%%%%%%%%%%%%%%%%%%%% F I G U R E %%%%%%%%%%%%%%%%%%%%%%%%%%%%%%%%%%%%%%%%%%%%%%%

%% file: CEBAF.tex
\subsection{The JLab CEBAF} \label{sec:CEBAF}

The HPS experiment utilizes beam from CEBAF at the JLab in Newport News, Virginia. CEBAF is oval shaped, consisting of two linacs connected by a pair of recirculating arcs, which enables injected beam 
to make multiple passes of the linacs — gaining \SI{2.2}{GeV} per pass for up to \num{5.5} passes — before extracting the beam into  one of four halls. Sub-harmonics of the \SI{1.497}{GHz}  beam may be simultaneously extracted into the different halls, allowing simultaneous operation of multiple experiments with high-rate (typically \SI{499}{MHz}) beam ~\cite{Leemann:2001dg}. Operation at the JLab CEBAF is fundamental to the success of the HPS experiment because it provides a very high repetition rate multi-\si{GeV} electron beam with low per-bunch charge. A higher per-bunch charge would spoil the clean tracking and vertexing needed for the displaced vertex search and lower current would require unacceptably long operations.

%The HPS experiment utilizes beam from CEBAF at the JLab in Newport News, Virginia.  CEBAF is oval shaped, consisting of two linacs connected by a pair of recirculating arcs, which enables injected beam to make multiple passes of the linacs --- gaining \SI{2.2}{GeV} per pass for up to \num{5.5} passes --- before extracting the beam into one of four halls. Sub-harmonics of the \SI{1.497}{GHz} beam may be simultaneously extracted into the different halls, allowing simultaneous operation of multiple experiments with high-rate (typically \SI{499}{MHz}) beam.~\cite{Leemann:2001dg}. Operation at the JLab CEBAF is fundamental to the success of the HPS experiment. The experiment requires a very high repetition rate multi-\si{GeV} electron beam with low per-bunch charge, together with precision hit timing in all subsystems, in order to screen out the high rate of background hits from scattered single electrons. A higher per-bunch charge would spoil the clean tracking and vertexing required for the displaced vertex search, while a lower current would require unacceptably long operations.  

%% file: BeamlineTarget.tex
\subsection{Hall B Beamline and Target} \label{sec:beamline}

\begin{figure*}[!htb]
     \centering
     \grinp[width=0.95\tw]{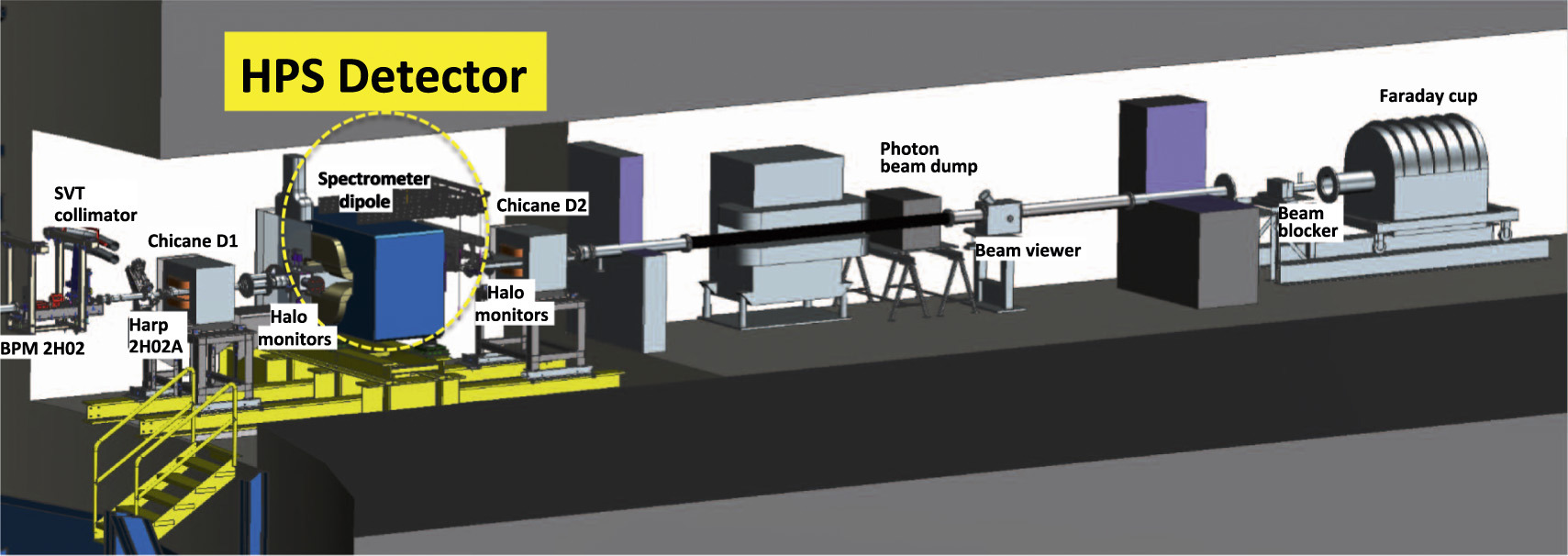}
     \caption{ Engineering rendering of the downstream alcove of experimental Hall-B, where the HPS apparatus is located.}
     \label{fig:hps_beam}
 \end{figure*}

The HPS apparatus operates in the downstream alcove of experimental Hall~B\cite{HPS:2016jta}, as shown in \Cref{fig:hps_beam}.
 %%%%%%%%%%%%%%%%%%%%%%%%%% F I G U R E %%%%%%%%%%%%%%%%%%%%%%%%%%%%%%%
%\begin{widetext}
%\end{widetext}
%%%%%%%%%%%%%%%%%%%%%%%%%% F I G U R E %%%%%%%%%%%%%%%%%%%%%%%%%%%%%%%
 The \SI{2.3}{GeV} electron beam is transported $\approx$\SI{57}{m} from the upstream Hall~B tunnel to HPS, passing through a number of quadrupole and dipole magnets that focus and steer the beam to the target. The extraordinary proximity of the SVT layers to the beam, as close as \num{500} microns between the edges of sensors and the center of the beam, places stringent requirements on the quality of the beam; a very small beamspot ($<$\SI{50}{\um} vertically) with vanishing low halo rate ($\lesssim${}$10^{-6}$ outside the Gaussian core) and excellent beam stability ($<$\SI{30}{\um} vertical variation). 

\begin{figure}[!htb]
     %\centering
     \grinp[width=0.45\tw]{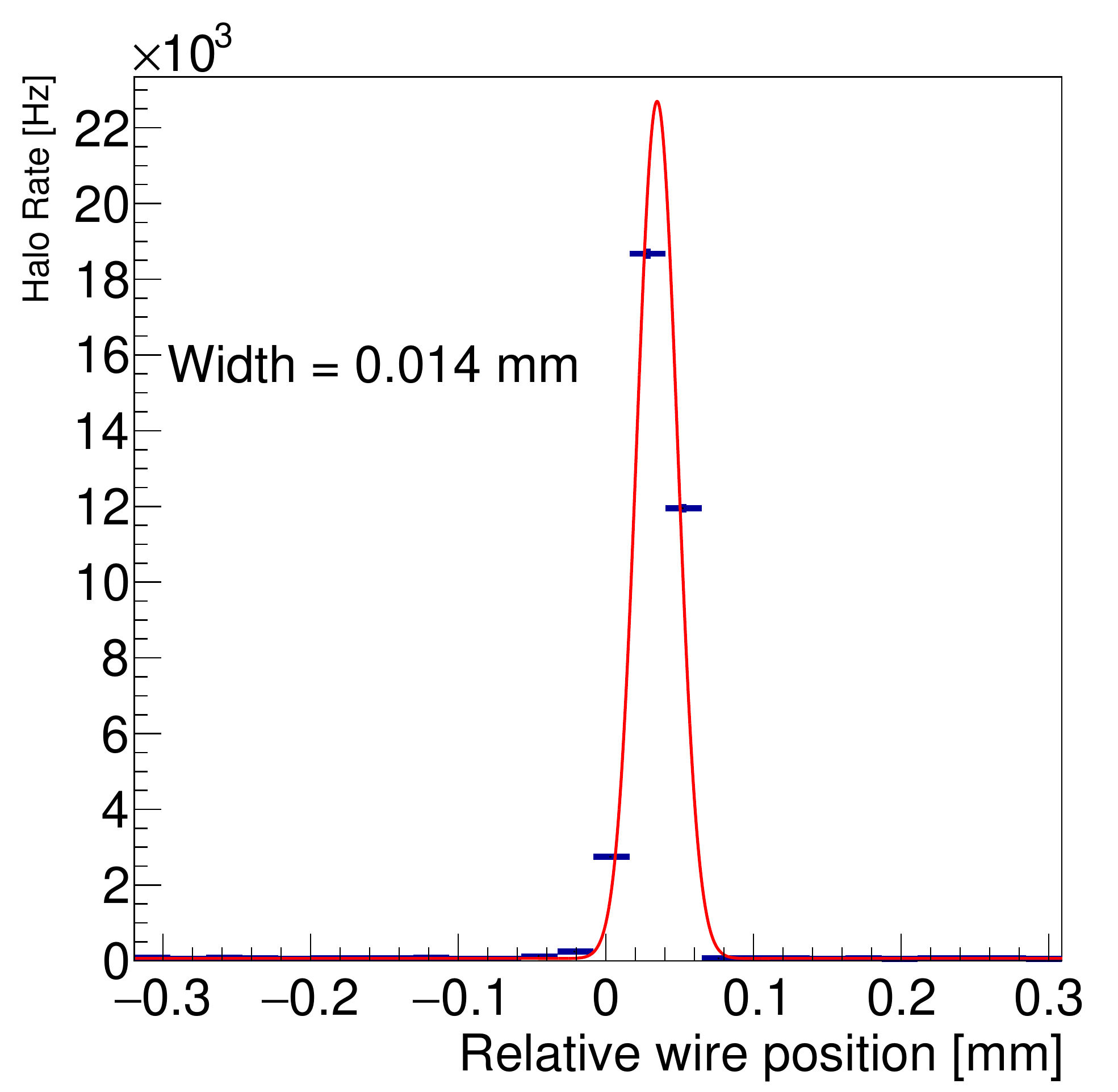}
     \caption{An example beam vertical profile obtained with the SVT wire during final beam tuning.}
     \label{fig:SVT_ScanYProfile}
\end{figure}

\begin{figure}[!htb]
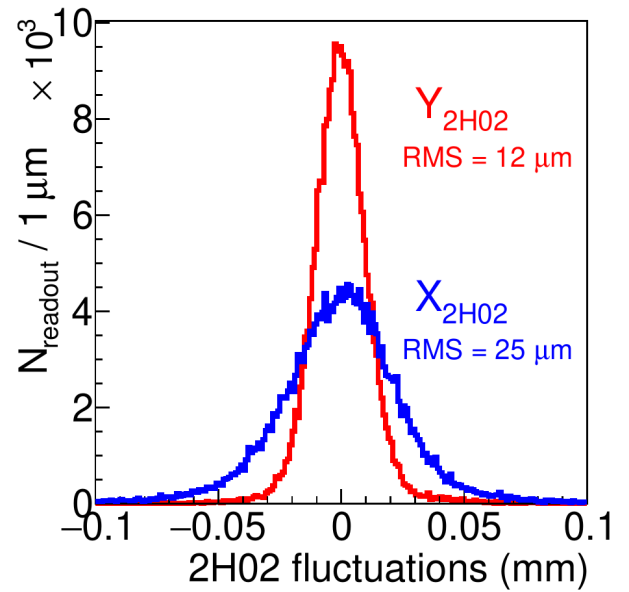

     %\centering
     \grinp[width=0.45\tw]{Figs/FinalPlotsInPaper/2H02_Y_Fluctuations.pdf}
     \caption{Distribution of the beam x and y positions reported by the 2H02 BPM, $\approx$\SI{2}{m} upstream of the target, over a period of roughly one hour.}
     \label{fig:2H02_fluctuations}
\end{figure}

Ensuring the safety of the SVT also requires multiple diagnostic and protection systems. During beam setup, the beam profile and position are measured by wire scanners (“harps”) located strategically along the beamline and used to tune the trajectory to produce the desired spot on the target. In addition, there are wires integrated into the movable structures of the SVT that are close to the target and precisely referenced to the positions of the silicon sensors that can be used to ensure the ideal profile and position of the beam. A typical scan of the beam with an SVT wire is shown in \Cref{fig:SVT_ScanYProfile}.
Beam position monitors (BPMs) are used to continuously monitor the transverse position of the beam at multiple locations during data-taking, and are tied to machine controls (orbit locks) to ensure the stability of the beam trajectory, as demonstrated by \Cref{fig:2H02_fluctuations}. 
A set of halo counters around the apparatus monitors background levels to detect any scraping of the beam upstream of or inside the apparatus. In addition to providing the data for harp scans, the halo counters are tied to the Fast Shut Down (FSD) system of CEBAF, and can trigger beam shutdown within a few milliseconds of exceeding settable thresholds. Finally, a collimator with a choice of several apertures directly upstream of HPS is used to protect the detector from large beam excursions during tuning and operations.

The target for the experiment is chosen to be as thin as possible to achieve the desired luminosity given the upper limit on beam currents in Hall~B.  We choose a thin target in order to minimize occupancy in the detector from multiple-scattered electrons and two-step processes in the target. The target system consists of a movable assembly with different thickness tungsten foils, in addition to carbon and polyethylene targets for calibration purposes. The data analyzed for this paper were taken with a \SI{4}{\um} tungsten foil, equivalent to approximately \num{0.125}\% of a radiation length.  The target, with respect to the coordinate system used for the experiment, was measured to be at z=\SI{-4.3}{mm}.

%% file: SVT.tex
\subsection{Silicon Vertex Tracker} \label{sec:svt}

\begin{figure}[tb]
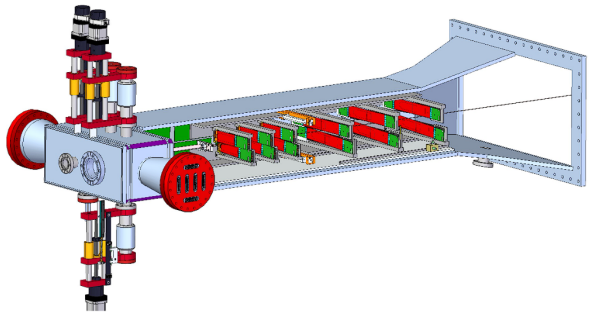

 \centering
 \grinp[width=0.45\tw]{Figs/FinalPlotsInPaper/svt-layout.png}
\caption{A diagram showing the SVT layout inside the vacuum box.}
 \label{fig:SVT_design}
\end{figure}

The SVT is a six layer, high precision, silicon tracking and vertexing detector responsible for estimating both the mass and decay position of $\epem$ pairs by measuring the momenta and trajectories of charged particles. The design of the SVT, shown in \Cref{fig:SVT_design}, is shaped by a few competing requirements.  First, $\aprime$ decay products have typical momenta $\lesssim E_\mathrm{beam}/2$, so multiple scattering dominates mass and decay length errors for any feasible material budget. Second, the signal yields for long-lived $\aprime$s are very small, so the rejection of prompt vertices must be exceedingly strong, better than $10^{-6}$, to reduce prompt background to the order of one event or less. Finally, as previously discussed, the passage of scattered and degraded primary beam through the apparatus creates a region of extreme occupancy and radiation in the same part of the detector that is critical for sensitivity to low-mass $\aprime$ that have decay products nearly collinear with the beam. This puts low-mass acceptance at odds with tracking and vertexing purity and the material budget for the detector, requiring careful design to allow the largest usable acceptance. A prototype detector, with many of the same general features and utilizing the same sensor design, is described in more detail in~\cite{testRunNim}.

The SVT employs radiation tolerant silicon microstrip sensors developed for the D\O\ RunIIb project~\cite{D0:2002hvi}, which allows the readout and cooling material to be placed outside the tracking volume. The sensors and their front-end readout electronics are cooled from the ends via their support structures to below \SI{-10}{\degreeCelsius} to extend their lifetime at peak fluences exceeding $10^{16}$ electrons/cm$^2$ (or \num{4e14}[1~\si{MeV}~neutron equivalent]/cm$^2$). The SVT is split into mirror-symmetric halves, above and below the plane of the scattered and degraded beam. As a result, the regions of high occupancy are small spots along the sensor edges, so that only a very short length of the edge strips see high occupancy. Long strips covering those regions have per-channel occupancies only a small factor larger than what pixels would experience. 

Each layer of the SVT consists of sensors placed back-to-back \SI{7.4}{mm} apart with a small stereo angle between them (\SI{100}{mrad} in front three layers; \SI{50}{mrad} in back three), so that 3D space points can be determined. Each half of the detector -- top and bottom -- is further divided into two separate structures -- front and back -- with three layers each. The first three detector layers are a single sensor in width and spaced \SI{10}{cm} apart along the beam direction, with the first layer just \SI{10}{cm} downstream of the target. The next three layers are two sensors wide and spaced \SI{20}{cm} apart beginning \SI{50}{cm} downstream of the target, where these double width layers improve acceptance for low-momentum particles. All four detector segments, with 36 sensors and 23004 channels total, are placed as close to the beam as backgrounds allow, with acceptance down to \SI{15}{mrad} above and below the beam plane with respect to the beam spot on the target. Since this places the active (passive) edges of the sensors in the first layer \SI{1.5}{mm} (\SI{0.5}{mm}) from the center of the beam, precision construction, alignment, and survey of the sensors are essential, and the structures holding the first three layers are movable, allowing them to be retracted from the beam during beam tuning. To eliminate displaced events and occupancy from beam-gas collisions, the SVT must operate inside the beam vacuum, and resides within a vacuum enclosure installed inside a dipole magnet with a downward-pointing central field of \SI{0.24}{T}.  

The sensors of the SVT are read out by APV25 ASICs \cite{French:2001xb} mounted on hybrid PCBs and wirebonded directly to the sensors.  Power, control, and monitoring of the hybrids, and clocking, control and digitization of APV25 samples are performed by a set of Front End Boards (FEBs) also located inside the SVT vacuum enclosure, to minimize the length of the analog cables and reduce the number of signals that must penetrate the vacuum barrier. Being in vacuum, the FEBs also require liquid cooling, which uses a separate system from the sensor modules to allow the temperatures of the two systems to be set independently.
Power and digital signals are passed from the FEBs via vacuum feedthroughs in a pair of flanges to the power supplies and central data acquisition system (DAQ) outside of the vacuum chamber. The central DAQ for the SVT, based on the Reconfigurable Cluster Element (RCE) architecture\cite{Herbst:2016prn}, connects to the data flange via \SI{50}{m} optical fibers, allowing it to be placed in a lower radiation environment. The SVT DAQ is capable of very high data rates, which is necessary to accommodate the torrent of irreducible trident backgrounds that must be accepted in order to search for rare $A'$ events.

\begin{figure}[tb]
 \centering
 \grinp[width=0.45\tw]{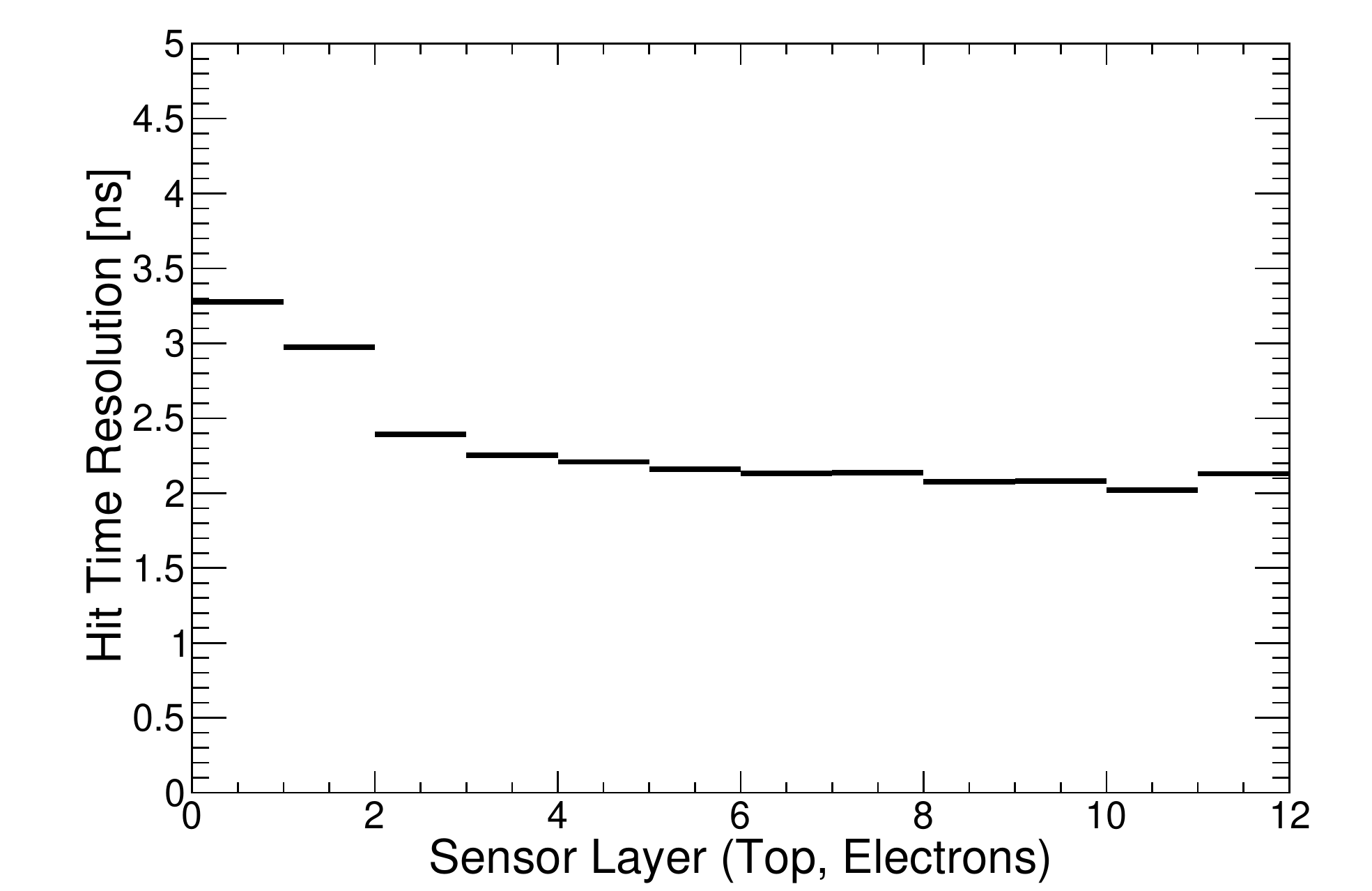}
\caption{The time resolution of SVT hits associated with tracks versus layer number.  }
 \label{fig:svt_timeres}
\end{figure}

\begin{figure}[tb]
 \centering
 \grinp[width=0.45\tw]{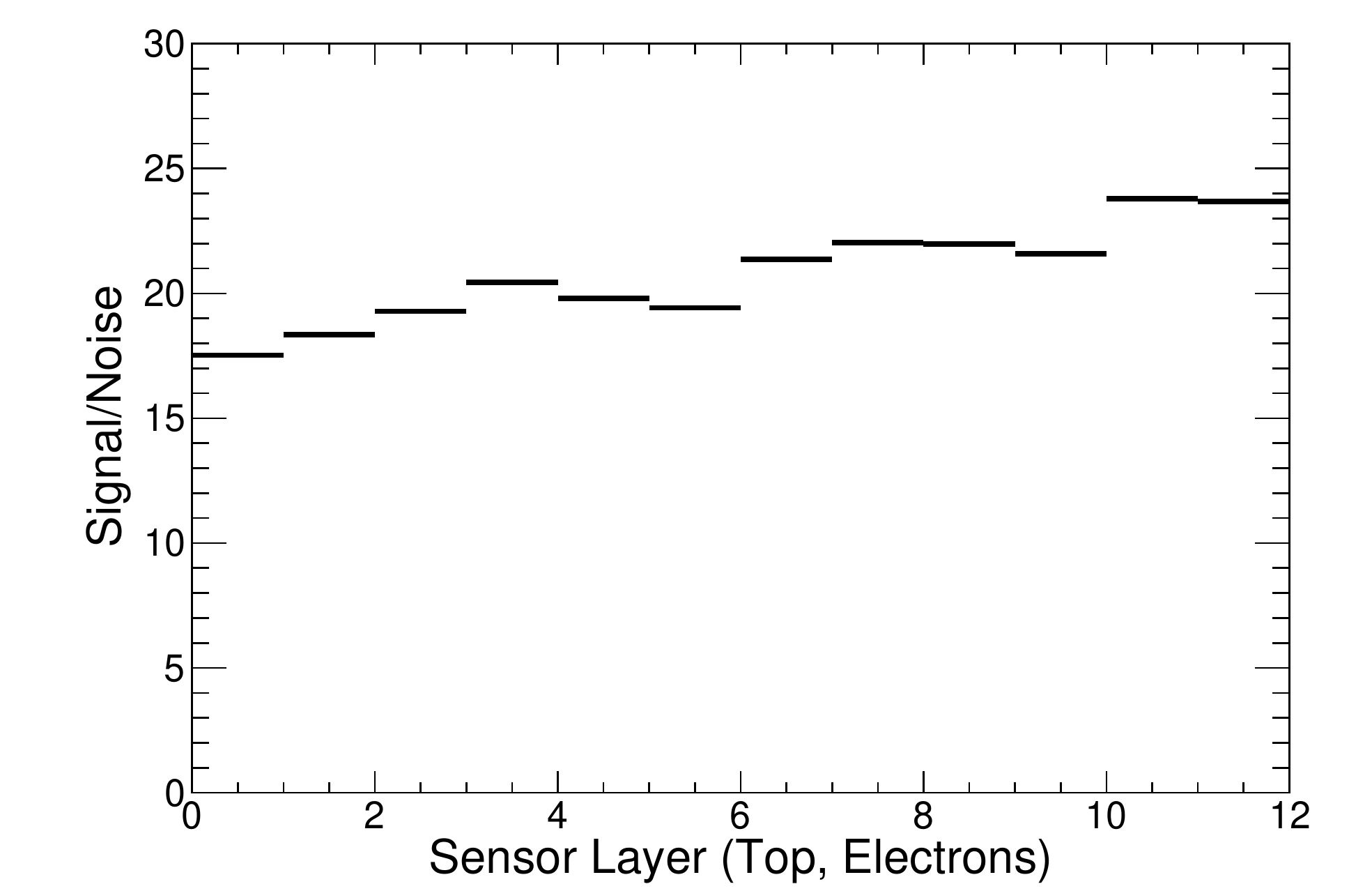}
\caption{The mean signal-over-noise of SVT hits associated with tracks versus layer number.  }
 \label{fig:svt_sovern}
\end{figure}

To further reduce occupancies for tracking, the CMS APV25 chip is used for readout in ``multi-peak'' mode, which records 6 samples of the signal development, allowing reconstruction of hit time with $\approx \SI{2}{ns}$ resolution --- near the level required to tag events in individual CEBAF bunches. \Cref{fig:svt_timeres} shows the time resolution versus layer number, with the inner, high-occupancy layers having slightly worse resolution than the back layers.  This is also reflected in the signal-to-noise of the sensors versus layer, shown in \Cref{fig:svt_sovern}.

%% file: ECal.tex
\subsection{Electromagnetic Calorimeter} 

The HPS Electromagnetic Calorimeter (ECal)~\cite{Balossino:2016nly} plays two critical roles. First, it provides a trigger for $\epem$ pairs with sufficient energy and time resolution to eliminate the overwhelming background of scattered single beam electrons. Second, it provides positive identification of electromagnetic energy deposits -- from electrons, positrons, or photons -- offline, with sufficient time resolution to tag them to a single CEBAF bunch, which can then be used to demand coincidence with tracks in the SVT. Like the SVT, the ECal must contend with extremely high rates and be relatively radiation tolerant in order to match the angular acceptance of the SVT as closely as possible.

\begin{figure}
    \centering
    \includegraphics[width=0.45\textwidth]{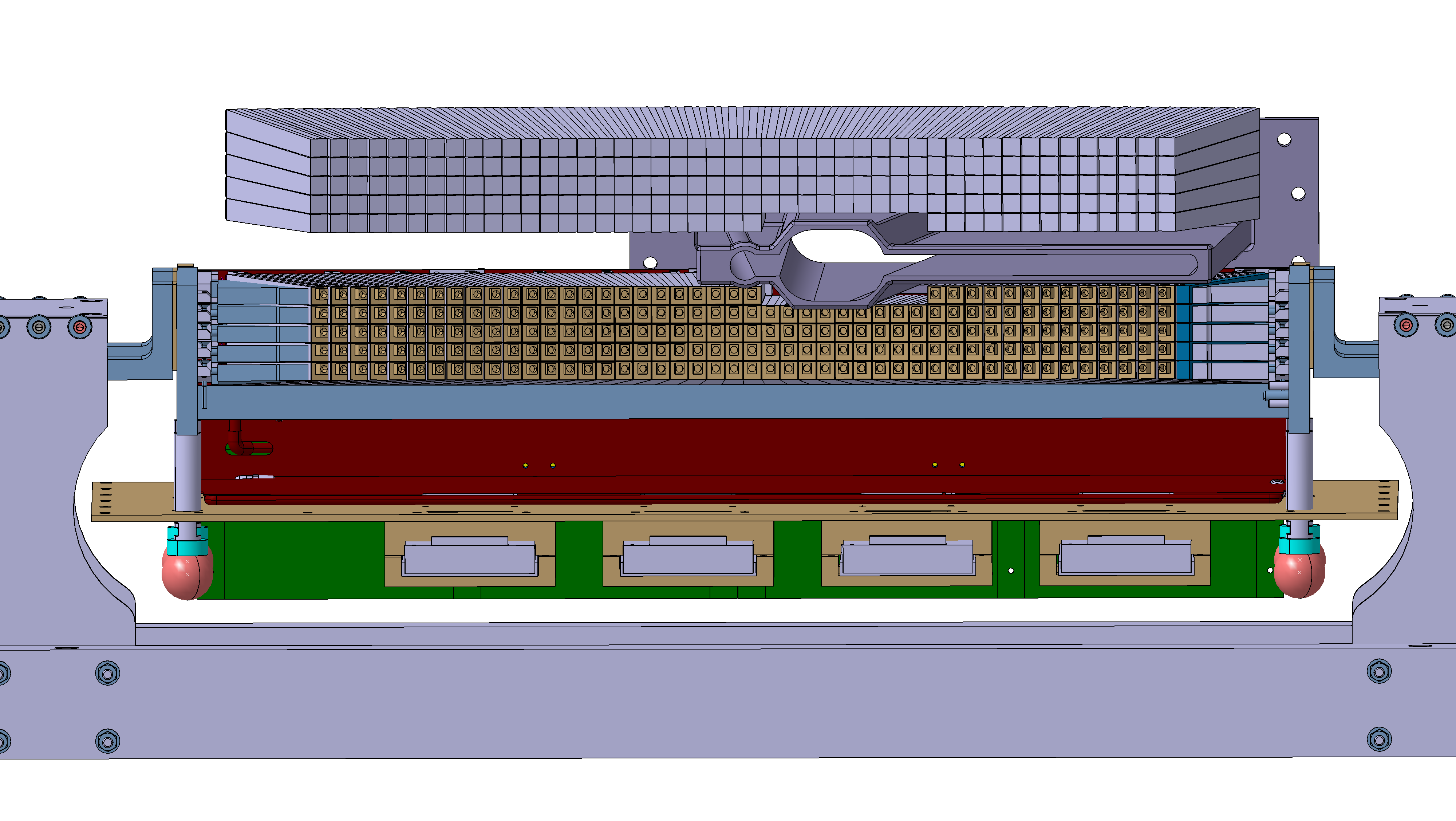}
    \caption{ECal crystal layout, as seen in the beam direction. For clarity, the top-half mechanical parts have been removed. For the bottom half, some mechanical elements such as the mother boards (in green) and the copper plates for heat shielding (in red) are visible. Between the two halves of ECal, the beam vacuum vessel is seen to be extended to the right to accommodate beam particles having lost energy through scattering or radiation.}
    \label{fig:HPS_ECal}
\end{figure}

The ECal meets these requirements through the use of 442 $\mathrm{PbWO_{4}}$ crystals arranged in two identical arrays --- placed symmetrically above and below the beam plane downstream of the SVT. The through-going degraded beam is transported between the two halves in a vacuum chamber to eliminate beam-gas backgrounds. Each half is a matrix of 5x46 $\mathrm{PbWO_{4}}$ crystals. From the first row of each half, 9 crystals are removed nearest the through-going beam as the rate of scattered beam electrons is intolerably high in that region, well in excess of \SI{1}{MHz}. The crystal layout and some mechanical elements of the ECal are shown in \Cref{fig:HPS_ECal}.
%%%%%%%%%%%%%%%%%%%%%%%%%%%%%%%%%%%%%%%%% F I G U R E %%%%%%%%%%%%%%%%%%%%%%%%%%%%%%%%%%%%%%%%%%%
%%%%%%%%%%%%%%%%%%%%%%%%%%%%%%%%%%%%%%%%% F I G U R E %%%%%%%%%%%%%%%%%%%%%%%%%%%%%%%%%%%%%%%%%%%
The ECal channels are read out via APD and \SI{250}{MHz} Flash ADC (FADC) boards which record samples of the pulses every \SI{4}{ns}. This provides a similar time window for triggers, whereas offline fitting of the FADC pulses provides a much better time estimate for ECal hits.

%% file: Trigger.tex
\subsection{Trigger System}

As outlined at the beginning of \Cref{sec:detector}, $\aprime$ production is peaked at small angles with respect to the beam direction, so the $\epem$ decay daughters are typically back-to-back relative to the beam direction~\cite{AprimeFixedTargetTheory}. As a result, when one daughter falls within the acceptance of the top half of the detector, the other will fall within the bottom acceptance.  Meanwhile, the vertical magnetic field of the spectrometer magnet will bend the electron and positron in opposite horizontal directions. Therefore, the primary trigger for the experiment is a ``pair trigger" in the ECal which requires energetic clusters in both halves (top and bottom) of the ECal, and with the two clusters displaced horizontally in opposite directions from the centerline according to their energies, since lower-energy particles will curve more in the magnetic field. 

\begin{figure*}
    \centering
    \grinp[width=0.95\tw]{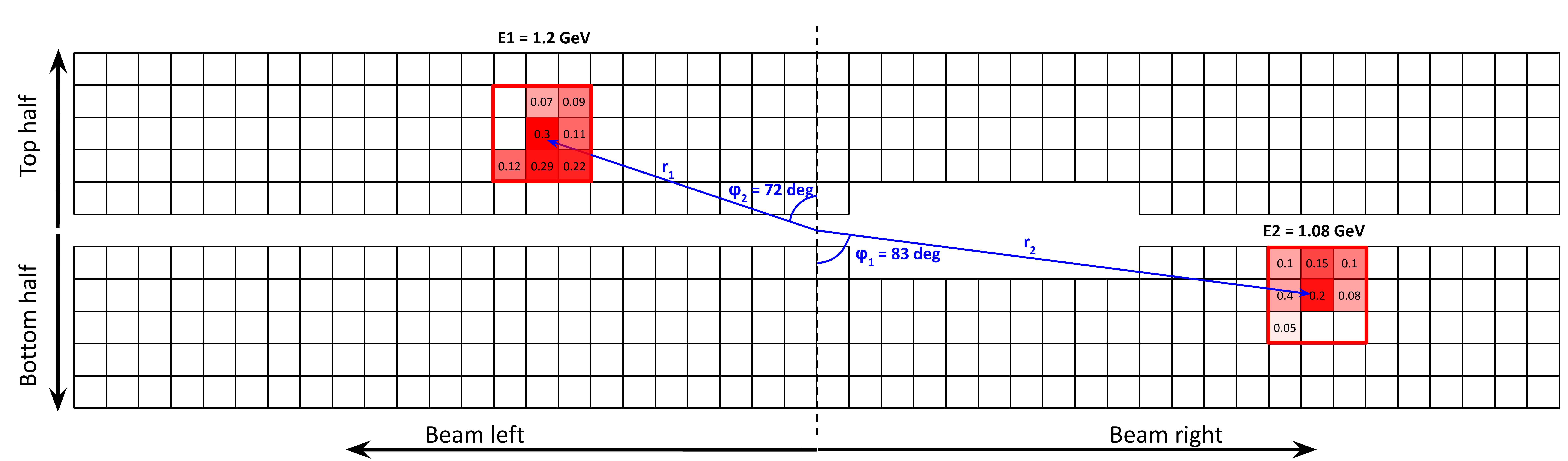}
    \caption{An illustration of an event satisfying trigger requirements. As described in the text, data for this analysis is collected using a pairs trigger that makes requirements on $|\phi_1-\phi_2|$ and the relationship between $r_{1,2}$ and $E_{1,2}$.}
    \label{fig:Pair1Trg}
\end{figure*}

Simulations showed that the two clusters in signal events are nearly back-to-back azimuthally, so the trigger requires that the azimuthal coplanarity of the two clusters is close to 0, as shown in \Cref{fig:Pair1Trg}. 
%%%%%%%%%%%%%%%%%%%%%%%%%% F I G U R E %%%%%%%%%%%%%%%%%%%%%%%%%%%%%%%
%\begin{widetext}
%\end{widetext}
%%%%%%%%%%%%%%%%%%%%%%%%%% F I G U R E %%%%%%%%%%%%%%%%%%%%%%%%%%%%%%%
The trigger also places a cut on the minimum cluster energies as a function of their horizontal displacement from the center-line of the ECal, according to $E + F\times r > E_{\rm threshold} $. Here E is the cluster energy, r is the distance of the cluster from the center of the calorimeter shown in  \Cref{fig:Pair1Trg}, and parameters $F$ and $\mathrm{E_{threshold}}$ are tuned using Monte Carlo simulation. This cut mostly eliminates the high rate of bremsstrahlung events with low-energy photons hitting close to the center of the ECal.

%% file: anaOverview.tex
\section{Analysis Overview} \label{sec:anaOverview} 

Our search for heavy photons uses two different techniques, outlined in detail below. The first is a traditional resonance search, where we search for a resolution-dominated resonance shape superposed on the copious $\epem$ invariant mass distribution which arises primarily from QED tridents. Heavy photons with relatively large coupling strengths have very short decay lengths, so appear prompt and would be detected in this search. The second is a vertex search for $\epem$ decay vertices significantly displaced from the target. The vertex search examines the observed decay length distribution mass-bin by mass-bin and looks for events beyond a cut where prompt backgrounds are expected to be small. Heavy photons with very small coupling strengths would have correspondingly large decay lengths, and would be detected in the vertex search. Thus HPS searches in two distinct regions of the heavy photon mass/coupling plane.
Both searches are performed partially blind, in the sense that all analysis cuts are frozen after inspecting 10\% of the data. The final analysis of the full data set, including this 10\%,  incorporates those cuts.

 %%%%%%%%%%%%%%%%%%%%%%%%%%%%%%%%%%%%%%%%%%%%%%%% F I G U R E %%%%%%%%%%%%%%%%%%%%%%%%%%%%%%%%%%%%%%%%%%%%%%%%%%%%
\begin{figure}[!htb]
    \centering
    \includegraphics[width=0.45\tw]{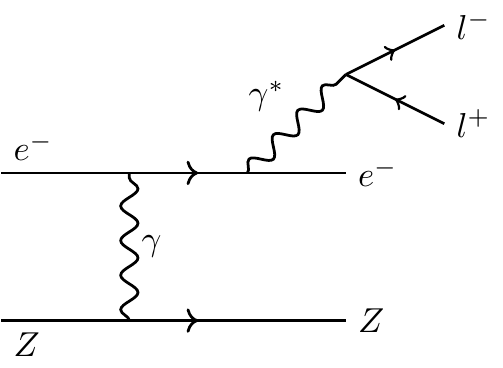}
    \caption{ Diagram of radiative trident production off the tungsten target }
\label{fig:rad_feynman}
\end{figure}
 %%%%%%%%%%%%%%%%%%%%%%%%%%%%%%%%%%%%%%%%%%%%%%%% F I G U R E %%%%%%%%%%%%%%%%%%%%%%%%%%%%%%%%%%%%%%%%%%%%%%%%%%%%
 
 %%%%%%%%%%%%%%%%%%%%%%%%%%%%%%%%%%%%%%%%%%%%%%%% F I G U R E %%%%%%%%%%%%%%%%%%%%%%%%%%%%%%%%%%%%%%%%%%%%%%%%%%%%
\begin{figure}[!htb]
    \centering
    \includegraphics[width=0.45\tw]{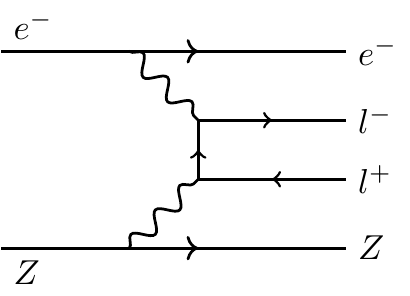}
    \caption{ Diagram of Bethe-Heitler trident production off the tungsten target.}
\label{fig:bh_feynman}
\end{figure}
 %%%%%%%%%%%%%%%%%%%%%%%%%%%%%%%%%%%%%%%%%%%%%%%% F I G U R E %%%%%%%%%%%%%%%%%%%%%%%%%%%%%%%%%%%%%%%%%%%%%%%%%%%%

Event selection and various data quality cuts are common to the two analyses, but not identical. The displaced vertex search, in particular, adopts special cuts to identify and eliminate long-lived backgrounds. The differences are detailed below.
Both analyses calculate their sensitivity to heavy photon production using the observed flux of $\epem$ pairs, which is predominantly due to two QED processes, trident production and wide-angle bremsstrahlung conversion (cWAB), which have a known relationship to the heavy photon production rate.  Trident production occurs via two processes, radiative (\Cref{fig:rad_feynman}) and Bethe-Heitler (\Cref{fig:bh_feynman}), and the interference between them.  Both analyses require $\epem$ pairs with total energy near that of the incident electron, as expected for heavy photon production. For tridents, this means that the observed pair likely excludes the recoil electron. For cWABs which convert in the target or first detector layer, the observed pair is usually the conversion positron and the recoil electron. Monte Carlo (MC) simulation of trident and cWAB production, incorporating their calculated cross-sections, reasonably accounts for the observed rate and momentum spectrum of $\epem$ pairs, demonstrating a good understanding of the sample composition. This procedure reduces dependence on experimental efficiencies. Using this MC estimation, the fraction of the observed events attributable to the purely radiative trident production diagram, which is proportional to heavy photon production, is determined. Hence, we calculate sensitivities to heavy photon production incorporating theoretical knowledge of trident and wide-angle bremsstrahlung cross-sections but normalized by the data.

Both analyses depend on knowing the experimental mass resolution and the invariant mass scale. For the resonance search, the mass resolution determines the width of the expected heavy photon resonance; the invariant mass scale, its exact position. For the displaced vertex search, the mass resolution determines what fraction of the signal appears in a given mass slice and how much background is included. M{\o}ller scattering results in $\ele \ele$ pairs of fixed mass for a given beam energy. Measuring the position and width of the M{\o}ller peak enables calibration of mass scale and resolution.

The measured decay length distributions for background (prompt) $\epem$ pairs arising from tridents and cWABs can be characterized by a broad Gaussian centered on the target location, with an exponential tail at large decay lengths. These features are the result of how the exiting $\pos$ and $\ele$ multiple Coulomb scatter, where resolution is dominated by scattering in the first detector layer.  Good agreement between Monte MC and data decay length distributions confirms our understanding of decay length resolution.

The following subsections review the data samples, detector calibration and event reconstruction, event selection, sample composition, and mass resolution for the two analyses.  The resonance search and displaced vertex search sections that follow discuss the specifics of each analysis in more detail.

%% file: DataSamples.tex
\subsection{Data Samples}

The results presented here use data collected during the 2016 Engineering Run, which operated on weekends during February~20--April~25 of 2016. All data used for analysis were collected at a beam energy of \SI{2.30}{GeV} with a current of \SI{200}{nA} on a Tungsten foil target \SI{4}{\um} ($\approx$0.125\%~$X_0$ equivalent) thick. The total luminosity of this dataset is \SI{10608}{nb^{-1}}, comprising 7.2 billion triggered events from a total charge on target of \SI{67.2}{mC}.

In addition to physics runs, a number of special runs were taken, such as field-off runs and runs with a trigger dedicated to collecting scattered single electrons over a wide range of scattering angles. Data from these runs were used to calibrate and align the ECal and SVT. 

In addition to experimental data, the analysis presented here makes use of MC simulation to understand some attributes of signal and background. MadGraph~\cite{MG5} is used to generate samples of $\aprime$ signal at a range of masses, as well as tridents, which include both Bethe-Heitler and radiative tridents (which are kinematically identical to signal) and their interference term, and converted WAB  events.
Monte Carlo of M{\o}ller scattering events is also used to study the mass resolution.  Beam backgrounds simulated using EGS5 \cite{Hirayama:2005zm}, predominantly scattered single electrons, are overlaid on all samples, distributed according to the time structure of the beam. Simulation of generated samples uses GEANT4 \cite{GEANT4:2002zbu} to model interactions with the detector, after which the detector response simulation and reconstruction are performed.

%% file: Recon.tex
\subsection{Detector Calibration and Event Reconstruction} \label{sec:reco} 

Raw data from the detector and simulation are reconstructed to produce the physics objects used  for analysis, which are reconstructed $\ele$ and $\pos$ as well as  $\aprime$ candidates consisting of reconstructed $\epem$ pairs emanating from a common vertex, which we refer to as ``V0 candidates''. The reconstruction of $\ele$ and $\pos$ is stepwise, taking place first separately in the ECal and the SVT, and then combining information from both subsystems. 

\subsubsection{ECal Calibration and Reconstruction}

The crystals of the ECal are small compared to the Moliere radius in PbWO$_4$, so in order to reconstruct and identify electrons and positrons in the calorimeter, the energy depositions in individual crystals must be calibrated and then combined, or ``clustered'', to provide a good estimate of the energy of incident electrons and positrons. 

\begin{figure}
    \centering
    \includegraphics[width=0.45\textwidth]{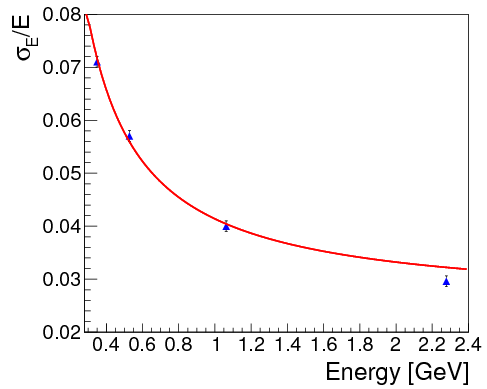}
    \caption{Energy resolution of the ECal as a function of energy. The three points below \SI{1.2}{GeV} were obtained from the 2015 Run, while the point at \SI{2.3}{GeV}, which benefits from electronics upgrades, was obtained from the 2016 Run using elastically scattered electrons, and was not used in the fit.  The energy resolution can be parameterized as $\frac{\sigma_E}{E}(\%)=\frac{1.6}{E}\bigoplus\frac{2.9}{\sqrt{E}}\bigoplus2.5$.}
    \label{fig:ECal_E_Resol}
\end{figure}

Calibration uses both minimum ionizing particles (MIPs) from cosmic ray events, as well as samples of scattered beam electrons collected with a special trigger, to determine the conversion of pulse height to energy. The simple clustering algorithm, which begins with a high-energy seed and iteratively adds adjacent crystals above a threshold, results in good energy resolution, as shown in \Cref{fig:ECal_E_Resol}. 
%%%%%%%%%%%%%%%%%%%%%%%%%%%%%%%%%%%%%%%%% F I G U R E %%%%%%%%%%%%%%%%%%%%%%%%%%%%%%%%%%%%%%%%%%%

\begin{figure}
    \centering
    \includegraphics[width=0.45\textwidth]{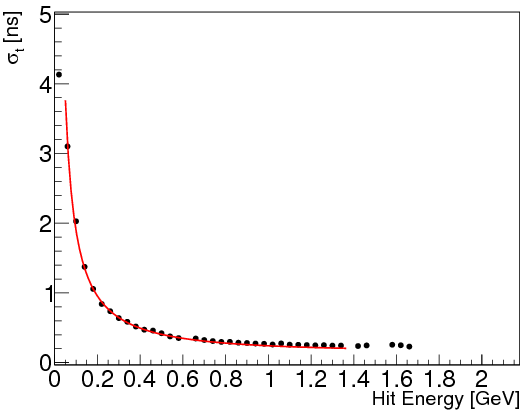}
    \caption{Time resolution of hits as a function of hit energy. The time estimate comes from a fit to samples of the APD output at \SI{4}{ns} intervals provided by the FADC readout used by the ECal.  The timing resolution can be parameterized as $\sigma_t = \frac{0.188}{E (GeV)}\bigoplus0.152~ns$.}
    \label{fig:ECal_t_Resol}
\end{figure}

%%%%%%%%%%%%%%%%%%%%%%%%%%%%%%%%%%%%%%%%% F I G U R E %%%%%%%%%%%%%%%%%%%%%%%%%%%%%%%%%%%%%%%%%%%
The pulse fit to the \SI{250}{MHz} FADC readout stream also results in excellent time resolution, as shown in \Cref{fig:ECal_t_Resol}.  More details may be found in~\cite{Balossino:2016nly}.
%%%%%%%%%%%%%%%%%%%%%%%%%%%%%%%%%%%%%%%%% F I G U R E %%%%%%%%%%%%%%%%%%%%%%%%%%%%%%%%%%%%%%%%%%%
%%%%%%%%%%%%%%%%%%%%%%%%%%%%%%%%%%%%%%%%% F I G U R E %%%%%%%%%%%%%%%%%%%%%%%%%%%%%%%%%%%%%%%%%%%

\subsubsection{SVT Calibration and Reconstruction}

The reconstruction of charged particle trajectories in the SVT detector starts with the formation of 3D space-points by combining the axial and stereo strip clusters on the two sides of each silicon module. In order to accept a 3D space-point, the two strip clusters' reconstructed times are required to be in a time window of \SI{12}{ns} from the trigger time and within \SI{16}{ns} of each other. 
Three 3D space points in selected SVT layers are then grouped together to form a track seed and an initial estimation of the track parameters is obtained by performing a helical fit under the assumption of a uniform magnetic field. 
The track-seed finding efficiency is maximized by choosing multiple combinations of the 3D space point triplets with different layer combinations to start the pattern recognition. Track-seeds are then extended by iteratively adding 3D space-points located on the other SVT layers and performing a global helical track fit selecting the track candidate with minimum $\chi^{2}$ during the procedure. At this stage, track candidates are required to have at least 5 associated 3D space points, momentum $p>\SI{100}{MeV}$ and track quality $\chi^{2}_{\rm 5hits} < 60$ and $\chi^{2}_{\rm 6hits} < 84$, for track candidates with 5 and 6 hits respectively. Track candidates are then refitted with the General Broken Lines (GBL) \cite{Keinwort:GBL1} algorithm to include the effects of  multiple scattering and refine the initial estimate of the track parameters. The GBL-refitted trajectories are also used for calibration and alignment of the SVT using Millepede II \cite{MPII}. The electron and positron particle candidates are then formed by requiring each reconstructed track to be associated with an ECAL cluster. 

Using  two final state particles reconstructed, one in each two detector volumes, vertices are then reconstructed using a global $\chi^{2}$ minimization algorithm \cite{BILLOIR1992139}. The final state particles used for vertex reconstruction are required to have an ECal cluster time difference within \SI{2.5}{ns} and the electron momentum $p_{\rm ele} < \SI{2.18}{GeV}$. Successfully reconstructed vertices are required to have a total momentum $p_{\rm vtx} < \SI{2.8}{GeV}$. 

\begin{figure}
    \centering
    \includegraphics[width=0.45\textwidth]{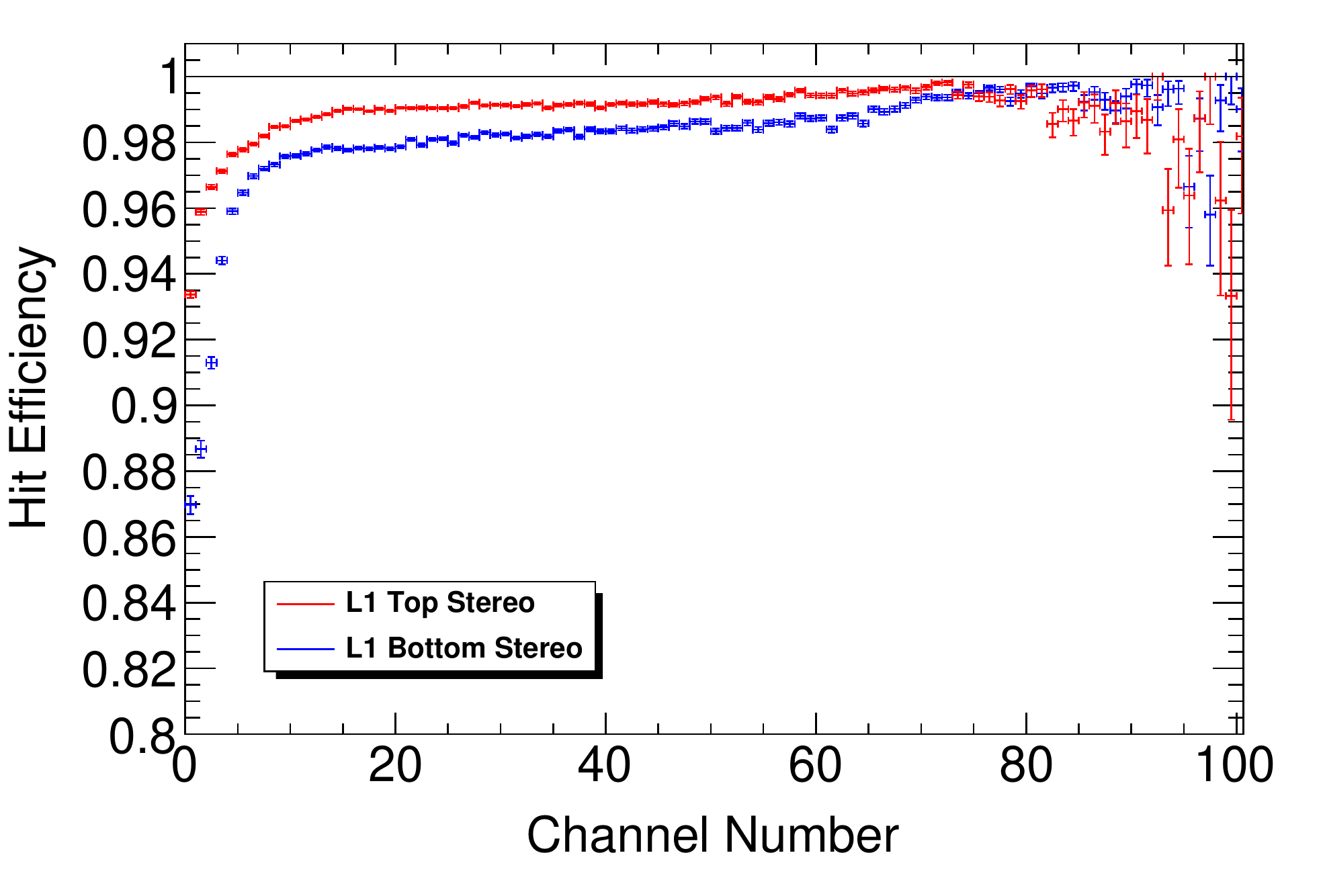}
    \caption{The hit-finding efficiency versus the extrapolated SVT channel number for two SVT layers at the front of the detector, one each in the top  (red) and bottom (blue) halves.  The nominal center of the electron beam at these layers is $\sim \SI{1.5}{mm}$ from channel zero (which is the edge of the active sensor).  The drop in efficiency is due to a combination of extrapolation error at the edge of the active volume and pile-up effects. }
    \label{fig:SVTHitEff}
\end{figure}

The performance of the SVT can be characterized by its tracking efficiency, momentum resolution, and vertex position resolution. Tracking efficiency is measured by selectively dropping hits in a particular layer from the track finding code, extrapolating the track as measured by the other layers to that layer, and measuring the fraction of times hits are found within a predicted region. Efficiencies are greater than 90\% in most of the SVT but are somewhat worse in the inner edges of the first two layers.  Dead channels also have noticeable effects. In the analysis below, tracking efficiency effects are included in critical simulations. The hit-finding efficiency for two of the first layers is shown in  \Cref{fig:SVTHitEff}. 

Momentum resolution is determined by measuring the momentum of elastically scattered beam electrons, which essentially have full beam energy. Henceforth these electrons will be referred to as FEEs (Full-Energy Electrons). Since the momentum resolution is dominated by multiple scattering effects in the SVT, determination of the momentum resolution of the highest momentum tracks suffices to characterize the resolution at all momenta.  The Monte Carlo does not accurately account for the observed momentum resolution, with simulation being better than reality. Accounting for this discrepancy is important in order to understand the actual invariant mass resolution, a critical parameter in the analysis. Procedures for doing so are described in \Cref{sec:massresolution}. 

The vertex position resolution of the tracker is easily measured by vertexing the copious trident signal, which originates at the known target position. The vertex resolution is well described by Monte Carlo simulation and is detailed in \Cref{sec:vtxAna}. The typical vertex resolution along the direction of the outgoing particles is on the order of \SI{1}{mm}. 

%% file: EventSelection.tex
\subsection{Event Selection} \label{sec:selection} 
The HPS experiment searches for $\aprime$s through their decays to $\epem$, so
 an event is required to contain at least one neutral, two-particle vertex 
 (called a V0). Due to the kinematics of $\aprime$ production,
 the electron and positron will almost always be in opposite
 halves of the HPS detector, so one track is required to be in the top half, the other in the bottom.
One of the particles must be positively charged, the other negatively charged. Each of
the particles must point to a cluster in the ECal.
 A V0 candidate is formed by fitting the two charged tracks to a vertex, following the procedures described in \cite{Hulsbergen:2005pu}: The vector momentum sum
of the electron and positron, $P_{\rm sum}$, 
 must meet the  condition $P_{\rm sum}< 1.2 \times P_{\rm beam}$, where $P_{\rm beam}$ is the
 beam momentum (\SI{2.3}{GeV} in this run). 

After the V0 candidates are formed, two V0 collections
are created. These are the “unconstrained V0 candidates” (UC) and “target-constrained V0 candidates”
 (TC). In these collections, a V0 particle is created and defined as the parent of the corresponding
 $\epem$ pair.   Including the TC in the vertex fit improves the angular resolution of the tracks and thus the invariant mass resolution.  
In the resonance search analysis, we use the TC V0s, where the
 $z$-coordinate of the vertex is constrained to be at the target position, and $x$-$y$ coordinates are constrained at
the beam spot coordinate.  The  $x$-$y$ coordinates of the beam spot at the target are obtained run-by-run from the average positions of UC V0s.  
The displaced vertex search analysis speciﬁcally searches for long-lived
 particles. Therefore, in the displaced vertex search analysis, the
 UC collection must be used.

\begin{figure}[!htb]
    \centering
    \includegraphics[width=0.45\tw]{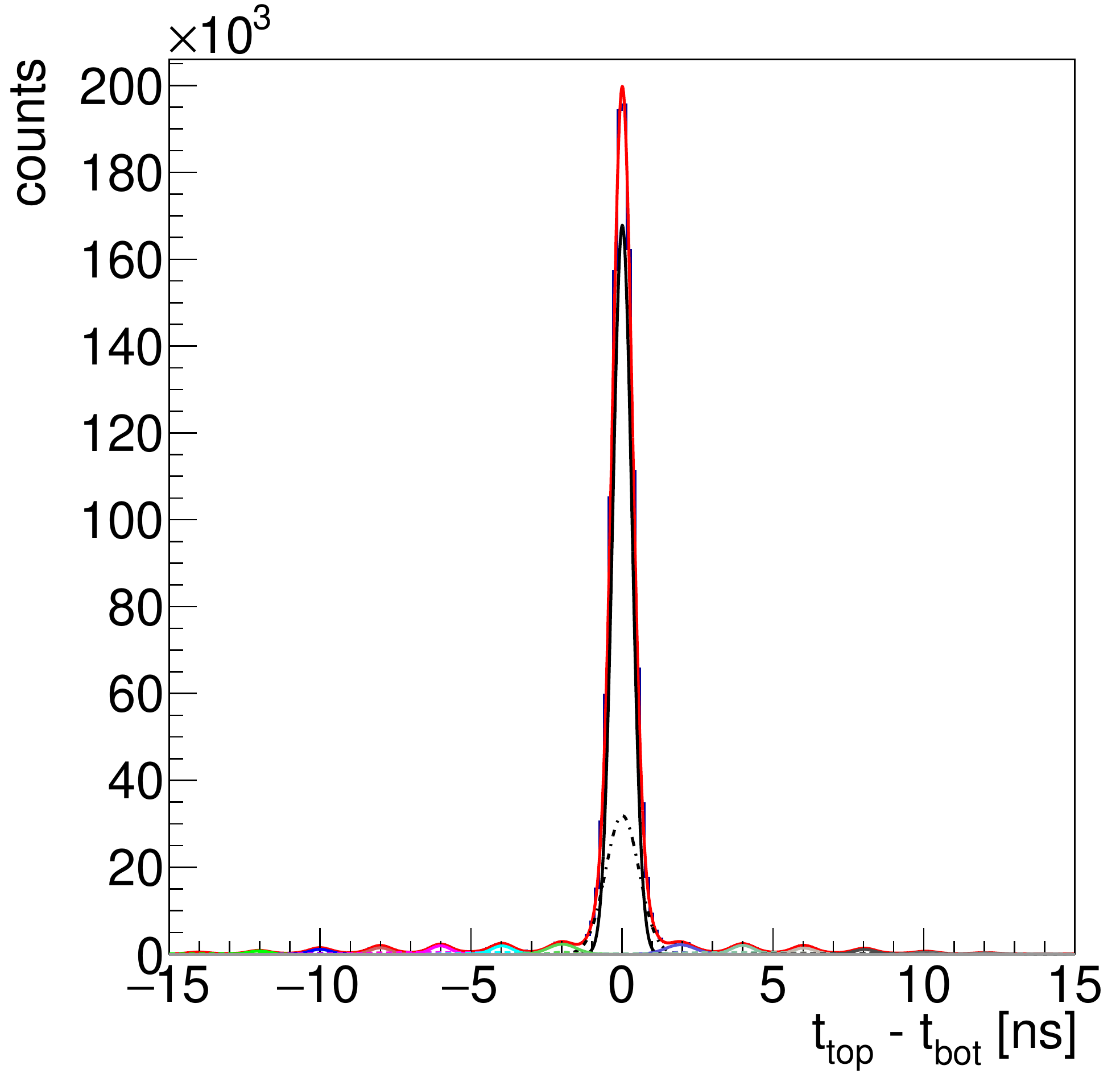}
    \includegraphics[width=0.45\tw]{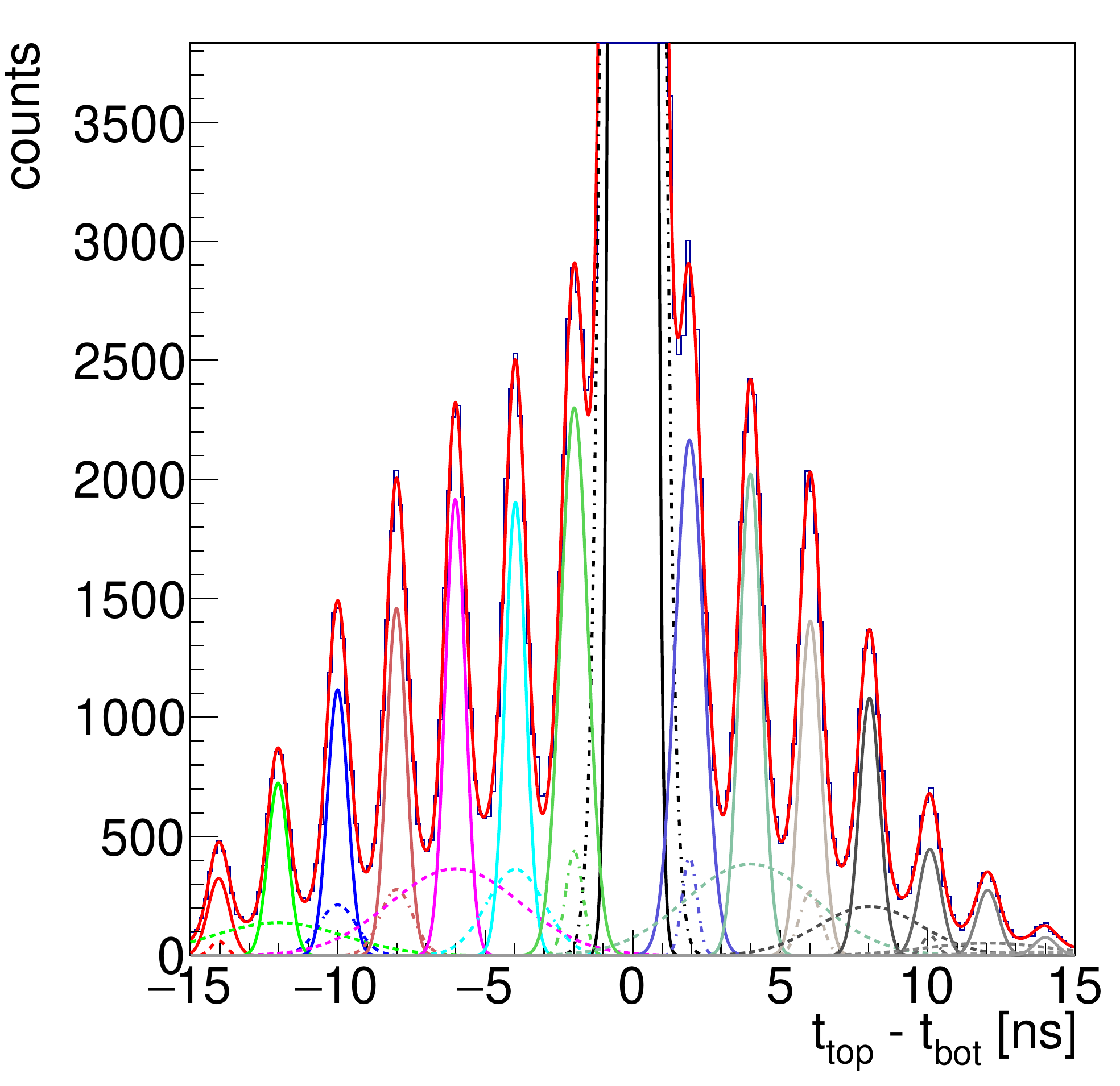}
    \caption{ Top and bottom cluster time difference, when the cluster energy sum is in the range $\SI{1.9}{GeV}$ to $\SI{2.4}{GeV}$. Bottom figure is the same as the top  
    with vertical axis adjusted to show peaks in tails.}
\label{fig:delta_t_cl}
\end{figure}

Further cuts on the V0 properties were imposed to minimize accidental backgrounds,
maximize the signal-to-background ratio of the radiative signal, and reduce physics backgrounds.
 Accidental backgrounds can be minimized by optimizing the cut on the time diﬀerence between the two ECal clusters. \Cref{fig:delta_t_cl} shows this cluster time diﬀerence, which is sharply peaked at 0. The bottom panel shows
 the same data, but with the vertical scale magniﬁed to show the structure in the tails, displaying peaks that occur 
 at multiples of \SI{2}{ns}, the spacing between CEBAF’s electron bunches. It shows that accidental coincidences created by particles between bunches occur at
 a low level. This distribution is ﬁt with the function given in \Cref{eq:clusterTimeDiffFitting} as a sum of peaks where each subpeak is parameterized as the
 sum of two Gaussian functions, one describing its core and another, wider and of lower amplitude, its tail. The
 ratio of the amplitudes of these two Gaussians is constrained to be the same for all peaks. The optimum time
interval is chosen to maximize the ratio $S/\sqrt{S+B}$ where $S$ is the
 integral of the central peak in the given $\pm \Delta t$ cut range, and $S+B$ is the integral of signal plus background.
 \begin{equation}
   \displaystyle F = \sum_{i = 0}^{N_{\mathrm{peak}}} a_{i}\cdot\left(\mathrm{Gauss}(x - \mu_{i}, \sigma_{1,i}) + b\cdot \mathrm{Gauss}(x - \mu_{i}, \sigma_{2,i}) \right)
    \label{eq:clusterTimeDiffFitting}
\end{equation}
For the resonance (vertex) search, the absolute value of the cluster time diﬀerence must be less than \SI{1.43}{ns} (\SI{1.45}{ns}).

 %%%%%%%%%%%%%%%%%%%%%%%%%%%%%%%%%%%%%%%%%%%%%%%% F I G U R E %%%%%%%%%%%%%%%%%%%%%%%%%%%%%%%%%%%%%%%%%%%%%%%%%%%%
 %%%%%%%%%%%%%%%%%%%%%%%%%%%%%%%%%%%%%%%%%%%%%%%% F I G U R E %%%%%%%%%%%%%%%%%%%%%%%%%%%%%%%%%%%%%%%%%%%%%%%%%%%%

\begin{figure}[!htb]
    \centering
      \includegraphics[width=0.45\tw]{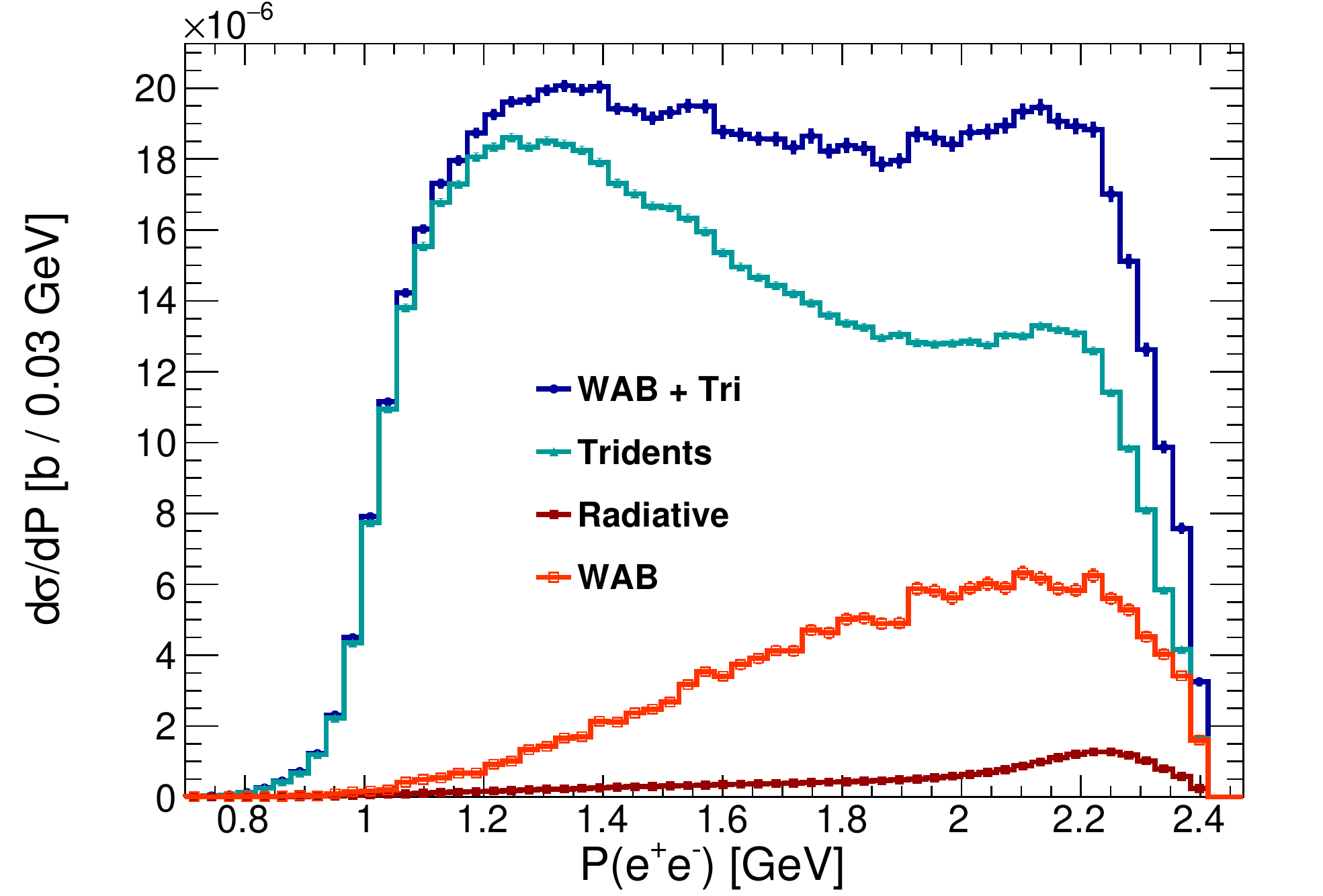}
    \caption{ Differential cross-section as a function of V0 momentum for different MC samples. Radiative tridents are presented by red markers, the WAB sample is represented by the orange-colored histogram, the trident sample is shown by the cyan histogram, and the blue histogram is the sum of WAB and tridents. }
\label{fig:PV0_MinCut}
\end{figure}

\Cref{fig:PV0_MinCut} shows the diﬀerential cross sections for the various physics processes that contribute to the event
 sample as a function of the V0 momentum.   Radiative tridents are peaked at high momenta, whereas
the full trident sample (which includes radiative and Bethe-Heitler tridents and their interference) and WABs are
 broadly enhanced at lower momenta. The sensitivity of the resonance search is proportional to the radiative 
 fraction, so a cut in the minimum V0 momentum that maximizes the ratio $N_{\rm rad}/\sqrt{N_{\rm tot}}$ is optimal. 
 For the resonance (vertex) search, this occurs at \SI{1.9}{GeV} (\SI{1.85}{GeV}).

 %%%%%%%%%%%%%%%%%%%%%%%%%%%%%%%%%%%%%%%%%%%%%%%% F I G U R E %%%%%%%%%%%%%%%%%%%%%%%%%%%%%%%%%%%%%%%%%%%%%%%%%%%%
 %%%%%%%%%%%%%%%%%%%%%%%%%%%%%%%%%%%%%%%%%%%%%%%% F I G U R E %%%%%%%%%%%%%%%%%%%%%%%%%%%%%%%%%%%%%%%%%%%%%%%%%%%%

Finally, a cut on the maximum V0 momentum reduces background from the cWABs, which extends beyond the beam
energy. For both the resonance and vertex searches, the maximum V0 momentum must be less than \SI{2.4}{GeV}.

\begin{figure}[!htb]
    \centering
     \includegraphics[width=0.45\tw]{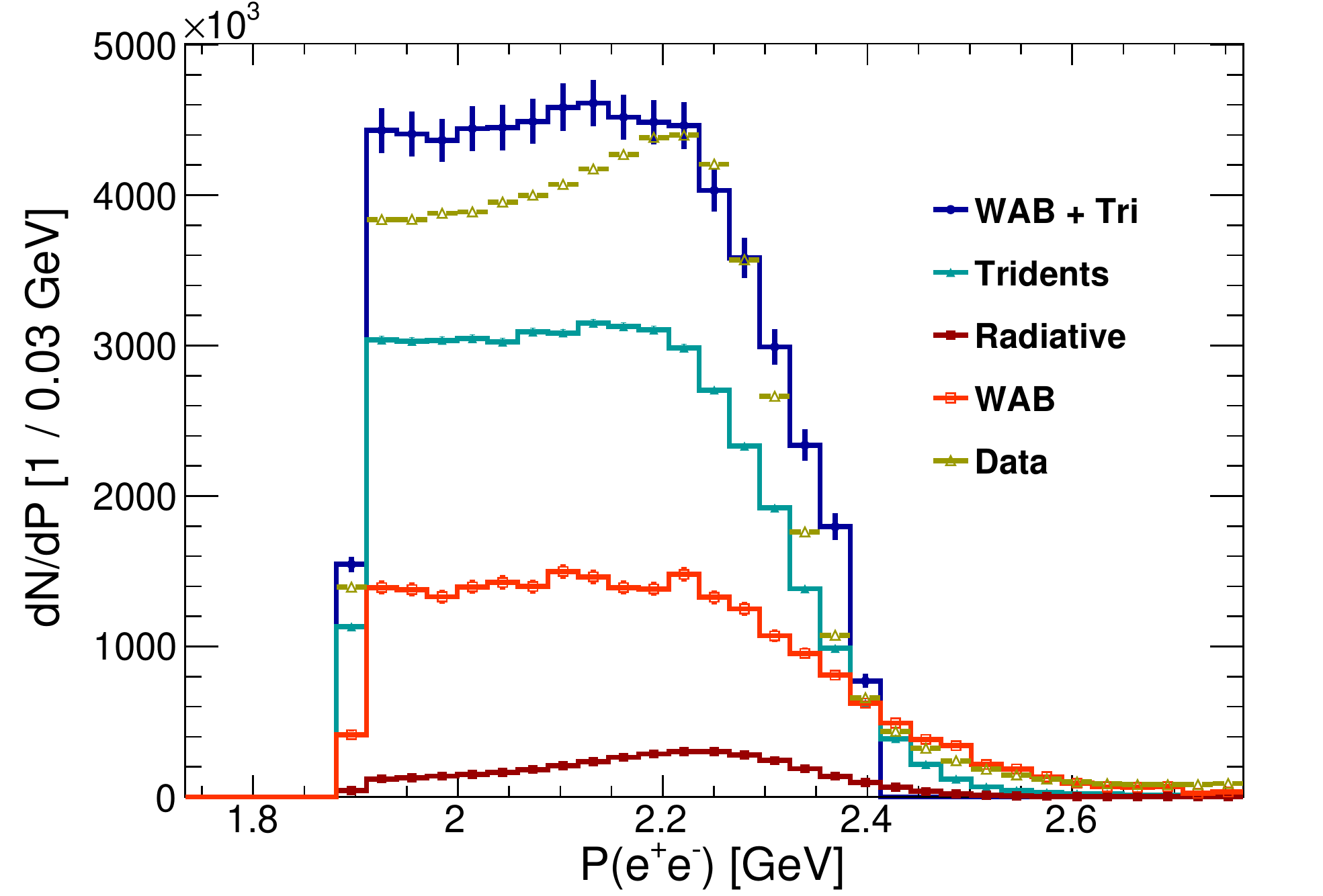}
    \caption{ Number of events as a function of V0 Momentum for the 10\% data sample,  shown in yellow,  and different MC samples. 
     Radiative tridents are presented by red markers, the WAB sample is represented by the orange-colored histogram, the trident sample is shown by the cyan histogram, and the blue histogram is the sum of WAB and tridents.}
\label{fig:PV0_MaxCut}
\end{figure}

\begin{figure}[!htb]
    \centering
     \includegraphics[width=0.4\tw]{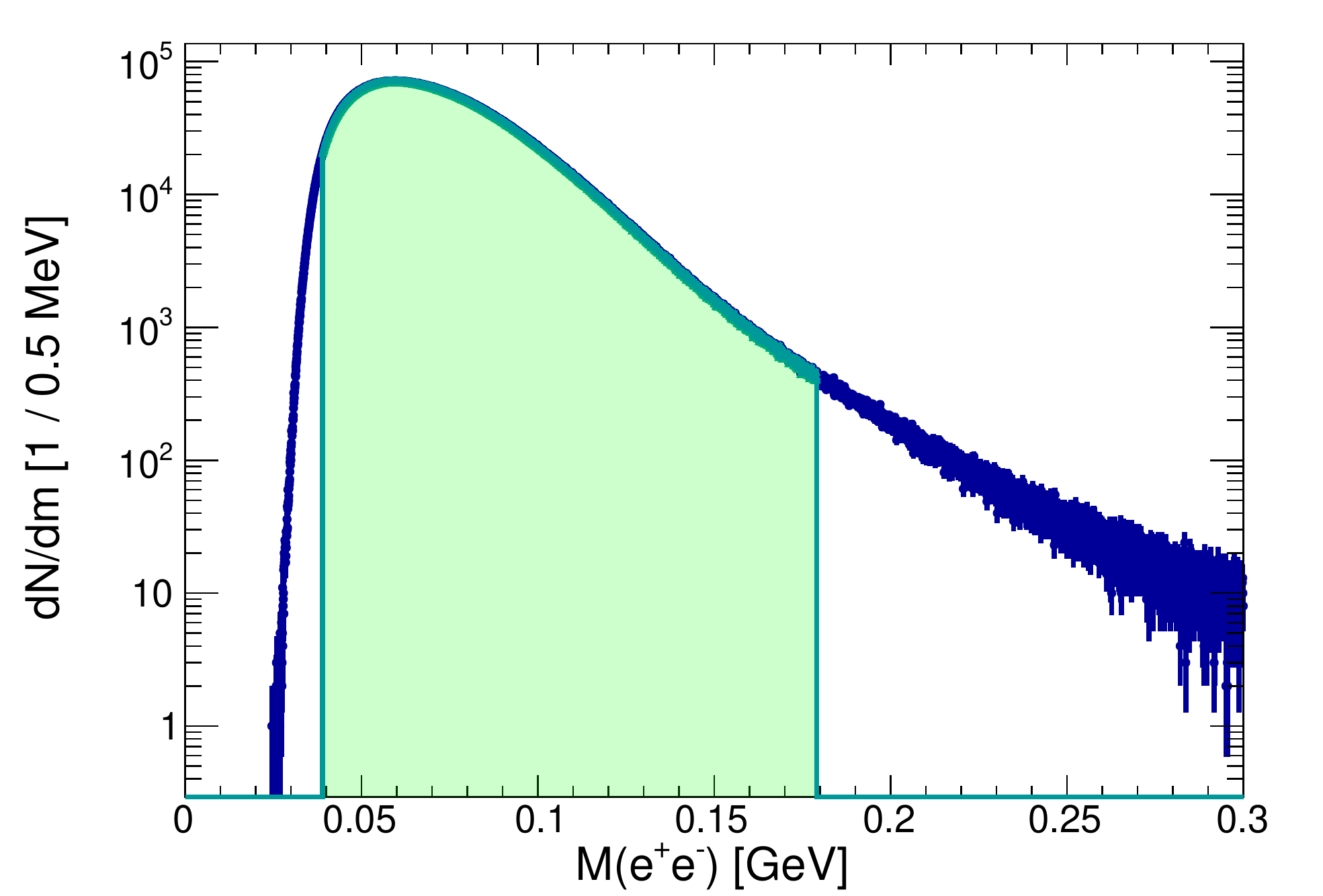}
%    \grinp[width=0.98\tw]{Figs/FinalMassDistribution.pdf}
    \caption{The mass distribution after all event selection cuts (described in the text). The highlighted area in green represents the range where the resonance search is performed.}
    \label{fig:FinalMass}
\end{figure}

\Cref{fig:PV0_MaxCut} compares the data with the Monte Carlo after all the above cuts except the cut on V0 momentum. The data and MC are in broad
 agreement, giving evidence that the sample composition is understood. At lower V0 momentum, the data fall
 below the MC, primarily because the trigger efficiency for low-energy clusters is not perfectly accounted for in the Monte Carlo. Momentum resolution
 eﬀects, also not perfectly accounted for, explain the data/Monte Carlo discrepancies at the high edge of the
 distribution. The invariant $\epem$ mass distribution for events passing these ﬁnal cuts is shown in \Cref{fig:FinalMass}. The
 highlighted region in green is the mass range where the resonance search was performed.  The mass range for the
 displaced vertex search is discussed in \Cref{sec:vtxAna}.

  %%%%%%%%%%%%%%%%%%%%%%%%%%%%%%%%%%%%%%%%%%%%%%%% F I G U R E %%%%%%%%%%%%%%%%%%%%%%%%%%%%%%%%%%%%%%%%%%%%%%%%%%%%
 %%%%%%%%%%%%%%%%%%%%%%%%%%%%%%%%%%%%%%%%%%%%%%%% F I G U R E %%%%%%%%%%%%%%%%%%%%%%%%%%%%%%%%%%%%%%%%%%%%%%%%%%%%

%%%%%%%%%%%%%%%%%%%%%%%%%% F I G U R E %%%%%%%%%%%%%%%%%%%%%%%%%%%%%%%
%\begin{widetext}
%\end{widetext}
%%%%%%%%%%%%%%%%%%%%%%%%%% F I G U R E %%%%%%%%%%%%%%%%%%%%%%%%%%%%%%%

%% file: radFrac.tex
\subsection{Sample Composition and Fraction of Radiative Rate} 
\label{sec:sampleCompsition}
While $\aprime$ events are primarily at high $x$, trident events cover the entire $x$ range and have a higher rate at low $x$.  The HPS detector accepts $\sim0.5 < x < 1$, and although events with $x < 0.8$ are not useful for $\aprime$ searches, they provide a high statistics sample for calibrations and sample composition studies.

It can be shown that the expected signal cross-section is \cite{AprimeFixedTargetTheory}:

\begin{equation}
    \left.\frac{\mathrm{d}\sigma_{A'}}{\mathrm{d}m} \right\vert_{m=m_{A'}} = \frac{3 \pi m_{A'} \epsilon^2}{2 N_{\mathrm{eff}} \alpha} \left.\frac{\mathrm{d}\sigma_{\gamma^*}}{\mathrm{d}m} \right\vert_{m=m_{A'}}
     \label{eq:sigRate}
\end{equation}
The luminosity, detector acceptance, and efficiency are factored out from this equation and it 
can be rearranged to give an equation to calculate the upper limit on $\epsilon^2$ via an 
upper limit on the signal rate. $N_{\mathrm{eff}}$ is the ratio of the sum of all branching 
ratios to the branching ratio of the electron-positron decay channel, and is
one for all masses in this search.  The differential cross section of the radiative trident process, 
$d\sigma_{\gamma^{*}}$, is taken at specifically the \Ap mass, and the notation indicating this will be dropped henceforth. This gives:

\begin{equation}
    \epsilon_{\mathrm{up}}^2 = \frac{2 \alpha N_{\mathrm{sig}}^{\mathrm{up}}}{3 \pi m_{A'} \frac{\mathrm{d}N_{\gamma^{*}}}{\mathrm{d}m} }
\end{equation}
where $N_{\mathrm{sig}}^{\mathrm{up}}$ is the upper limit on the number of signal events observed in the data. \Cref{sec:bh_methodology} discusses in detail how this upper limit is set. The focus of this section is to present how the differential rate $\mathrm{d}N_{\gamma^{*}}/\mathrm{d}m$ is evaluated in the analysis.

The differential $\gamma^{*}$ rate is only defined theoretically and is not something that can be directly extracted 
from the data. We start by defining the radiative fraction as:
\begin{equation} \label{eq:radFracDef}
    f_{\mathrm{rad}} = \frac{ \frac{\mathrm{d}N_{\gamma^{*}}}{\mathrm{d}m}}{ \frac{\mathrm{d}N_{\rm bkg}}{\mathrm{d}m}} = \frac{ \frac{\mathrm{d}N_{\gamma^{*}}}{\mathrm{d}m}}{ \frac{\mathrm{d}N_{\rm tri}}{\mathrm{d}m} + \frac{\mathrm{d}N_{\rm wab}}{\mathrm{d}m} }.
\end{equation}
where $N_{\rm tri}$ and ${N}_{\rm wab}$ are the number of trident and WAB events, respectively.  

Using this definition the equation for $\epsilon_\textrm{up}^2$ can be rewritten as:
\begin{equation} \label{eq:eps2full}
    \epsilon_\textrm{up}^2 = \frac{2 \alpha N_{\mathrm{sig}}^{\mathrm{up}}}{3 \pi m_{A'} f_{\mathrm{rad}}. \frac{\mathrm{d}N_{\rm bkg}}{\mathrm{d}m} }
\end{equation}
It is important to note that the differential background rate in the denominator of \Cref{eq:radFracDef} is with respect to the reconstructed mass, while the numerator is with respect to the true $\gamma^{*}$ mass. This definition is chosen so \Cref{eq:eps2full} will use the true mass of the signal while the differential background rate is with respect to the reconstructed mass. This also corrects the systematic uncertainty from events migrating into other mass bins due to resolution effects. The differential background rate in \Cref{eq:eps2full} is extracted directly via the fit to data described in \Cref{sec:bh_methodology}. The radiative fraction \Cref{eq:radFracDef} is constructed entirely via Monte Carlo simulations, so the differential background rate with respect to the reconstructed mass in this equation uses the simulated background rate. Finally, reconstruction of e+e- mass for the signal Monte Carlo results in both a peaking component when the radiative pair is used, but also a diffuse component in the case that the recoiling electron is incorrectly associated with the positron.  Since the signal model in our fit corresponds only to the peaking component, only the contribution from reconstructed events where the e+e- is the radiative pair are used for the numerator of \Cref{eq:radFracDef}.

\begin{figure}[tb]
 \centering
 \grinp[width=0.45\tw]{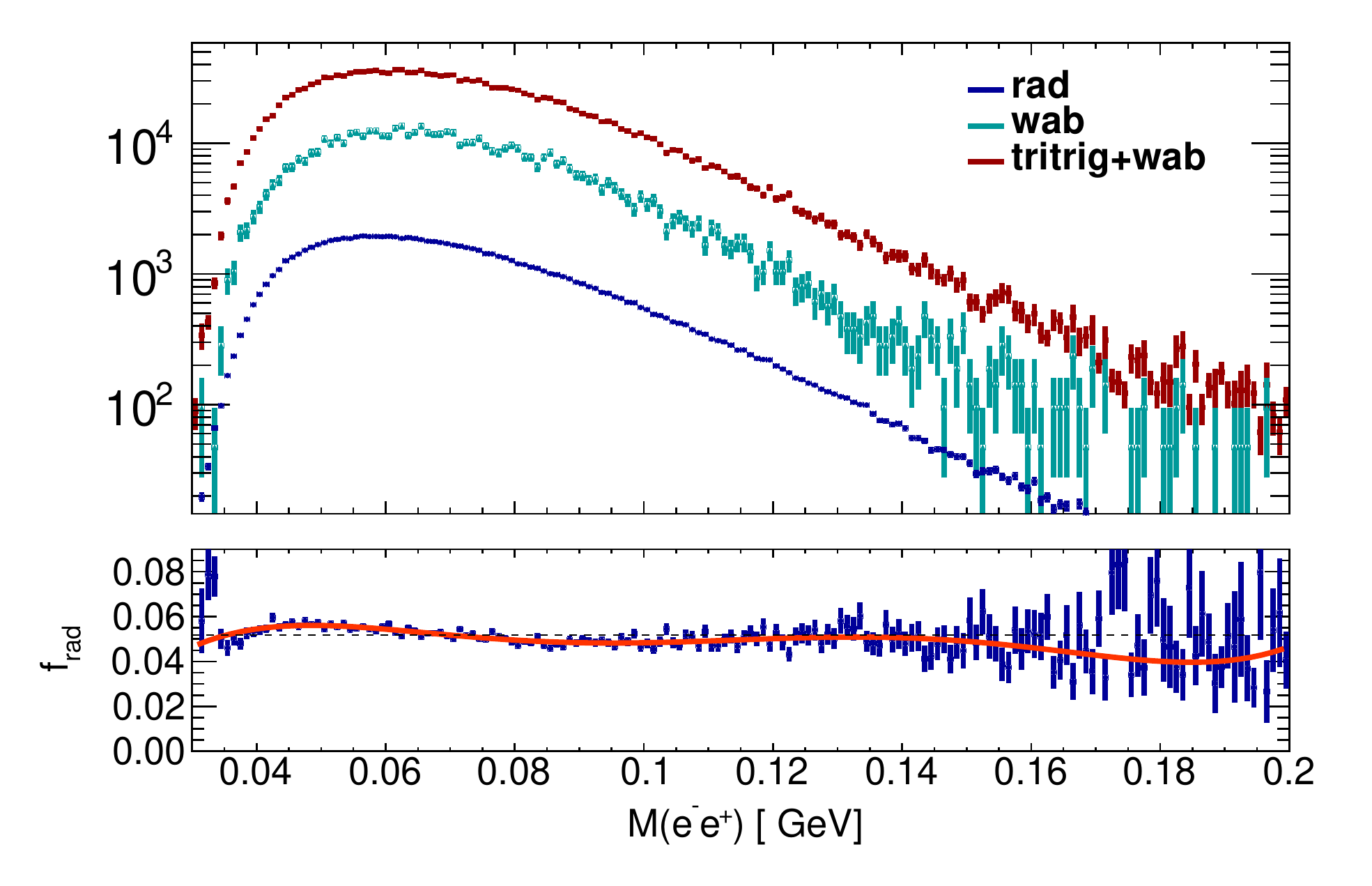}
\caption{The differential rates in units of \si{\per\mega\eV} for all the MC samples that go into the radiative fraction for the target-constrained vertex fit used in the resonance search analysis. The radiative component is a function of the true mass of the MC generated $\gamma^*$ for the event. The WAB and trident components (labeled "tritrig" in plot) are a function of the invariant mass of the selected V0 candidate. The bottom panel shows  the radiative fraction. The red line is the fifth order polynomial fit, and the dotted line is the constant fit.}
 \label{fig:radFrac}
\end{figure}

\Cref{fig:radFrac} shows the radiative fraction versus invariant mass for the resonance search selection. The corresponding plot for the displaced vertex search is shown in \Cref{sec:vtxAna}.

%% file: massReso.tex
\subsection{Invariant Mass Resolution} \label{sec:massresolution}
Searching for a resonance peak on top of a large background requires accurate knowledge of its width. The width of the expected \Ap signal is
dominated by the experimental resolution, so it is critical that the mass resolution is well understood. The mass resolution for observed 
\Mlr events is compared to Monte Carlo simulations, which are then tuned to get agreement. This tuned Monte Carlo is then used to derive 
the expected mass resolution for all masses of interest to the analyses. These steps are detailed in this section. 

The Monte Carlo is used to evaluate the mass resolution using  simulations of \Ap signal at several ﬁxed mass points. These generated signal events are processed through the GEANT4 simulation chain with a full detector model. Since the natural width of the \Ap is signiﬁcantly smaller than 
 the detector resolution (by more than a factor of 1000), 
 the observed width of the signal shape is determined solely by the mass resolution. 

\subsubsection{Using the \Mlr Resonance to Calibrate the MC Mass Resolution}

\begin{figure}
    \centering
    \grinp[width=0.45\tw]{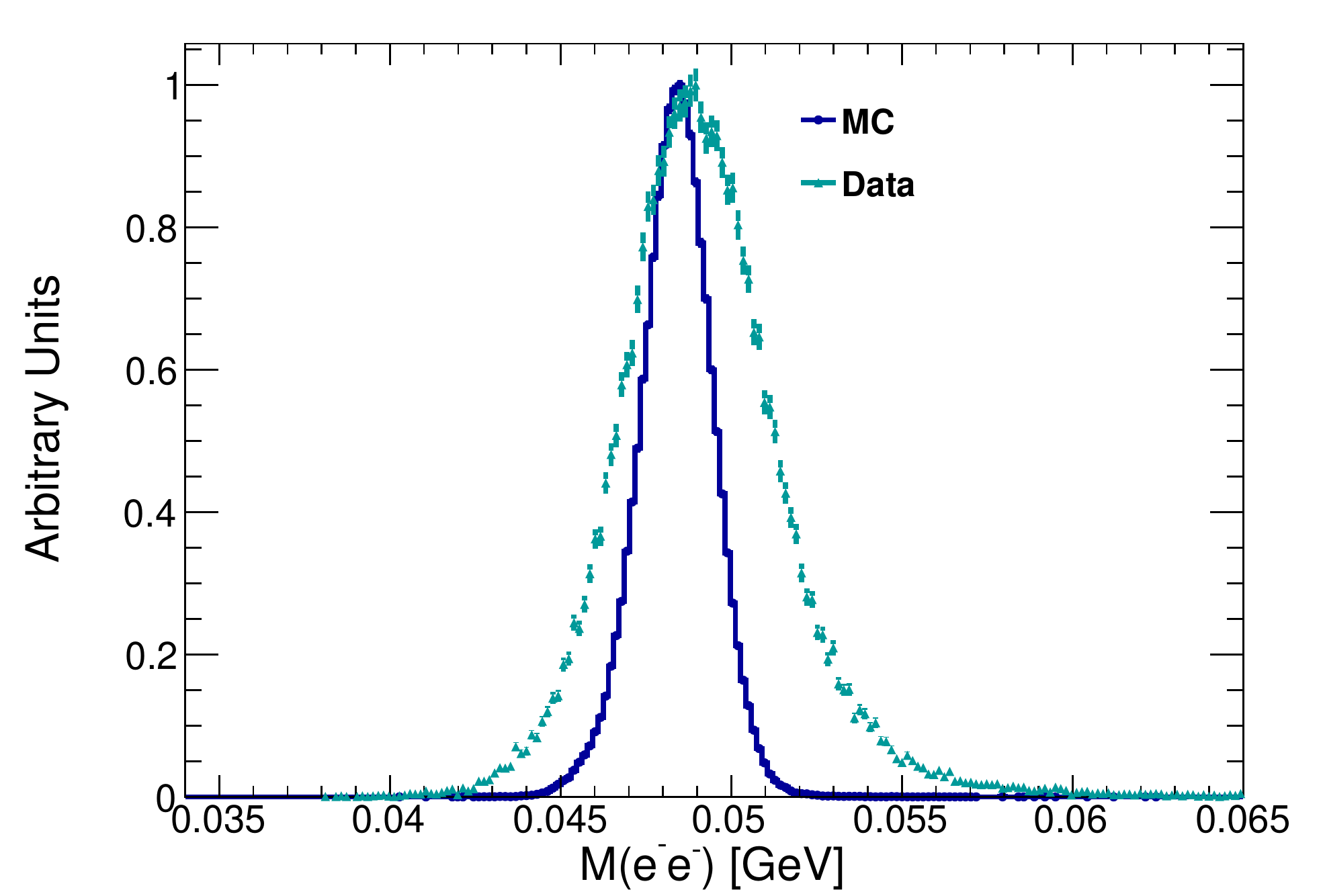} 
%    \grinp[width=0.45\tw]{Figs/Mlr_Mass_Data_MC.pdf} 
    \caption{Mass distribution of \Mlr events from data and MC overlaid. Histograms are scaled to have the same maximum value. The cyan line represents data, while the blue line represents MC.}
    \label{fig:MlrMass_MC_Data}
\end{figure}

The \Mlr process $\mathrm{e^{-}e^{-}\rarr e^{-}e^{-}}$ provides a direct measurement of the mass resolution  
since the center of mass energy of a beam electron and an electron at rest is equal to the invariant mass of the ﬁnal state electrons (called the \Mlr mass). A beam energy of \SI{2.3}{GeV} will have a \Mlr mass of \SI{48.5}{MeV}. 
Just like the \Ap process, the observed width of the \Mlr invariant mass is dominated by detector resolution. 
Furthermore, the \Mlr and \Ap final states both have particles of equal mass which will multiple scatter in the detector material essentially identically. The mass resolution for $\mathrm{e^{+}e^{-}}$ and $\mathrm{e^{-}e^{-}}$ ﬁnal states is expected to be nearly equivalent at the same invariant mass. 
\Cref{fig:MlrMass_MC_Data} shows the $\mathrm{e^{-}e^{-}}$ invariant mass distributions of \Mlr events in the data (cyan) and MC (blue).  These histograms have been scaled to have the same maximum bin value. Note the mass resolution in data is about a factor of two worse than in MC.

%%%%%%%%%%%%%%%%%%%%%%%%%%%%%%%%%%%%%%%%%%%%%%%%%%%% F I G U R E %%%%%%%%%%%%%%%%%%%%%%%%%%%%%%%%%%%%%%%%%%%%%%%%%%%%%%
%%%%%%%%%%%%%%%%%%%%%%%%%%%%%%%%%%%%%%%%%%%%%%%%%%%% F I G U R E %%%%%%%%%%%%%%%%%%%%%%%%%%%%%%%%%%%%%%%%%%%%%%%%%%%%%%

The \Mlr mass is written in terms of the momenta of the final state particles ($P_1$ and $P_2$) and the angle theta between them ($\theta$), neglecting the mass squared terms: 
 \begin{equation}
    M = 2\sqrt{P_{1}P_{2}}\cdot \sin\frac{\theta}{2}.
\end{equation}
%%%%%%%%%%%%%%%%%%%%%%%%%%%%%%%%%%%%%%%%%%%%%%%%%%%%%% E Q U A T I O N %%%%%%%%%%%%%%%%%%%%%%%%%%%%%%%%%%%%%%%%%%%%%%%%%%%%%%%%%%%%%%%%%%
This formula demonstrates the source of the discrepancy in the \Mlr mass resolutions in data and MC is modeled as discrepancies in the momentum resolution and/or angular resolution. 

\subsubsection{Momentum Resolutions With Full Energy Electrons}
\label{sec:momSmear}

\begin{figure}[!htb]
    \centering
     \grinp[width=0.45\tw]{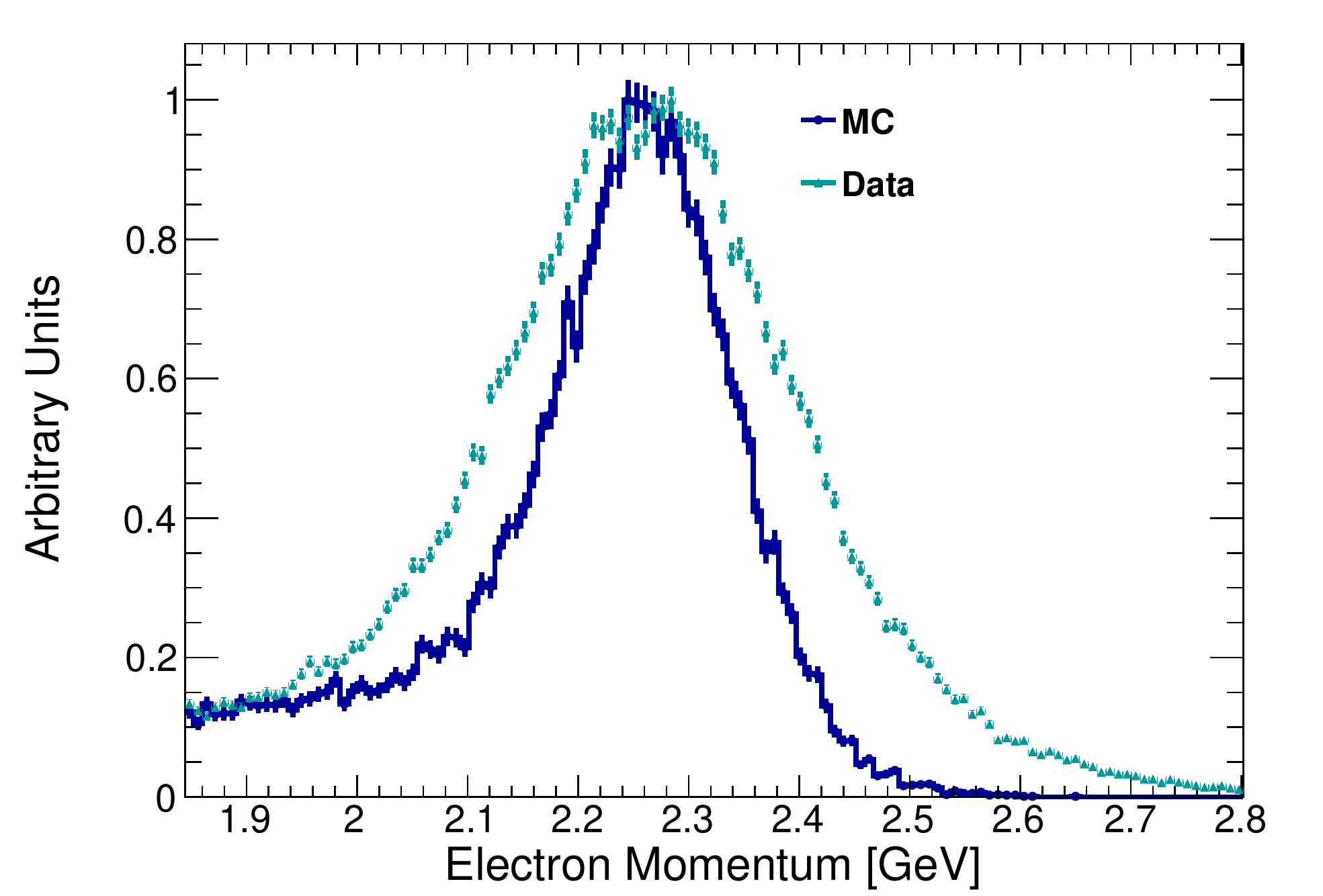} 
%    \grinp[width=0.45\tw]{Figs/FEE_P_DataMC_Overlaid.pdf}
    \caption{FEE momentum distributions for 6 hit tracks in the bottom half of the tracker. The cyan line represents data and the blue line represents the un-smeared MC.}
    \label{fig:FEE_Bot5hitsDataMC}
\end{figure}

Elastically scattered full beam energy electrons (FEEs) provide an experimental check of the momentum scale and resolution. Since the electron is so light compared to a tungsten nucleus, it loses nearly zero energy in elastic interactions.  Consequently, elastically scattered beam electrons are expected to appear as a single peak in the electron momentum distribution. The width of this peak is a measurement of the momentum resolution at the beam energy. As it is natural to expect better momentum resolution for 6 hit tracks compared to 5 hit tracks, these resolutions are measured 
separately. The top and bottom tracks are also separated because the two detector halves are not expected to have systematically identical misalignments. \Cref{fig:FEE_Bot5hitsDataMC} shows FEE peaks for 
6 hit negative tracks in the bottom half of the tracker, where the cyan line is from data and the blue is MC. 
In this particular case, the data resolution is a factor of 1.6 times worse than the MC resolution. Over all the categories, the momentum resolution 
in data is worse than that in MC by factors ranging from 1.3 to 1.6. 
 
Adding additional momentum smearing can bring the MC and data mass into agreement. The smearing coefficients for  
each MC category (bot/top/5-hit/6-hit) are parameterized by: 
\begin{equation}
   \displaystyle \Sigma_{\mathrm{smear}} \equiv \frac{\sigma_{\mathrm{smear}}}{P_{\mathrm{MC}}} = \sqrt{ \left(\frac{\sigma_{\mathrm{data}}}{\mu_{\mathrm{data}}}\right)^{2} - \left(\frac{\sigma_{\mathrm{MC}}}{\mu_{\mathrm{MC}}}\right)^{2} }.
   \label{eq:smeatEquation}
\end{equation}
where $\sigma_{\mathrm{smear}}$ is the factor by which an MC electron with a given momentum ($P_{\mathrm{MC}}$) is smeared.  The data and MC FEE momentum resolutions are $\sigma_{\mathrm{data}}$ and $\sigma_{\mathrm{MC}}$, respectively.
Finally, $\mu_{\mathrm{data}}$ and $\mu_{\mathrm{MC}}$ are the mean values of the FEE momentum peaks.  The momentum resolution discrepancy between data and 
 MC is assumed to be independent of momentum. This is expected since $\mathrm{\sigma(p)/p}$ is nearly constant over all relevant momenta, being 
multiple scattering dominated.

\begin{figure}[!htb]
    \centering
     \grinp[width=0.45\tw]{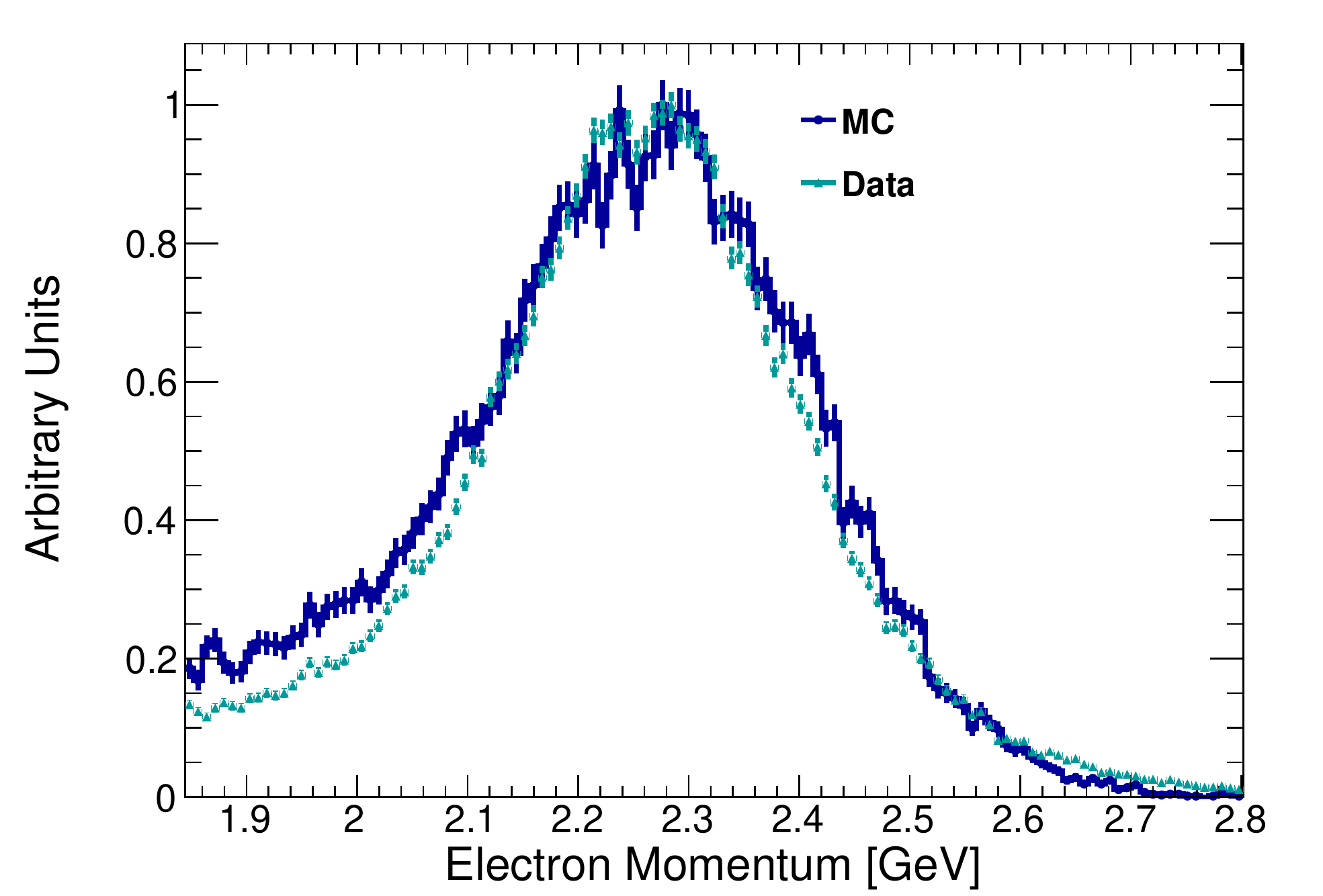} 
    \caption{FEE momentum distributions for 6 hit tracks in the bottom half of the tracker. The cyan line represents data and the blue line represents the smeared MC momentum.}
    \label{fig:Mom_MC_Data_ScCm}
\end{figure}

The MC tracks are then smeared with the appropriate $\Sigma$, depending on the category.  
\Cref{fig:Mom_MC_Data_ScCm} compares the smeared MC momentum distribution in blue with data in cyan. The mean of the MC distribution has been shifted slightly so that the peaks overlap for ease of comparison. The matching between MC and data for other categories is comparable. 
In all cases, there is good agreement between data and the smeared MC distributions. Accordingly, smearing is applied to all 
the tracks from the \Mlr and \Ap MC samples. 

%%%%%%%%%%%%%%%%%%%%%%%%%% F I G U R E %%%%%%%%%%%%%%%%%%%%%%%%%%%%%%%
%%%%%%%%%%%%%%%%%%%%%%%%%% F I G U R E %%%%%%%%%%%%%%%%%%%%%%%%%%%%%%%

%%%%%%%%%%%%%%%%%%%%%%%%%% F I G U R E %%%%%%%%%%%%%%%%%%%%%%%%%%%%%%%
%%%%%%%%%%%%%%%%%%%%%%%%%% F I G U R E %%%%%%%%%%%%%%%%%%%%%%%%%%%%%%%

\subsubsection{Recalculated Mass after MC Momentum Smearing}\label{sec:RecalculMass}

The \Mlr mass is recalculated using the smeared electron momenta. The mass taking into account the smeared momenta is 
 expressed in terms of the unsmeared mass, using \Cref{eq:MassSmear}. 

\begin{equation}
M(\mathrm{ee})^{\mathrm{smear}} = \sqrt{\frac{P_{1,\mathrm{smear}}}{P_{1,\mathrm{rec}}}\frac{P_{2,\mathrm{smear}}}{P_{2,\mathrm{rec}}}}\cdot M(\mathrm{ee})
\label{eq:MassSmear}
\end{equation}

Here, $M(\mathrm{ee})^{\mathrm{smear}}$ is the smeared mass, $P_{1,\mathrm{smear}}$ ($P_{2,\mathrm{smear}}$) is the smeared momentum of 1st (2nd) particle, $P_{1,\mathrm{rec}}$ 
 ($P_{2,\mathrm{rec}}$) is the reconstructed (unsmeared) momentum of the 1st (2nd) particle, $M(\mathrm{ee})$ is the unsmeared target constrained mass.  

\begin{figure}[!htb]
    \centering
     \grinp[width=0.45\tw]{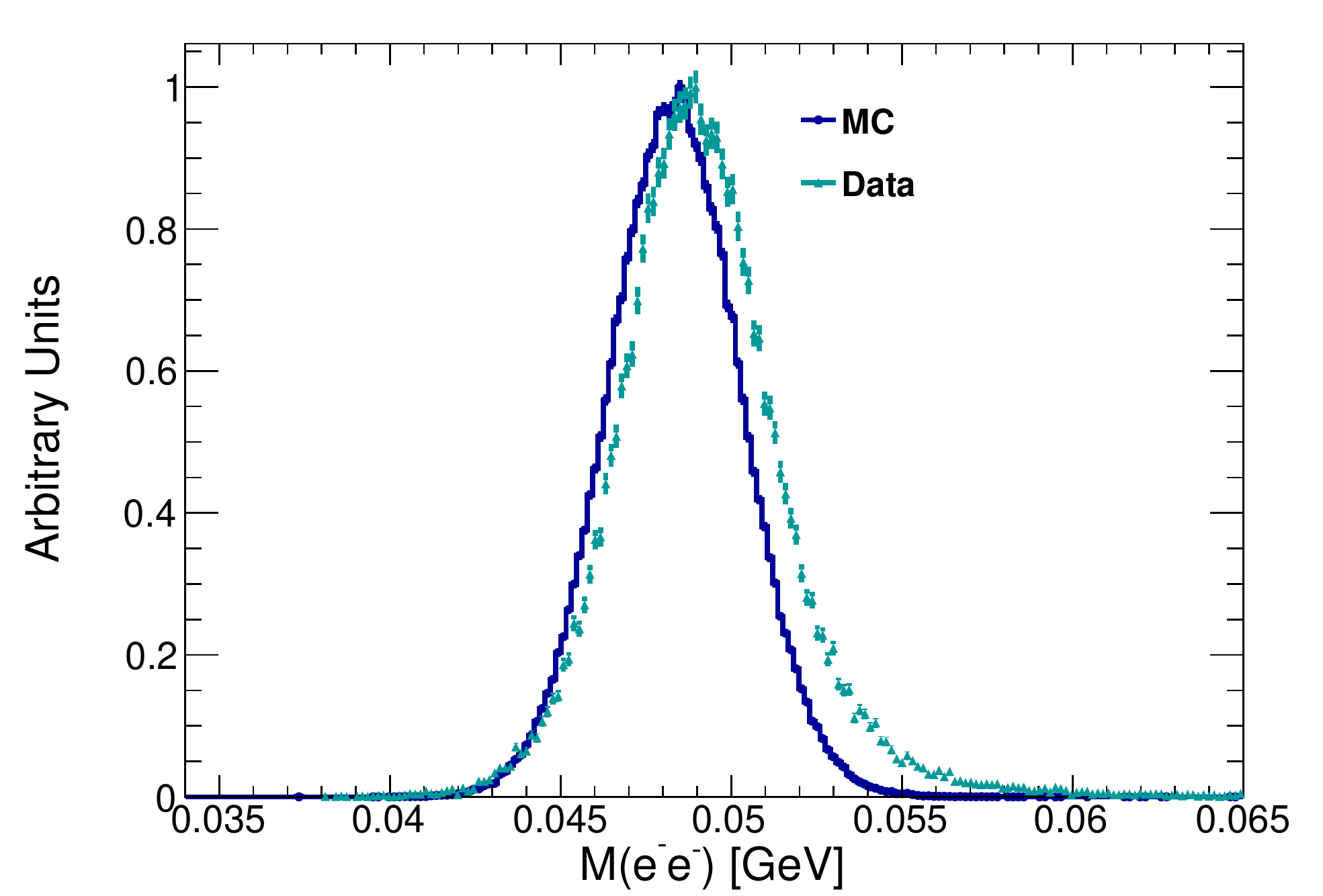} 
    \caption{Smeared mass distribution of \Mlr MC events (blue), and \Mlr events in data (cyan)}
    \label{fig:MlrScSmMass_MC}
\end{figure}

 After smearing the mass with \Cref{eq:MassSmear}, the smeared mass of \Mlr events shown in \Cref{fig:MlrScSmMass_MC} (blue) 
 is obtained.
 Incorporating smearing, the mass resolution discrepancy is reduced from about a factor of 2 to about 6\%.

%%%%%%%%%%%%%%%%%%%%%%%%%% F I G U R E %%%%%%%%%%%%%%%%%%%%%%%%%%%%%%%
%%%%%%%%%%%%%%%%%%%%%%%%%% F I G U R E %%%%%%%%%%%%%%%%%%%%%%%%%%%%%%%

\subsubsection{Parametrizing the \texorpdfstring{\Ap}{A'} Mass Resolution}\label{sec:ApMassParametrize}

\begin{widetext}
\begin{figure*}[!htb]
    \centering
    \grinp[width=0.99\tw]{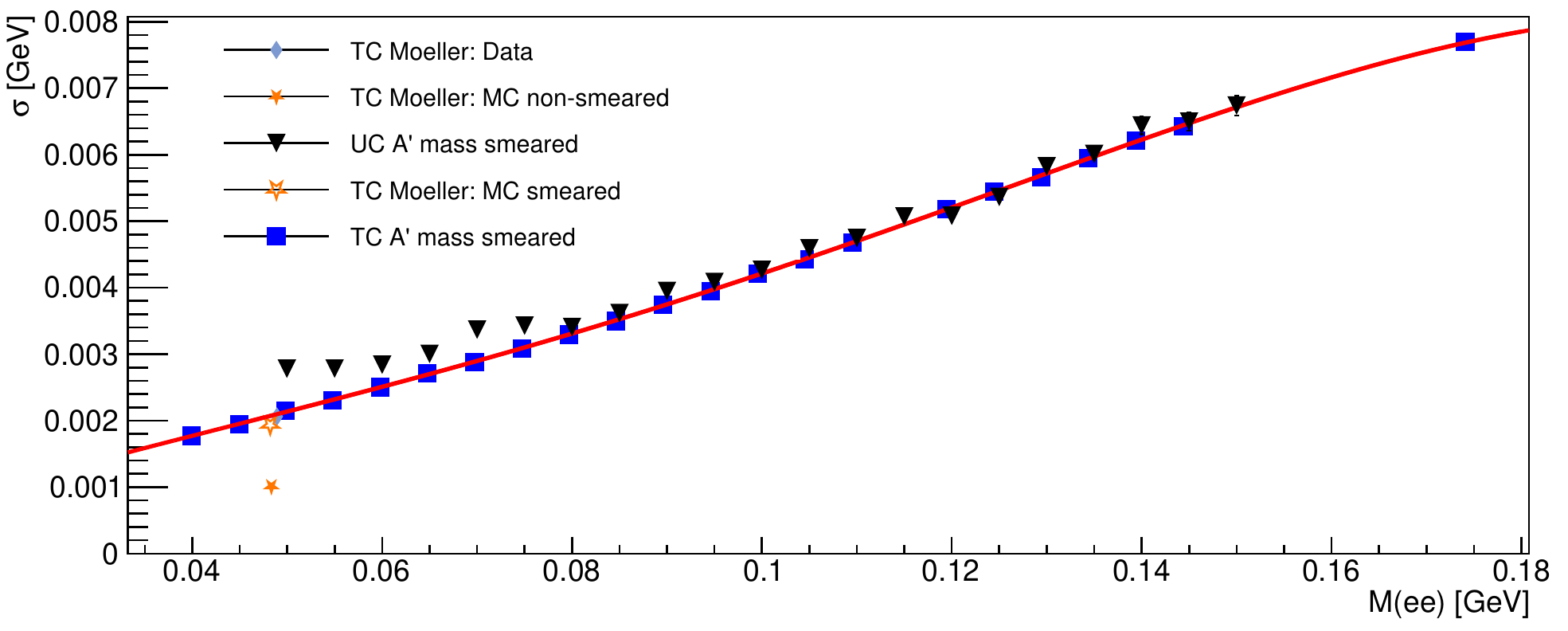}
    \caption{Mass resolutions. Filled (empty) star markers represent the unsmeared (smeared) target constrained MC \Mlr mass resolution. Diamond-shaped markers show the target constrained \Mlr mass resolution from data. 
    Filled squares show the smeared target constrained \Ap mass resolutions, while the filled triangles represent unconstrained \Ap mass resolutions. 
    The solid curve in red is the fit over target constrained \Ap mass resolutions.}
    \label{fig:MassResolParam}
\end{figure*}
\end{widetext}

We study the expected mass resolution for $\mathrm{A^{\prime}}$s of various masses using a collection of simulated \Ap samples with masses ranging from \SI{40}{MeV} to \SI{175}{MeV}, 
with the momenta of the $\mathrm{e^{-}}$ and $\mathrm{e^{+}}$ tracks smeared with the procedure described above. 
 The smeared mass distributions of all the \Ap MC samples are ﬁt with a Gaussian function to obtain the mass resolutions. These smeared \Ap mass resolutions and the 
\Mlr mass resolutions are shown in \Cref{fig:MassResolParam}. The vertical axis shows the mass resolution, and the horizontal axis represents the mean value, the mass, of the Gaussian fit.

%%%%%%%%%%%%%%%%%%%%%%%%%% F I G U R E %%%%%%%%%%%%%%%%%%%%%%%%%%%%%%%
%%%%%%%%%%%%%%%%%%%%%%%%%% F I G U R E %%%%%%%%%%%%%%%%%%%%%%%%%%%%%

%% file: ResonanceSearch.tex
%%%%  June 9, 2022 --- MG
%%%%  included replacement text from JJ
%%%% old text is commented out in-line

\section{Resonance Search} \label{sec:bumphunt} 

%This section describes the resonance search analysis.
%Event selection has been described above in \Cref{sec:selection}. 
%There the invariant $\epem$ mass distribution for events passing these final cuts was shown in \Cref{fig:FinalMass}. 
%The region  highlighted in green is the mass range where the resonance search was performed.
This section describes the resonance search technique, 
systematic uncertainties, and final physics results. All Monte Carlo momenta and masses used 
in this section are smeared according to the procedure described in \Cref{sec:momSmear}.

\subsection{Statistical Analysis} \label{sec:StatAnalysis}
If an \Ap exists within the acceptance of HPS, it will manifest itself as an excess in the $\textrm{e}^{+}\textrm{e}^{-}$ invariant mass spectrum (a ``bump'').
The excess is expected to take the form of a Gaussian centered at the mass of the \Ap ($m_{A'}$) with a width equal to the mass resolution for that point as discussed in \Cref{sec:massresolution}.

However, since the mass of the \Ap is not known, it is necessary to search for it at all possible masses. To do this, 
HPS employs a resonance search over a mass range of \SI{39}{MeV} to \SI{179}{MeV}, in steps of \SI{1}{MeV}, using a maximum likelihood 
fit ratio to test the background-only hypothesis at each mass hypothesis. The full methodology of this process is discussed in detail in this section.

\subsubsection{Resonance Search Methodology}\label{sec:bh_methodology}
First, a fit window is selected centered on each mass hypothesis. The width of this window is chosen carefully so as not to introduce a bias in the signal yield and to minimize the signal yield uncertainty due to the background shape uncertainty. An exception occurs when the mass hypothesis is near the edge of the invariant mass distribution, and the fit window extends into a region where there are no reconstructed events. In these cases, the window is shifted such that the lower (upper) edge is at the lowest (highest) mass event, which results in the window no longer being centered on the mass hypothesis.

The probability density function for this window is defined by \Cref{eq:windowProbDensity}.
\begin{equation}
P(m_{\mathrm{e}^{+}\mathrm{e}^{-}}) = \mu \cdot \phi(m_{\mathrm{e}^{+}\mathrm{e}^{-}}|m_{A'}, \sigma_{m_{A'}}) + 10^{\mathrm{L}_{N}(m_{\mathrm{e}^{+}\mathrm{e}^{-}}|\vec{t})}
\label{eq:windowProbDensity}
\end{equation}
Where $m_{\mathrm{e}^{+}\mathrm{e}^{-}}$ is the $\mathrm{e}^{+}\mathrm{e}^{-}$ invariant mass, $\mu$ is the signal yield, $\phi(m_{\mathrm{e}^{+}\mathrm{e}^{-}}|m_{A'}, \sigma_{m_{A'}})$ 
is a Gaussian probability distribution describing the signal shape, and $\mathrm{L}_{N}(m_{\mathrm{e}^{+}\mathrm{e}^{-}}|\vec{t})$ is a Legendre polynomial 
of the first kind of order $N$ with coefficients (also the nuisance parameters) $\vec{t}=\langle t_0,t_1,...,t_N\rangle$ used as the background model. We used order 5 polynomials at low mass and order 3 above \SI{66}{MeV}. The fit window width is an integer multiple of the mass resolution varying from 6 to 10 depending on the mass.

\begin{figure}[!htb]
    \centering
    \includegraphics[width=0.48\tw]{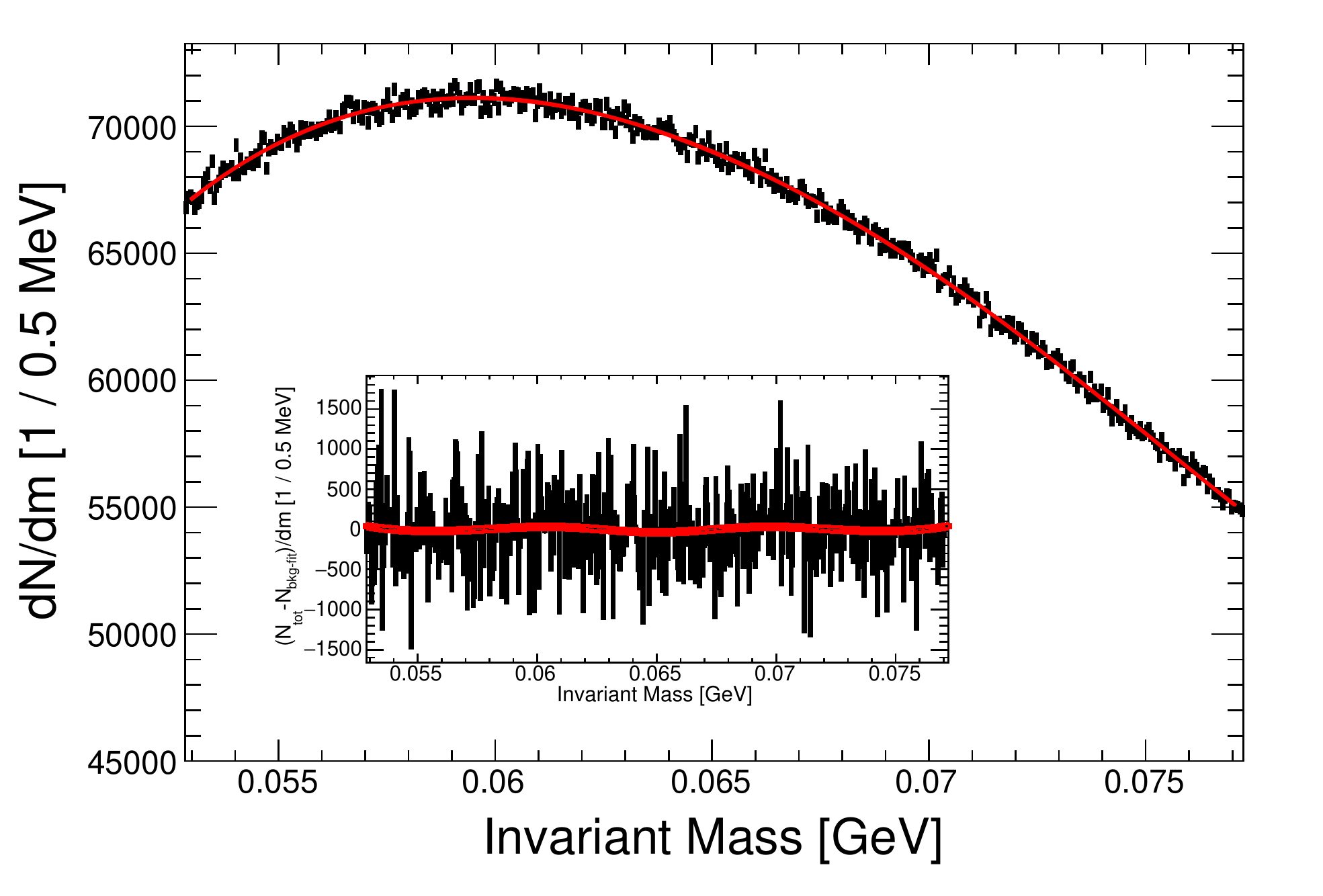}
    \caption{The invariant mass distribution in an example fit window with data (black) with the background-only fit overlaid (red).  Inset:  The data subtracted by the background-only fit (black) and the signal-plus-background fit subtracted by the background only fit (red).  
}
\label{fig:exMassSlice}
\end{figure}

An example fit window is shown in Figure \ref{fig:exMassSlice} centered at a mass hypothesis of \SI{65}{MeV} with a resolution of $\SI{2.7}{MeV}$. The Figure also shows the raw invariant mass distribution from data overlaid with the fit described above for background only.  The inset to Figure  \ref{fig:exMassSlice} shows the data after subtraction of the background fit overlaid with the signal-plus-background fit, also subtracted by the background-only fit.    
 %%%%%%%%%%%%%%%%%%%%%%%%%%%%%%%%%%%%%%%%%%%%%%%% F I G U R E %%%%%%%%%%%%%%%%%%%%%%%%%%%%%%%%%%%%%%%%%%%%%%%%%%%%

A typical t-test is performed at each mass as a search for evidence of a bump from some potential signal as discussed in \cite{BumpHuntMath}. We use a test statistics, similar to that used in\cite{Aad:2012an}. Since we consider more than one mass hypothesis, expected random fluctuations will produce a lower p-value somewhere in the search space as we search more masses. To account for this, it is necessary to estimate a correction due to this effect, commonly known as the ``look-elsewhere effect''. If all search regions were independent of each other this correction could simply be approximated for p-values much less than 1 by:
\begin{equation*}
    p_{\mathrm{global}} = N_{\mathrm{reg}} p_{\mathrm{local}}
\end{equation*}
where $N_{\mathrm{reg}}$ is the number of independent search regions investigated \cite{lookElsewhereEffect}. In this case, since the raster size of the search is \SI{1}{\mega\eV} and this is less than the mass resolution, the mass hypotheses are not independent; therefore, $N_{\mathrm{reg}}$ is not simply the number of mass points in the search. The number of search regions is approximated via
\begin{equation*}
    N_{\mathrm{reg}} \approx \frac{W}{\sigma_{\mathrm{ave}}}
\end{equation*}
where $W$ is the width of the full search window in mass and $\sigma_{\mathrm{ave}}$ is the average mass resolution in the window. It is found for this search that $N_{\mathrm{reg}} \approx 30$. A summary of the search p-values is shown in \Cref{fig:p_ValueWithSystematics}, from which we conclude we do not have evidence of a resonance. We then set an upper limit on the signal yield using a 95\% $\mathrm{CL}_s$ limit as described in \cite{BumpHuntMath} and \cite{CLsLimit}.

\subsection{Systematic Uncertainties}
\label{sec:Uncertainties}
There are two categories of uncertainties in this analysis: the uncertainty of our estimate of the mass resolution and that in estimating the radiative fraction. The two main contributors to the mass resolution uncertainty are our understanding of the target position and the momentum resolution of the apparatus. We estimate the uncertainty due to the mass resolution by varying the smearing coefficients extracted to replicate the mass resolution observed at the \Mlr  mass according to their statistical uncertainties. We simulate the experiment with the target position at $\pm \SI{0.5}{mm}$ and compare the resulting mass distributions. We then add the two uncertainties in quadrature at each mass independently and choose the largest uncertainty across the entire spectrum, which is 3.4\%. We account for this uncertainty by performing the final fit to the data 10,000 times while varying the signal shape width by this amount and selecting the 84\% quantile of the results.

The uncertainty of the radiative fraction has two contributions, from mismodeling the detector in MC and from uncertainties in the cross-sections used to scale the rate of each of the components of the radiative fraction. Efficiencies, momentum resolution, and acceptance of the final selection were varied in MC simulations to study the detector mismodeling uncertainty contribution. It was found that these effects introduce an uncertainty less than 1\% on the radiative fraction. The first component of uncertainty from cross-section scaling is from the uncertainty in their evaluation by MadGraph, which we evaluated to be roughly 7\% in total. The last component of uncertainty comes from our modeling of the rate of accidental track coincidences.  After adding this in quadrature with the uncertainty from our evaluation of cross-sections the total uncertainty on the radiative fraction is determined to be at most 7.4\%. This is accounted for by simply scaling the radiative fraction down by this amount. 

\subsection{Results}
\label{sec:Results}
\begin{figure}[!htb]
    \centering
    \includegraphics[width=0.48\tw]{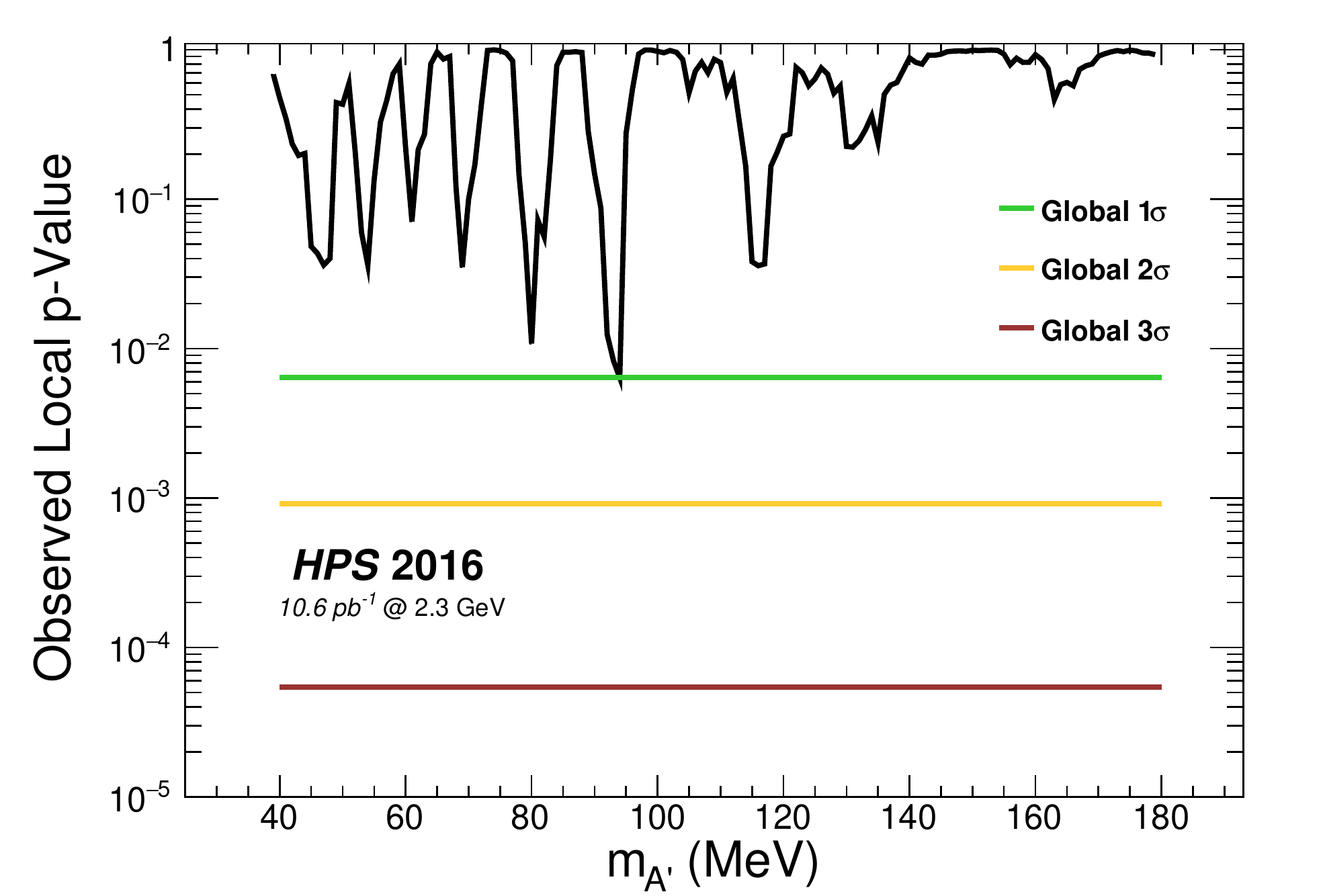}
    \caption{ The local p-values produced by a resonance search analysis of 100\% of the HPS 2016 data set. Here, the green line represents the global $\mathrm{1\sigma}$ threshold, the orange line represents the global $\mathrm{2\sigma}$ threshold, and the red line represents the global $\mathrm{3\sigma}$ threshold. Mass resolution systematics are included in this plot.
}
\label{fig:p_ValueWithSystematics}
\end{figure}

\begin{figure}[!htb]
    \centering
    \includegraphics[width=0.48\tw]{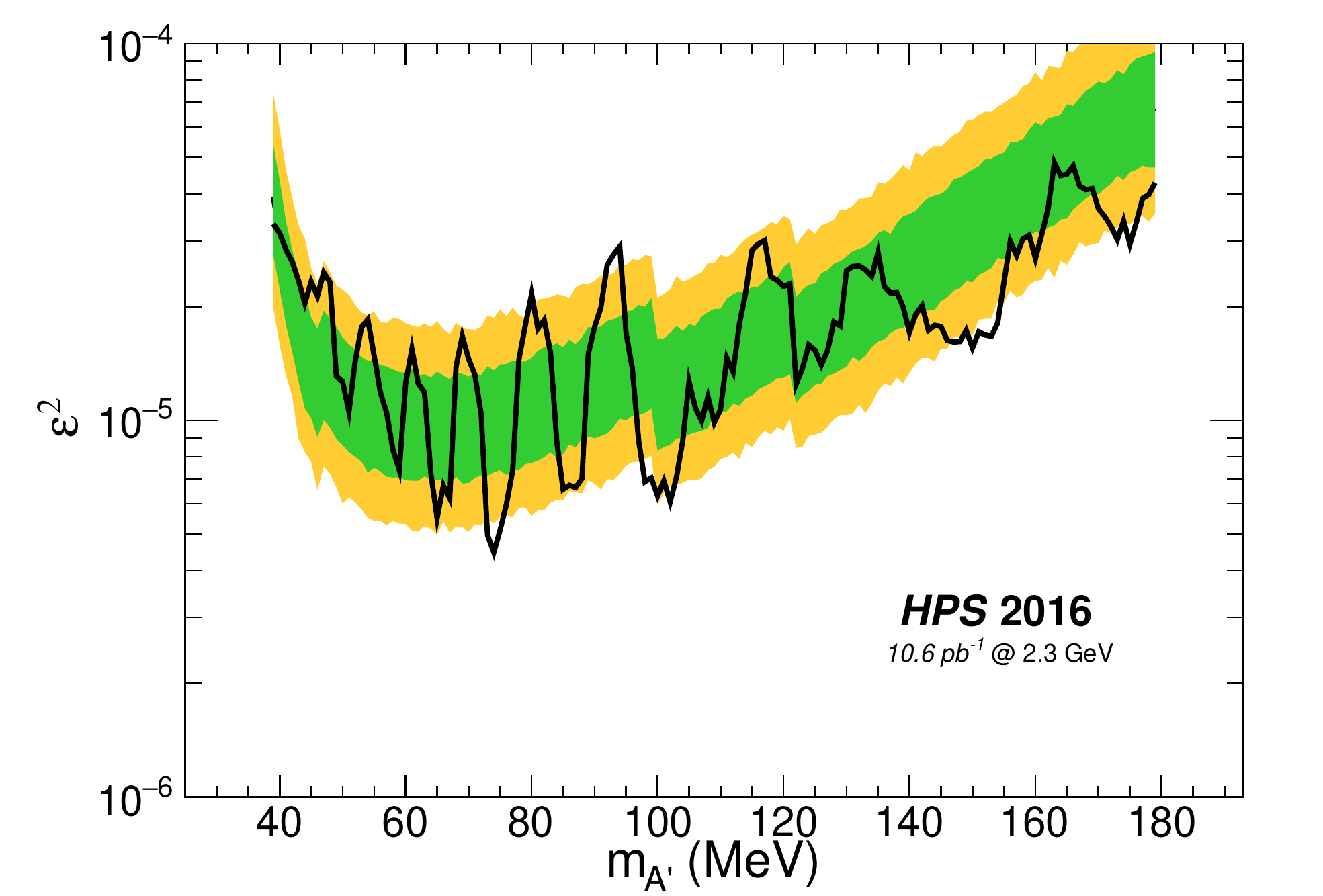}
    \caption{The $\epsilon^{2}$ upper limit produced by the resonance search analysis of 100\% of the HPS
2016 data set, including all systematic uncertainty effects. The green band represents the 68\%
quantile range while the orange band represents the 95\% quantile range. The steps seen in the expected limit happen at masses where the background model polynomial order and/or window size change.
}
    \label{fig:UpperLimitWithSystematics}
\end{figure}

We calculated p-values in the mass range $m\mathrm{(e^{+}e^{-}}) \in(\SI{39}{MeV} - \SI{179}{MeV})$ with the method described in  \Cref{sec:bh_methodology}.
Our search for a resonance failed to reject the null hypothesis at every searched mass point.
The smallest local p-value $\mathrm{ = 6.38\times 10^{-3}}$ is observed for the mass $m\mathrm{(e^{+}e^{-}}) = \SI{94}{MeV}$. 
After accounting for the look-elsewhere effect \cite{lookElsewhereEffect},
the global p-value (\Cref{fig:p_ValueWithSystematics}) corresponds to about $\mathrm{1\sigma}$.
 %%%%%%%%%%%%%%%%%%%%%%%%%%%%%%%%%%%%%%%%%%%%%%%% F I G U R E %%%%%%%%%%%%%%%%%%%%%%%%%%%%%%%%%%%%%%%%%%%%%%%%%%%%
 %%%%%%%%%%%%%%%%%%%%%%%%%%%%%%%%%%%%%%%%%%%%%%%% F I G U R E %%%%%%%%%%%%%%%%%%%%%%%%%%%%%%%%%%%%%%%%%%%%%%%%%%%%
\Cref{fig:p_ValueWithSystematics} shows the p-values for the searched mass hypotheses.
For each mass hypothesis, we then calculate upper limits on $\epsilon^{2}$ (\Cref{fig:UpperLimitWithSystematics}) with the method described in  \Cref{sec:sampleCompsition}.
 %%%%%%%%%%%%%%%%%%%%%%%%%%%%%%%%%%%%%%%%%%%%%%%% F I G U R E %%%%%%%%%%%%%%%%%%%%%%%%%%%%%%%%%%%%%%%%%%%%%%%%%%%%
 %%%%%%%%%%%%%%%%%%%%%%%%%%%%%%%%%%%%%%%%%%%%%%%% F I G U R E %%%%%%%%%%%%%%%%%%%%%%%%%%%%%%%%%%%%%%%%%%%%%%%%%%%%
\Cref{fig:UpperLimitWithSystematics} shows the upper limit results from the HPS 2016 data set and includes all systematic uncertainty effects discussed in \Cref{sec:Uncertainties}. The green band represents the 68\% quantile range while the orange band represents the 95\% quantile range of limits set on an ensemble of background-only simulation.
This analysis of the $\mathrm{e^{+}e^{-}}$ mass distribution in the range \SI{39}{MeV} to \SI{179}{MeV} did not  yield any statistically significant variations from the background-only hypothesis; therefore we report upper limits on $\epsilon^{2}$ in the searched mass range. 
The integrated luminosity of the reported run is insufficient to cover new territory in the
dark photon  parameter space (see Figure \ref{fig:fullResult} and references cited there). HPS does exclude dark photon production over a region of mass and coupling, but this region has already been excluded by previous experiments. 

%% file: DisplacedVertexSearch.tex
\section{Displaced Vertex Search} \label{sec:vtxAna}

The goal of the displaced vertex analysis is to search
for long-lived $\aprime$s produced in the target that decay to
$\epem$ pairs in the range 1--\SI{10}{cm} downstream. These
rare signal processes must be distinguished from a large
number of prompt QED tridents which can appear to originate
downstream of the target because of detector resolution, scattering
effects, or tracking errors. Consequently,
this search is limited by the vertex position resolution of HPS, anomalous scatters, 
 and the quality of the tracking. For incident electron energies of a few GeV, the vertex position resolution is dominated by multiple scattering in the tracker, particularly in the ﬁrst layers. 

\begin{figure}
    \centering
    \includegraphics[width=.45\textwidth]{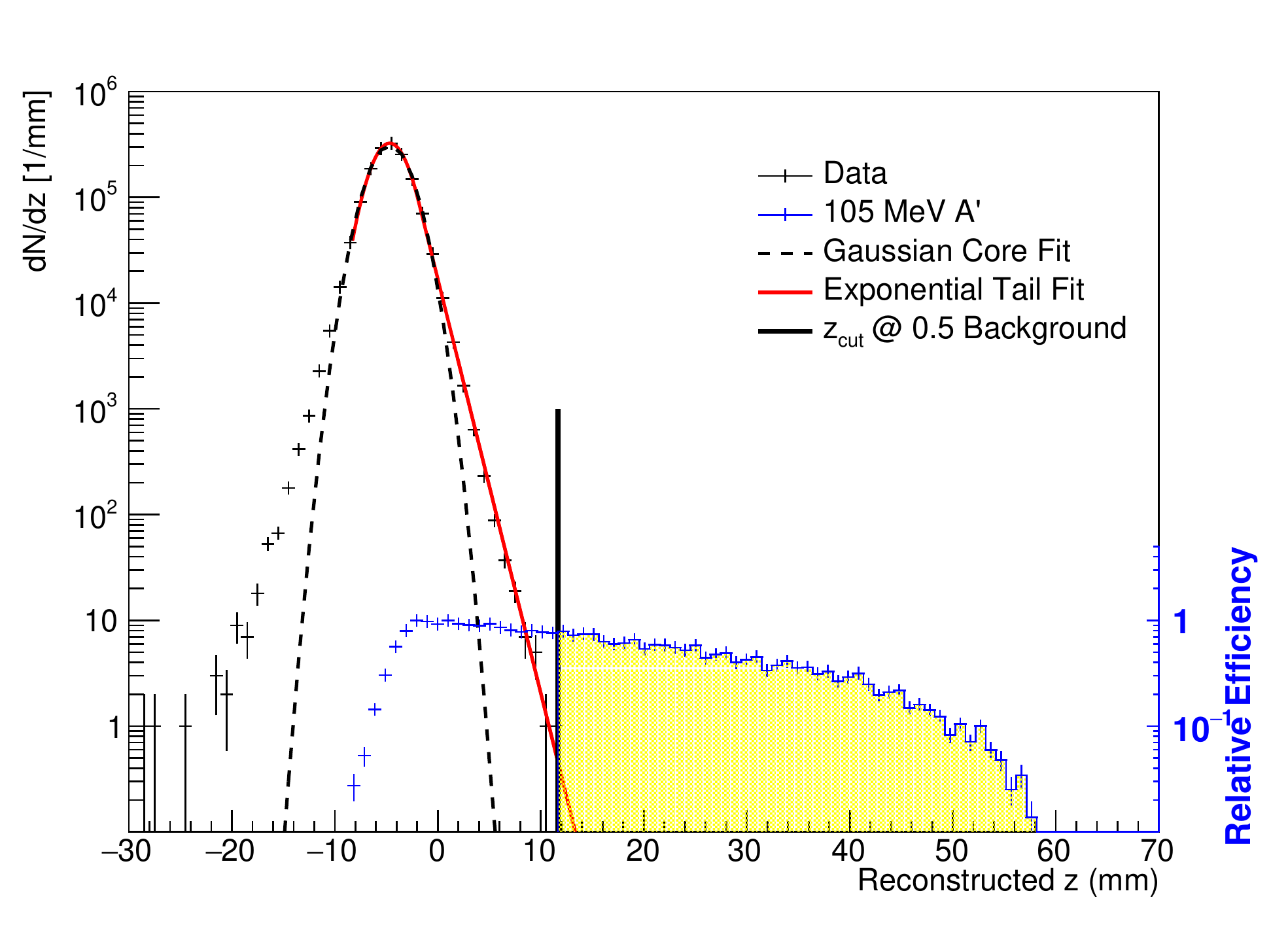}
    \caption{The vertex distribution along the beam
direction for the full data set (black) and a simulated
\SI{105}{MeV} $\aprime$ with a uniform decay in $z$ (blue) in a $105 \pm \SI{4.7}{MeV}$ mass slice.  See text
for details.}
    \label{fig:mass_slice}
\end{figure}

The basic principle of the analysis is illustrated in \Cref{fig:mass_slice}, which shows the vertex distribution (in the coordinate along the beam
 axis, $z$) for reconstructed $\epem$ pairs
in the invariant mass slice of $105 \pm \SI{4.7}{MeV}$. The black
distribution shows data, which are composed entirely
of prompt backgrounds. The blue distribution shows the shape of the
acceptance from a simulated \SI{105}{MeV} $\aprime$, assuming
a decay uniform in $z$. The actual normalization and decay distribution of the $\aprime$ distribution
is dependent on $\epsilon^2$, and is, in general, very small compared to the background. 
Note that the background is well characterized by a Gaussian peak centered on
 the target location, with a power law tail on its high side. The search is
 conducted at values of $z$ beyond which 0.5 background events are expected from
 an exponential fit to the tail, which we call $z_{\rm cut}$. 
Since a near-zero background region is necessary to search
for a very low signal rate, every decay downstream of $z_{\rm cut}$ (the yellow region) is
 considered as a signal candidate. This search is performed using
mass slices over the entire mass range considered in this
analysis.

The following sections describe the event selection, analysis technique, and results of the displaced vertex search.  

\subsection{Event Selection}

In addition to the cuts that select V0s described in \Cref{sec:selection} above, the displaced vertex search, which depends critically on tracking, imposes several additional cuts on track and vertex quality.
 For both the electron and positron, the diﬀerence between the track time and the associated ECal cluster time is required to be less
 than \SI{4}{ns}, to reduce accidental backgrounds.
 Tracks are required to have a $\chi^2/{\rm d.o.f.} < 6$ along with
 a minimum momentum of \SI{0.4}{GeV} to eliminate those
 that arise from particles that suﬀer very large hard
 scattering in the tracker. Electron tracks are required
 to have a momentum magnitude less than \SI{1.75}{GeV} in
 order to remove contamination from FEEs
 whereas positron tracks have no such
 requirement. The unconstrained vertex ﬁt is required to have $\chi^2< 
 10$ to reduce $\epem$ pairs that are inconsistent with originating 
from a single decay vertex.

The final set of cuts, described in the next section, is imposed
 to separate the prompt background that falsely reconstructs downstream of the target from true long-lived particles. These cuts are aimed at
 eliminating nearly all backgrounds arising from prompt
 sources, leaving a clean signal region beyond the $z_{\rm cut}$.

\subsection{Reducing High $z$ Backgrounds}\label{sec:apvertexcuts}

To reduce prompt backgrounds that reconstruct at
large $z$, the so-called “high $z$ background”, additional
cuts beyond the event selection cuts described above
 must be employed. Most of the
high $z$ background results from a prompt track scattering in the first two layers of the tracker (both the active and inactive detector material)
 or from mis-reconstructed tracks. There are several
handles that can be used to distinguish between a true
displaced vertex and a high $z$ background. In general,
a true displaced vertex will have a good vertex $\chi^2$; will
project back to the beam spot; and will be composed of
tracks that each have large vertical impact parameters.
These conditions are rarely true for high $z$ backgrounds.
In addition, to guard against high $z$ backgrounds due to
mis-tracking, the so-called “isolation cut”, described below, 
is implemented. All these cuts have been designed to
eliminate most high $z$ background events while having
minimal impact on the efficiency to detect the $\aprime$ signal. They were tuned
using a 10\% sample of the data.

\begin{figure}
    \centering
    \includegraphics[width=.45\textwidth]{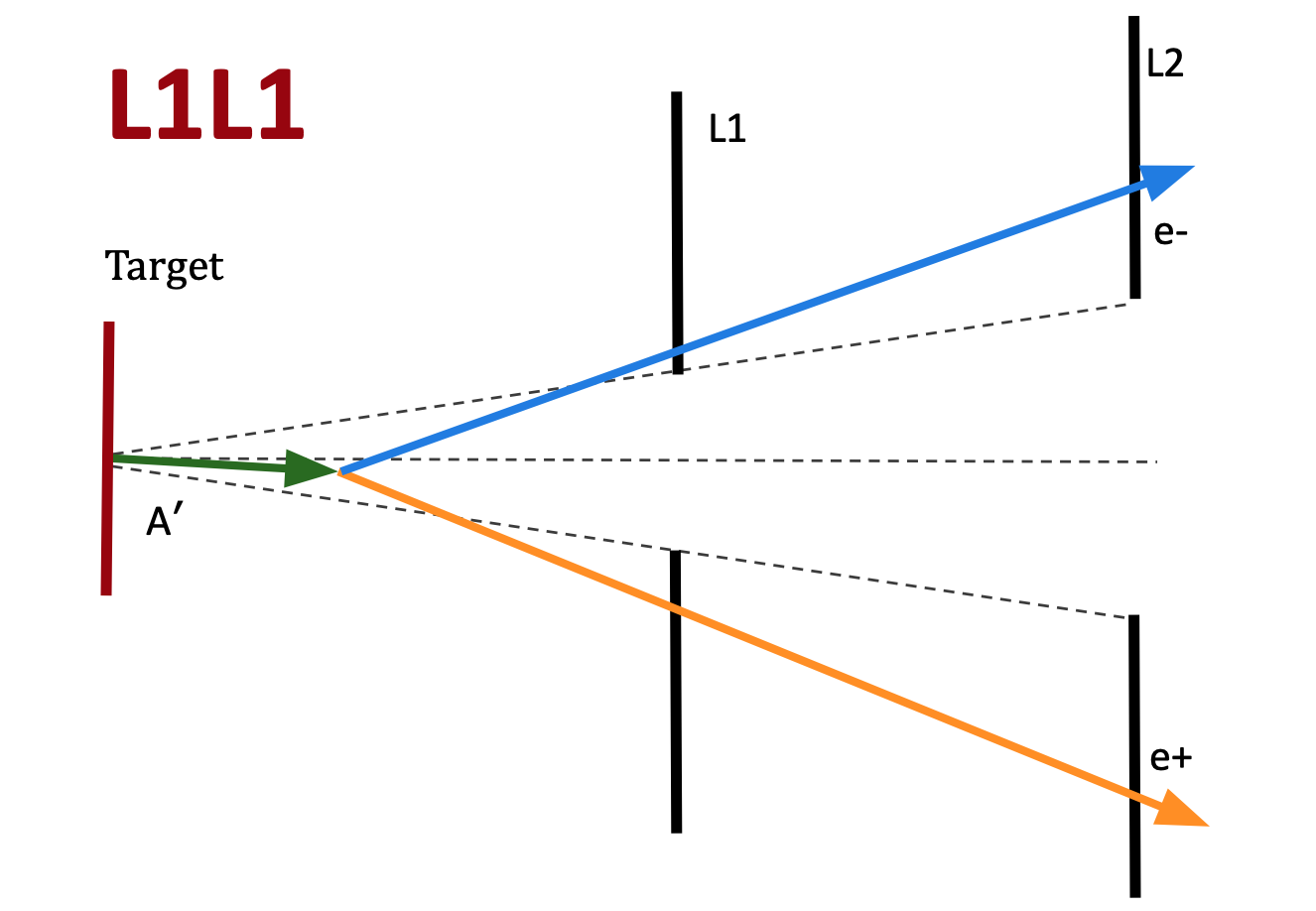}
    \includegraphics[width=.45\textwidth]{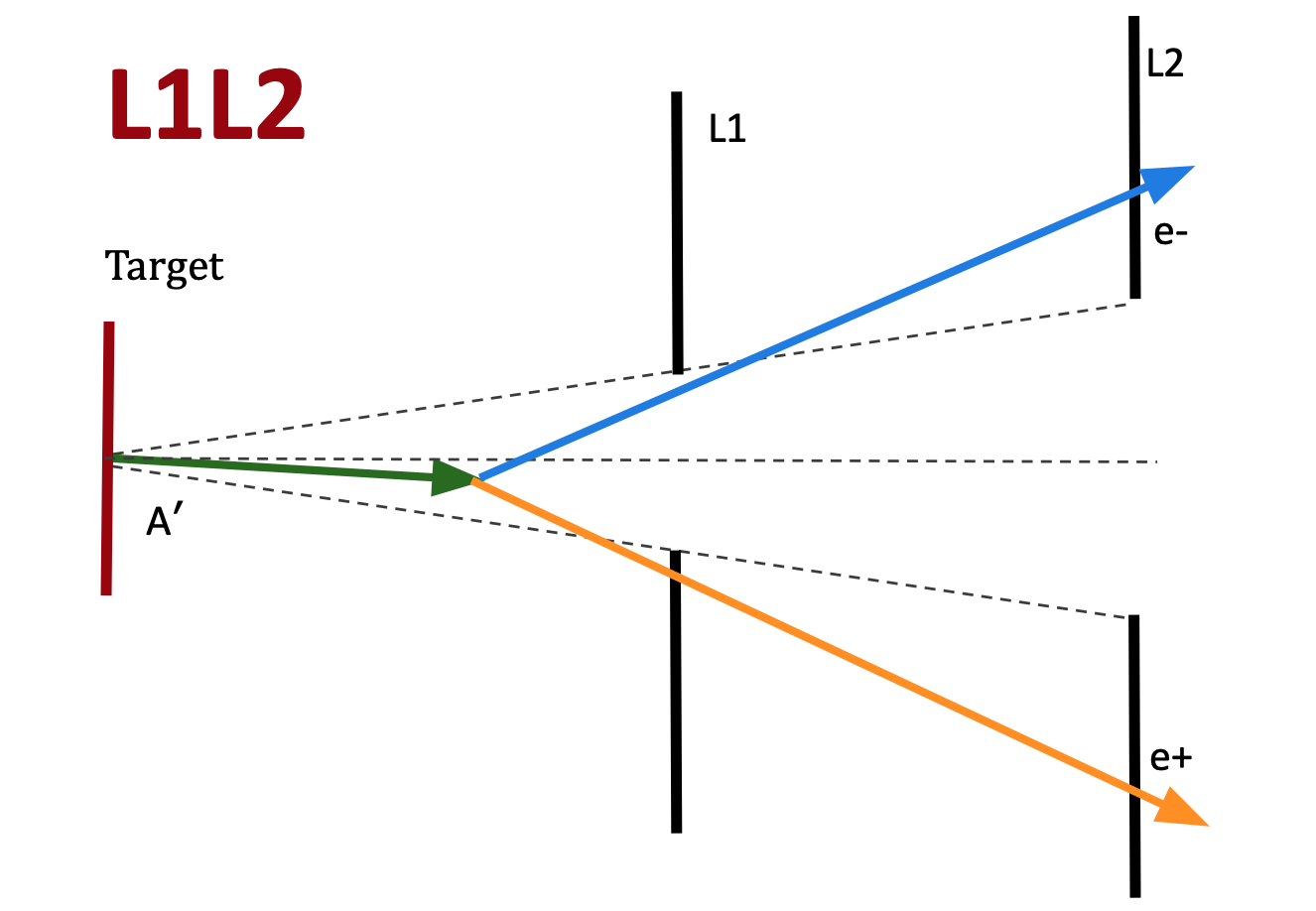}
    \caption{ Top: A schematic of a relatively short $\aprime$
decay length in which both daughter particles have a
layer 1 hit. This is referred to as the “L1L1” category.
Bottom: A schematic of a relatively long $\aprime$ decay
length in which one of the daughter particles misses
layer 1 (but hits L2) and the other daughter particle
hits layer 1. This is referred to as the “L1L2” category.
}
    \label{fig:L1L1_L1L2_schem}
\end{figure}

\begin{figure}
    \centering
      \includegraphics[width=.45\textwidth]{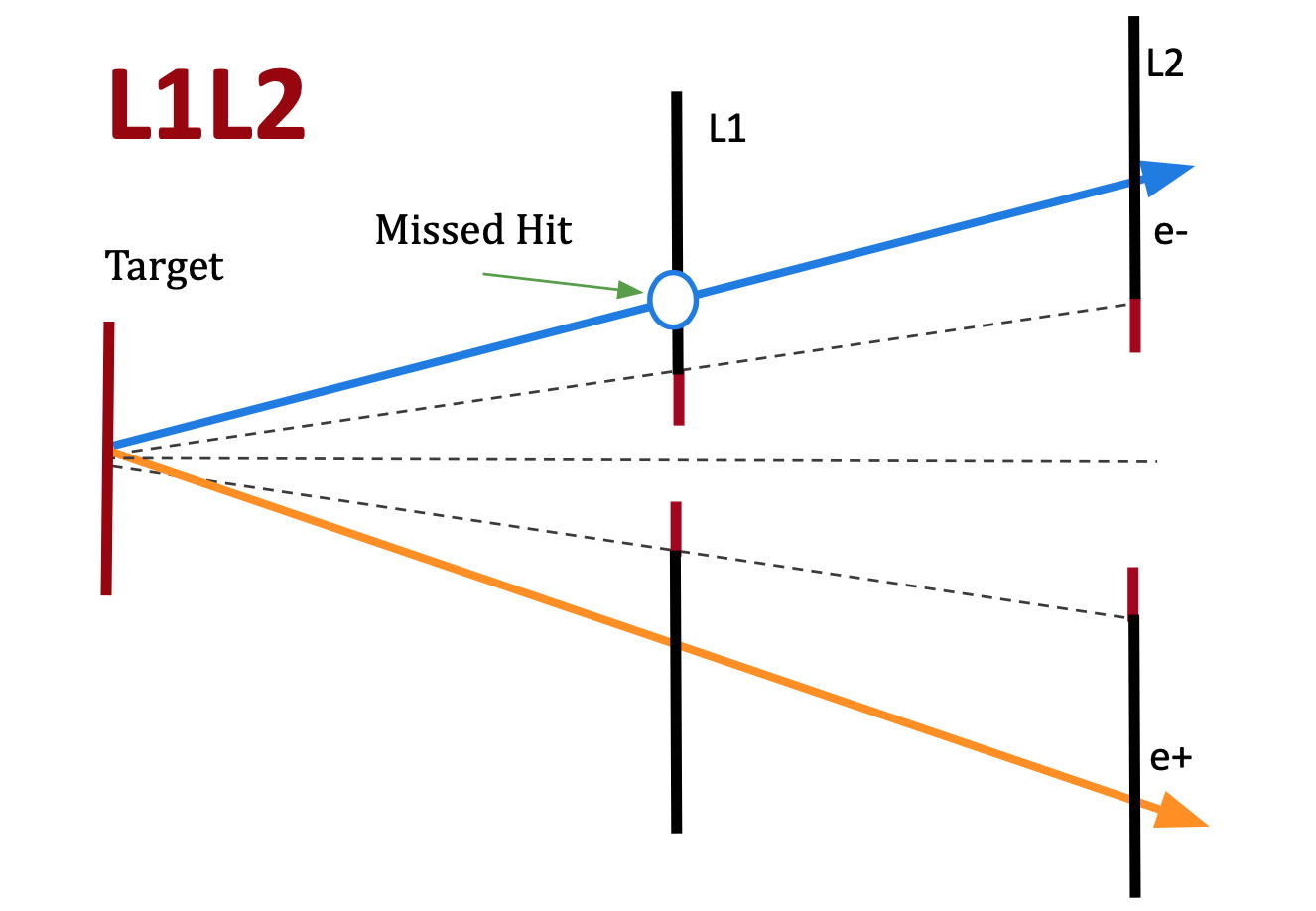}
    \includegraphics[width=.45\textwidth]{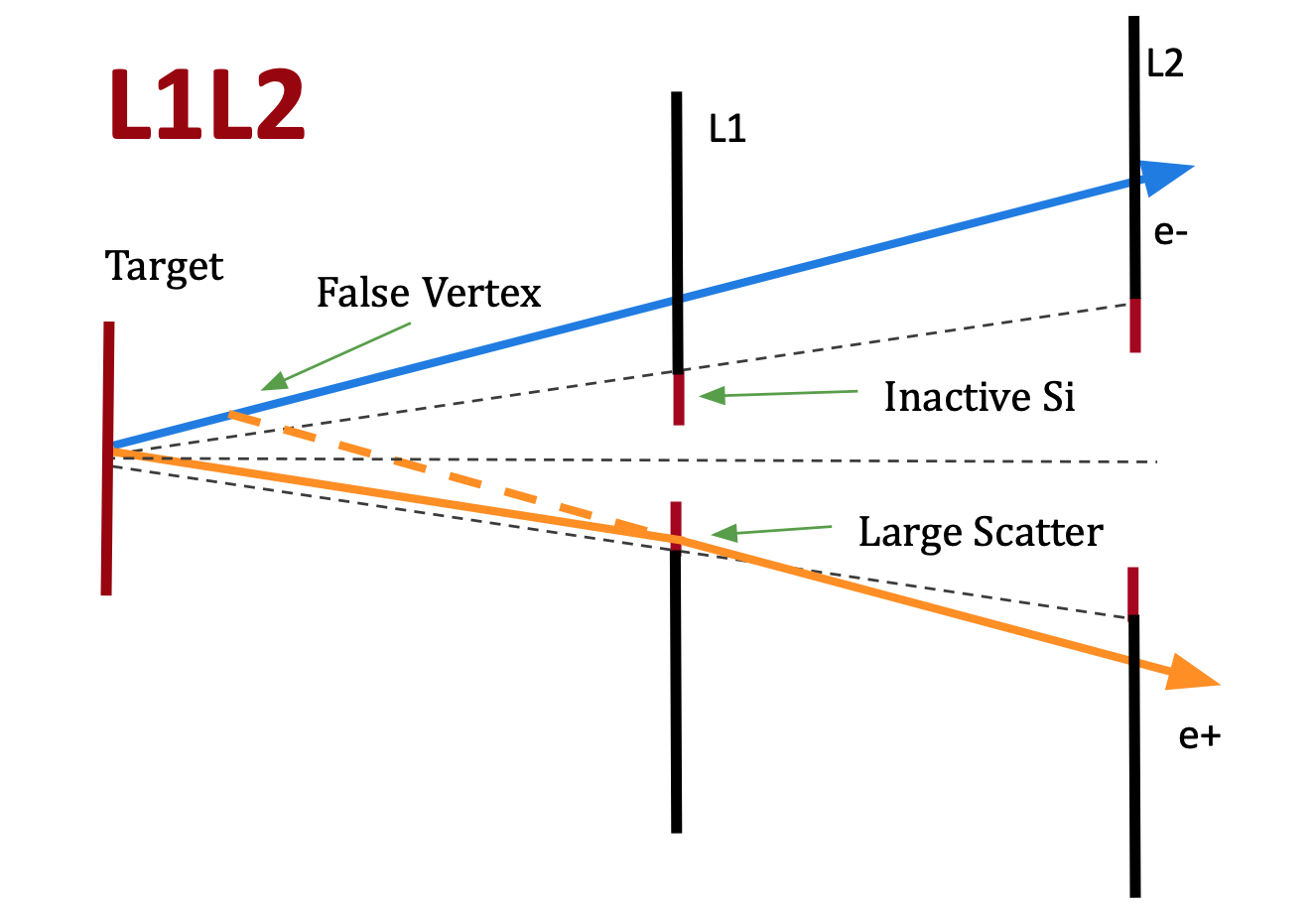}
    \caption{ Top: A schematic of a prompt background
process that has a hit ineﬃciency in layer 1 and is
placed in the L1L2 category. Bottom: A schematic of a
prompt background process in which one of the
daughter particles scatters away from the beam in the
inactive silicon of layer 1 and into the acceptance of the
tracker. This process is placed in the L1L2 category and
also reconstructs a false vertex downstream of the
target.
}
    \label{fig:L1L2_background}
\end{figure}

An $\aprime$ with a relatively short decay length will have layer 1
(L1) hits for both daughter particles, whereas an $\aprime$ with
a longer decay length may have one or both of these particles 
miss L1 due to geometrical acceptance effects as
shown in \Cref{fig:L1L1_L1L2_schem}. For prompt processes, two effects 
may cause particles to “miss” the first layer. 
First, hit detection inefficiencies in L1 may
cause particles to be undetected even though the particle 
traverses the active sensor plane. Second, particles from the target
 can interact with or convert in the
inactive material in L1, resulting in no L1 hit, but
scatter into the acceptance of the downstream layers and
be detected. These effects are illustrated in \Cref{fig:L1L2_background}. Consequently,
 the analysis is divided into several mutually
exclusive categories based on which layer has the first hit
for each of the two daughter particles. If both particles
have an L1 hit, the event is placed in the so-called
“L1L1” category. If exactly one particle hits L1 and
the other particle misses L1 but hits layer 2 (L2), the event
is placed in the “L1L2” category. If both particles miss
L1 but have their first hits in layer
2, the event is placed in the “L2L2” category. These are
the only three possible categories since the tracking algorithm requires at least 5 hits on a track in the 6 layer SVT.
For the purposes of this analysis, only the L1L1 and L1L2
categories are used. The probability of $\aprime$s populating
the L2L2 category requires such long lifetimes and correspondingly low
rates that much more luminosity is required to see them.
 The L2L2 category will add significance to future analyses that 
incorporate detector upgrades and have larger integrated luminosity.

The L1L1 and L1L2 categories are analyzed separately for several
reasons. First, the vertex position resolution is highly dependent
on which layer is hit first. The closer the first hit is
to the target, the better the vertex position resolution.  Second,
the nature of the backgrounds varies in the different categories.
 In the L1L1 category high $z$ backgrounds are typically due to mis-tracking and
multiple scattering in the active region of L1 sensors, whereas backgrounds in the L1L2 and L2L2 categories 
are typically due to hit inefficiency
effects, multiple scattering in both active and inactive
regions of L1, converted WABs, mis-tracking, and
even trident production in L1.
The following cuts are implemented for both L1L1 and L1L2 to reduce 
the high $z$ backgrounds.

 \subsubsection{  V0 Projection to the Target}
The V0 position is projected back to the target location at $z=\SI{-4.3}{mm}$ using the V0 momentum vector direction.
There its $x$-$y$ coordinates are compared to those of the beam spot.
 The position of the beam center is corrected for
run-by-run variations and then the projected vertex position is required to be
within a $2\sigma$ elliptical region of the mean beam position in $x$-$y$ space.

 \subsubsection{  Isolation Cut }
Mis-reconstructed tracks are tracks that contain at
least one hit that is not created by the particle responsible
 for the majority of the other hits on the track.
For instance, a track can reconstruct a real particle
trajectory but include a spurious hit from a beam electron,
recoil electron, converted photon, or noise hit.  When this mis-reconstructed
hit is in L1 and is closer to the beam than the true
hit, it can result in a vertex that appears downstream of
the target, often significantly downstream, i.e., one that
appears signal-like. 

\begin{figure}
    \centering
    \includegraphics[width=.45\textwidth]{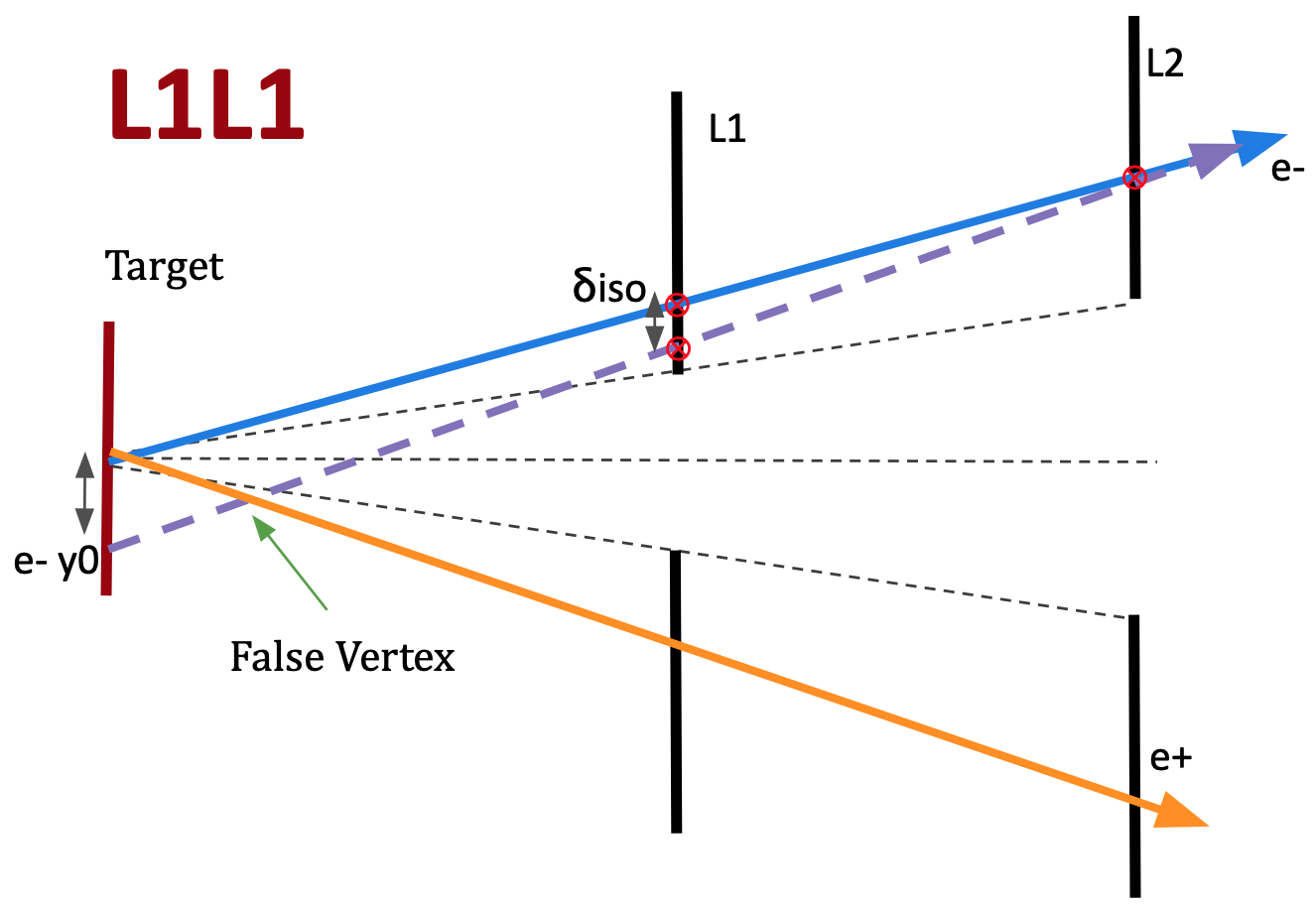}
    \caption{A geometric picture of the isolation cut comparing the distance between the nearest hit away from the beam $\delta_{\rm iso}$ and the longitudinal impact parameter of the track $y_0$. The correct track is in blue and the incorrect track found by the tracking algorithm is in dashed purple. This can result in a falsely reconstructed vertex downstream of the target.
    }
    \label{fig:iso_cut}
\end{figure}

Mis-reconstructed hits in L1 often occur as a result of scattering in L2; those in 
L2, if L1 is missing, from scattering in L3.
Such scattering can cause the track to extrapolate to
the incorrect hit and occurs at a significant
enough rate that it needs to be mitigated. The isolation
cut provides a simple test to see if substituting a nearby
hit in the innermost layer would give a track that is more consistent
with coming from the beam spot, and is thus more likely the correct hit.
 If such a hit is found, the event is eliminated. The
isolation cut compares the distance between the hit associated with the track and that closest to it in the direction away from the
beam, called the isolation value $\delta_{iso}$, to the track vertical
impact parameter $y_0$, as shown in \Cref{fig:iso_cut} for the L1L1 case. Multiple
scattering and beam size effects complicate this cut. 
Reconstruction errors on the impact parameter $\Delta y_0$ are
comparable to the beam size (both $\sim \SI{100}{\micro m}$) so both must 
be accounted for. The isolation cut for the L1L1 
category and  for L1L2 tracks that pass through L1 is as follows:
\begin{equation}
     \delta_{\rm iso} + \frac{1}{2} \ (y_0 - n_{\sigma} \ \Delta y_0)> 0. 
     \label{equ:iso_final_simple}
\end{equation}

For the L1L2 tracks that miss L1, it is:
\begin{equation}
    \delta_{\rm iso} + \frac{1}{3} \ (y_0 - n_{\sigma} \ \Delta y_0)> 0. 
     \label{equ:iso_L1L2_simple}
 \end{equation}

The factor of 1/2 in \Cref{equ:iso_final_simple} is the ratio of the distance between the
first two layers and the distance between L2 and the target. The factor of 1/3
in \Cref{equ:iso_L1L2_simple} is the ratio of the distance between L2 and layer 3 and the distance from layer 3 to the target, appropriate when the first hit is in L2. Multiple scattering is taken into account with the error term $n_{\sigma}$, where $n_{\sigma}$
is selected to be 3 and $\Delta y_0$ is the combination of the projected impact parameter resolution and the beam size. 
A Monte Carlo study shows that
the cut eliminates most high $z$ background due to mis-tracking but has minimal impact on signal efficiency.

  \subsubsection{  Impact Parameter Cut }
\begin{figure}
    \centering
    \includegraphics[width=.45\textwidth]{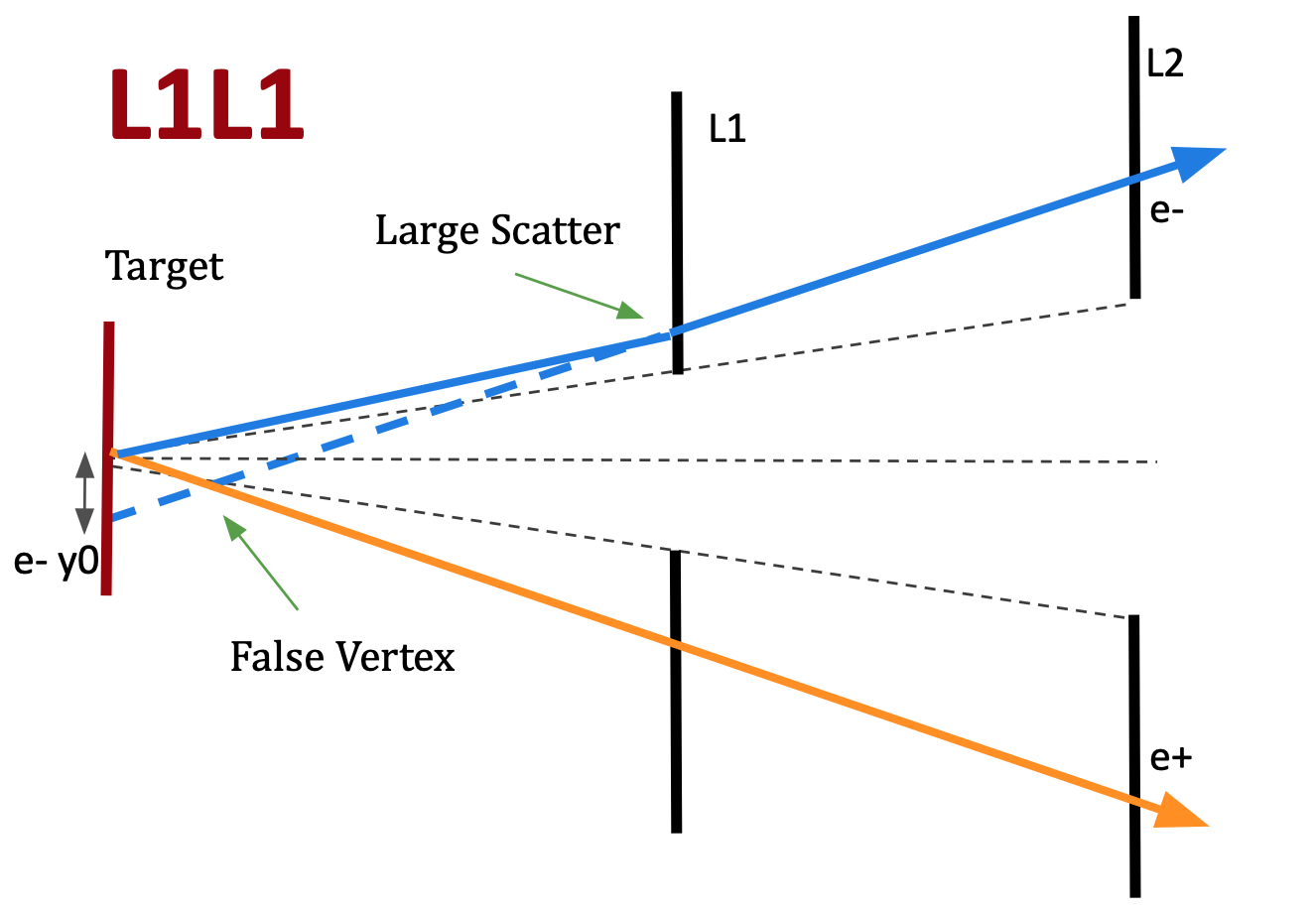}
    \caption{ Prompt background that falsely
reconstructs at a large $z$ due to an $\mathrm{e}^-$ particle with a
large scatter away from the beam plane in L1 of the
SVT. The corresponding $\mathrm{e}^+$ does not have a large
scatter and the track points back near the primary. A
cut on the impact parameter can eliminate such
backgrounds.  
}\label{fig:z0_schem}
\end{figure}

For the $\aprime$ signal, a true displaced vertex will have
large vertical impact parameters ($y_0$) for both its electron
and positron tracks. Furthermore, these impact parameters
 are correlated with $z$, increasing with increasing $z$.
For prompt background that reconstructs at large $z$, this is usually
not the case. Instead, it is likely that just one particle has
a large scatter away from the beam plane (and thus
a large impact parameter) and the other particle is either
consistent with coming from the beam spot or has an impact
 parameter smaller than is expected from signal. With a cut
on the impact parameters of both $\epem$ tracks, such backgrounds 
can be eliminated. This concept is illustrated in \Cref{fig:z0_schem}.
The impact parameters for signal display correlated
bands in the $y_0$–$z$ space,  average $y_0$ increasing as $z$ does. This
correlation is approximately linear for the masses of interest
 in this analysis, so the cut depends linearly on $z$.
Since larger mass $\aprime$s have larger decay angles on average, they will
also have larger impact parameters. Thus, the
cut is also parameterized as a function of mass. Both the
electron and positron are required to satisfy the impact
parameter condition. Before imposing the cut, the $y$
position of the beam is corrected for changing beam conditions.

\begin{figure}
    \centering
    \includegraphics[width=.45\textwidth]{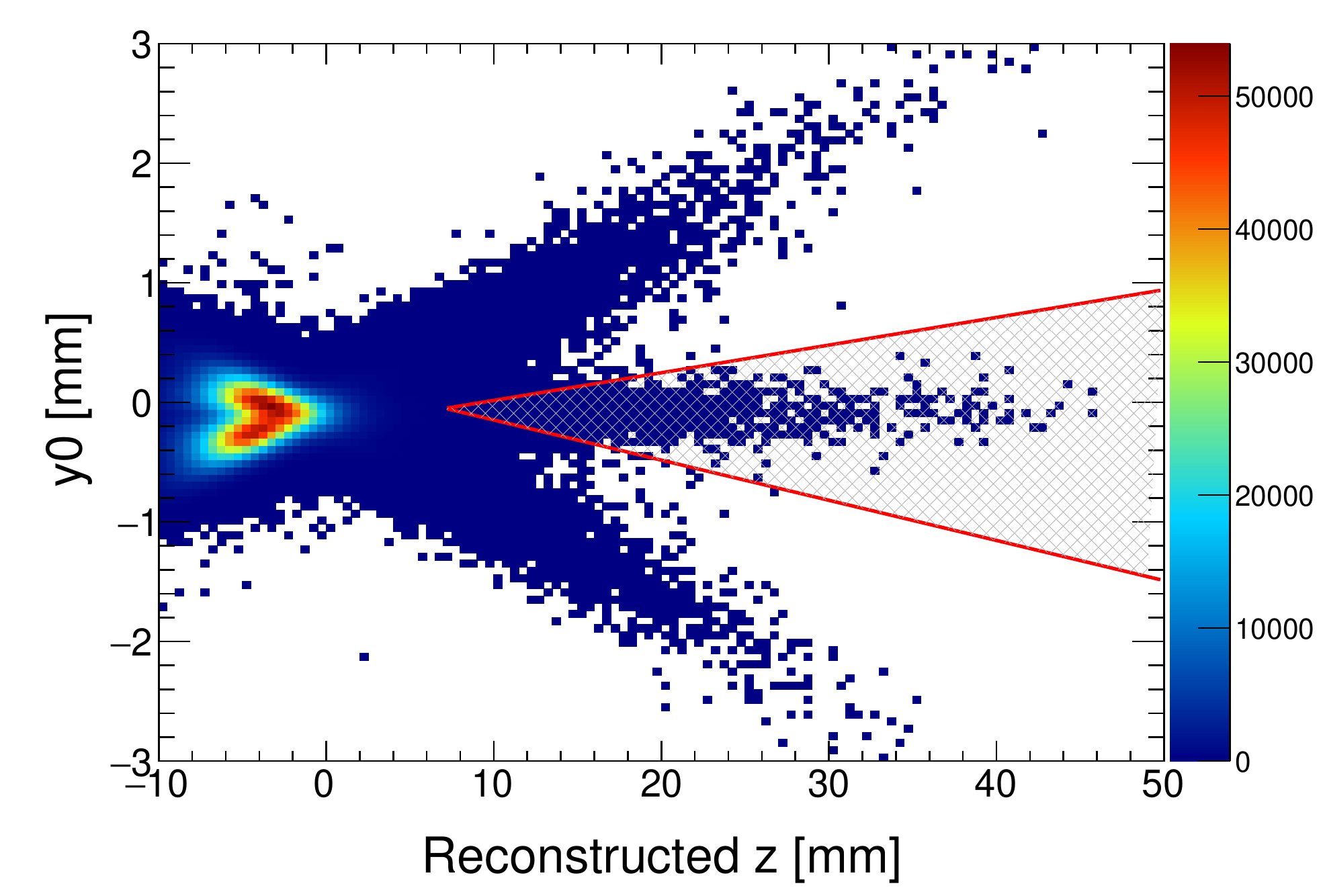}
    \includegraphics[width=.45\textwidth]{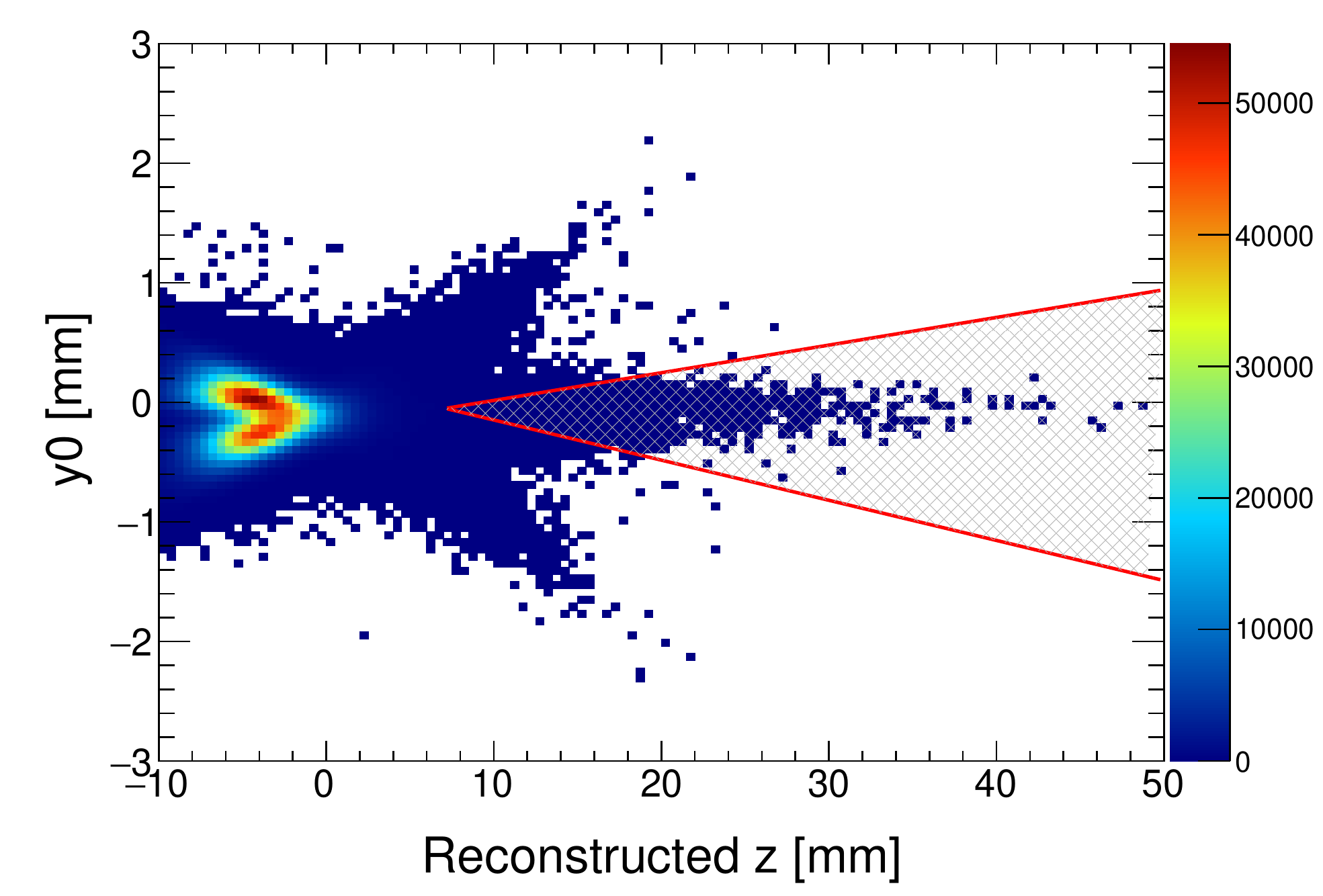}
    \includegraphics[width=.45\textwidth]{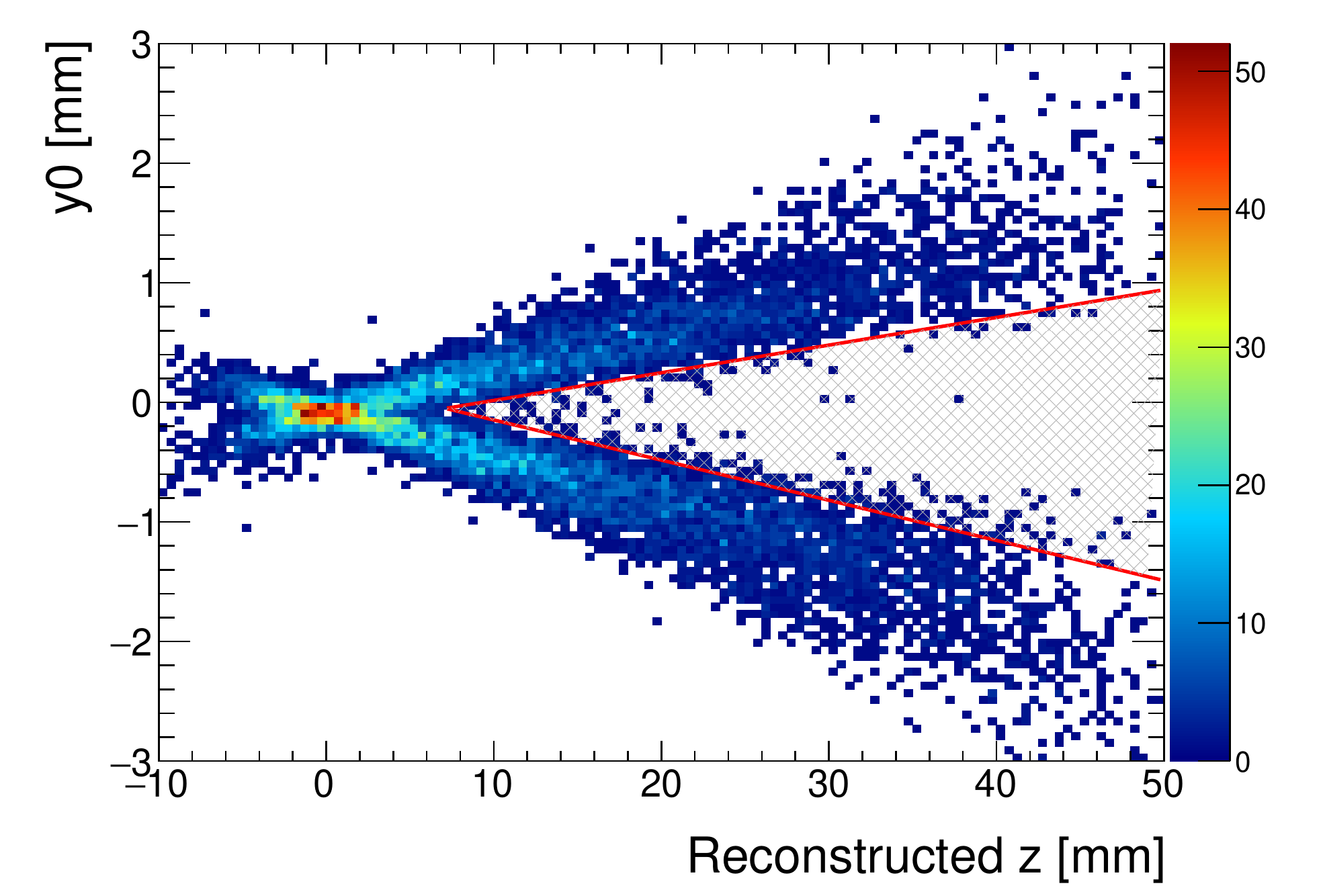}
    \caption{Top:  Example distributions of positron y0 versus reconstructed vertex Z position for L1L1 data prior to any impact parameter cut.  Middle: Electron  y0 versus reconstructed vertex Z position after the positron impact parameter selection.  Bottom:  Electron y0 versus reconstructed vertex Z position for displaced 80 MeV $\aprime$ MC (after selection not shown as only a few events are removed).    The shaded region represents the area removed by the impact parameter cut.  
}\label{fig:z0_vs_vz}
\end{figure}

Figure \ref{fig:z0_vs_vz} shows  y0 versus 
reconstructed vertex z distributions from data for positrons (before any impact parameter selection)
and electrons (after making the selection on positrons).  
Event selection requires both electron
and positron impact parameters pass the cut. Imposing the positron impact
parameter cut removes events which typically have electron impact
parameters outside the excluded zone, so many events in that region
are also removed, as seen in the electron plot.
The bottom plot in this figure shows
the displaced 80 MeV $\aprime$ MC before impact parameter cut.  Very few signal events are removed 
so we do not show plots after selection. 
The data for this Figure was selected around 80$\pm$5 MeV
to have similar kinematics as the $\aprime$.  

 \subsubsection{  Removing Tracks with Shared Hits and Selecting Single V0s }
The last step is to remove both tracks with shared hits and
events with multiple V0 particles. In the reconstruction,
 tracks are allowed to share hits with other tracks, and
these shared hits can be from hits from another particle.
 There is evidence in both data and MC that tracks
with shared hits may produce high $z$ background events. 
To eliminate this possibility, tracks
that share any hits with any other track are eliminated.
The final requirement is that each event must have
exactly one V0 candidate that passes all previous cuts.
This will prevent there being multiple candidate vertices
in an event, which is extremely unlikely \emph{a priori}. 

\begin{figure}[!htb]
    \centering
    \includegraphics[width=.45\textwidth]{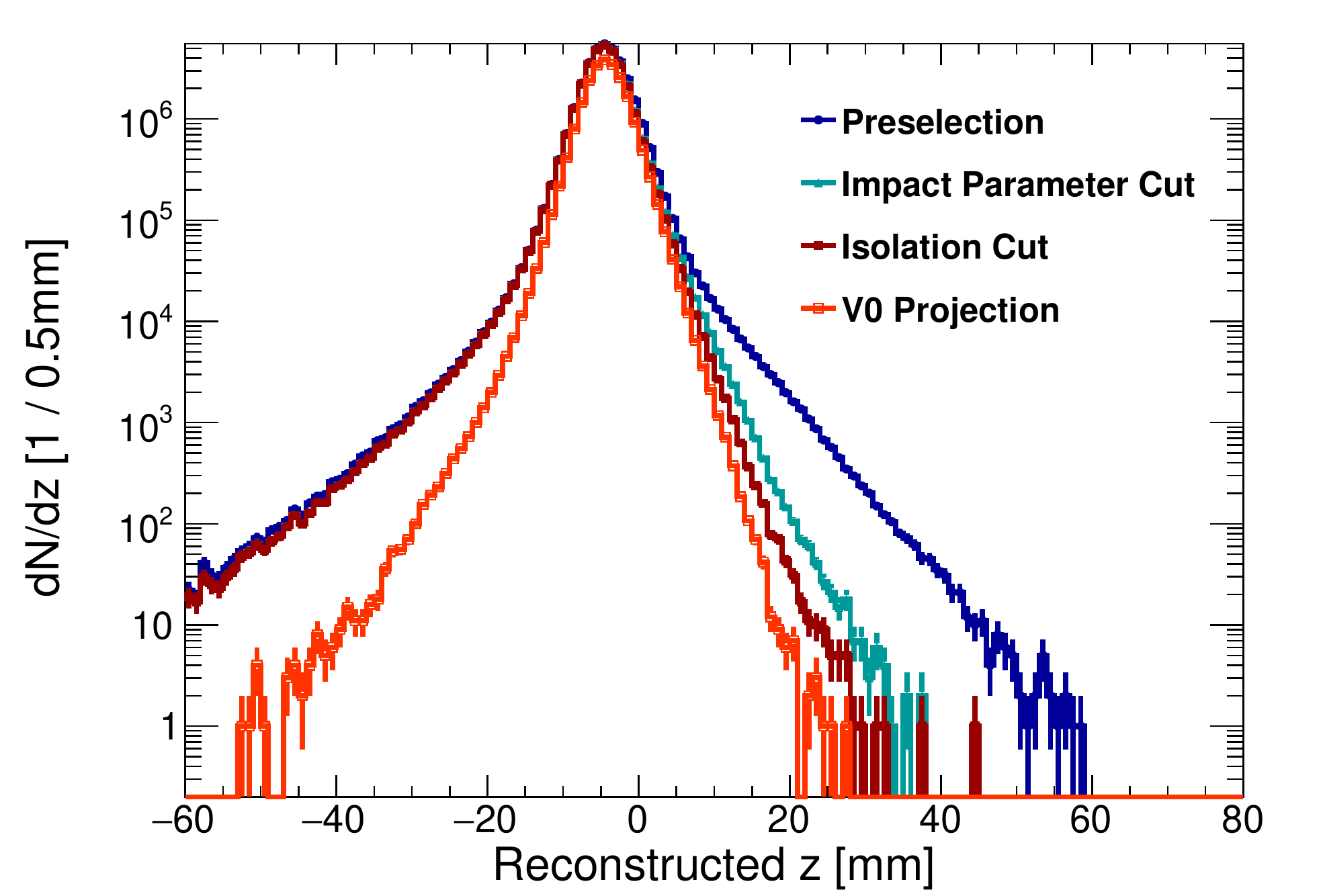}
    \includegraphics[width=.45\textwidth]{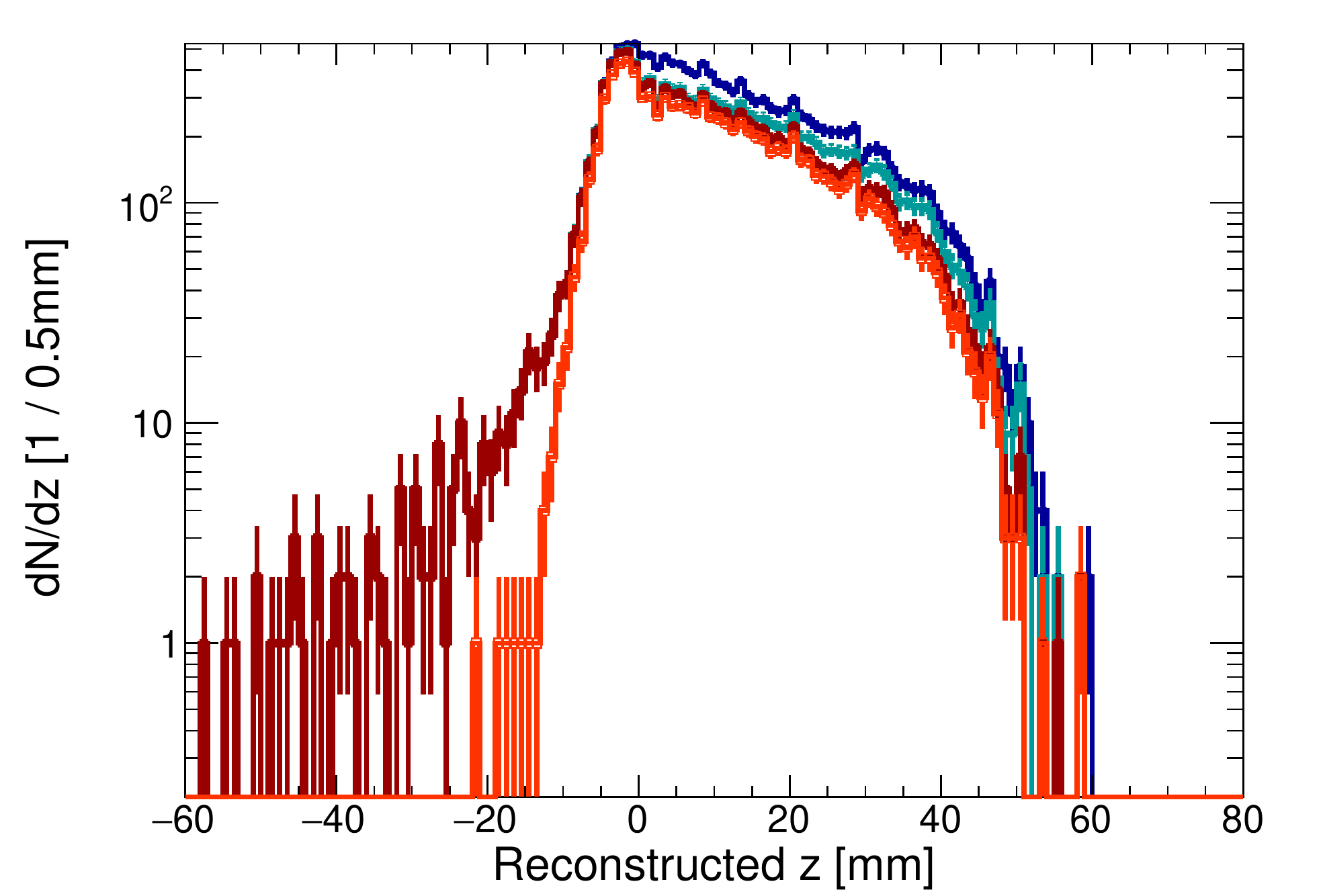}
    \caption{
    Top: The cut flow for data in the L1L1 category. Bottom: The cut flow  for \SI{80}{MeV} displaced $\aprime$ MC in the L1L1 category. }
    \label{fig:tightcutflow_L1L1}
\end{figure}

\begin{figure*}
    \grinp[width=0.95\tw]{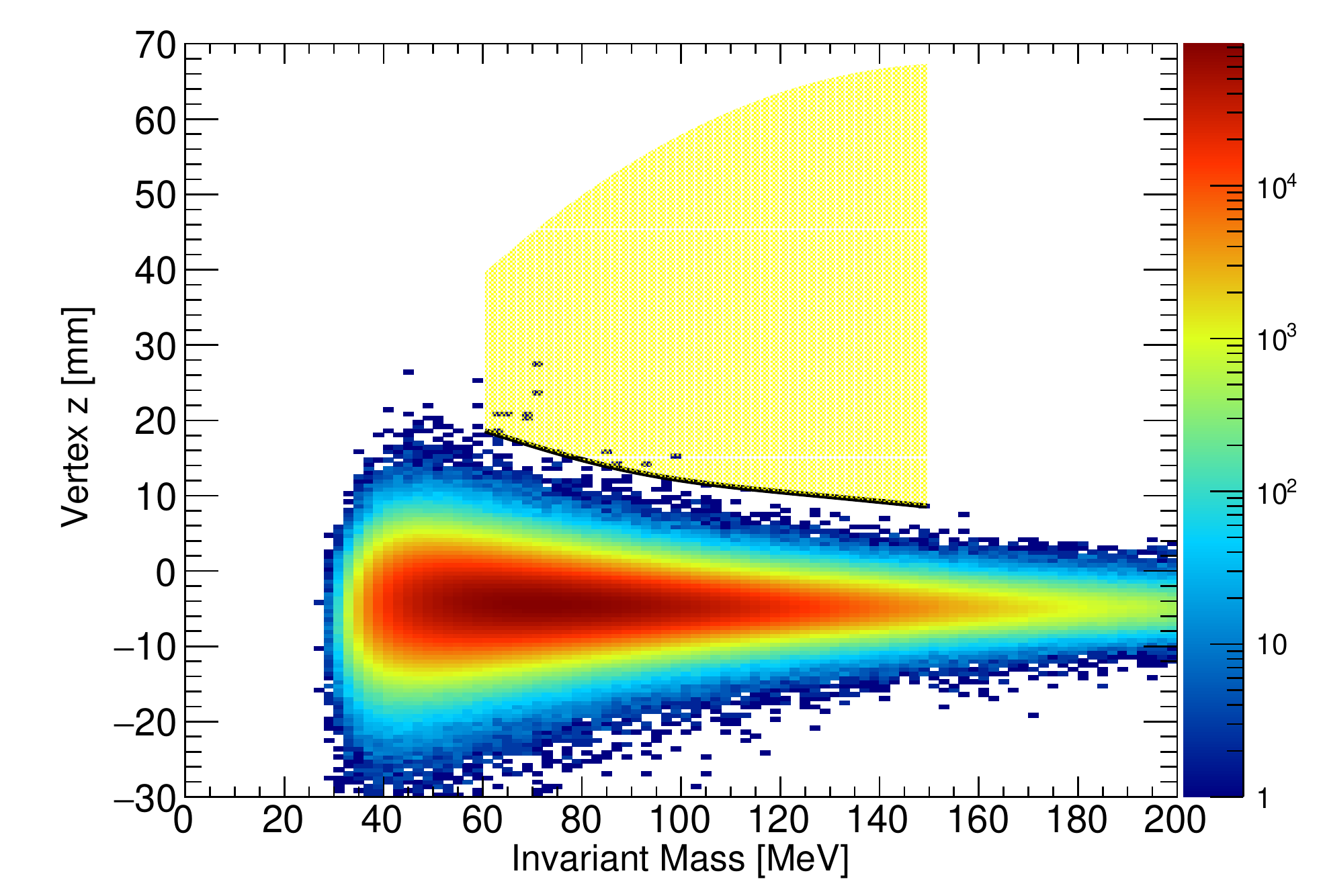}
 \caption{
    	The final selection for the L1L1 category is plotted as reconstructed $z$ vs reconstructed $\epem$ mass for the full data set.  The black line shows the value of $z_{\rm cut}$ versus mass and the yellow shaded region is, roughly, the region of sensitivity to $\aprime$ events.   
    }\label{fig:singleV0_2D}
\end{figure*}

\begin{figure}[!htb]
    \centering
    \includegraphics[width=.45\textwidth]{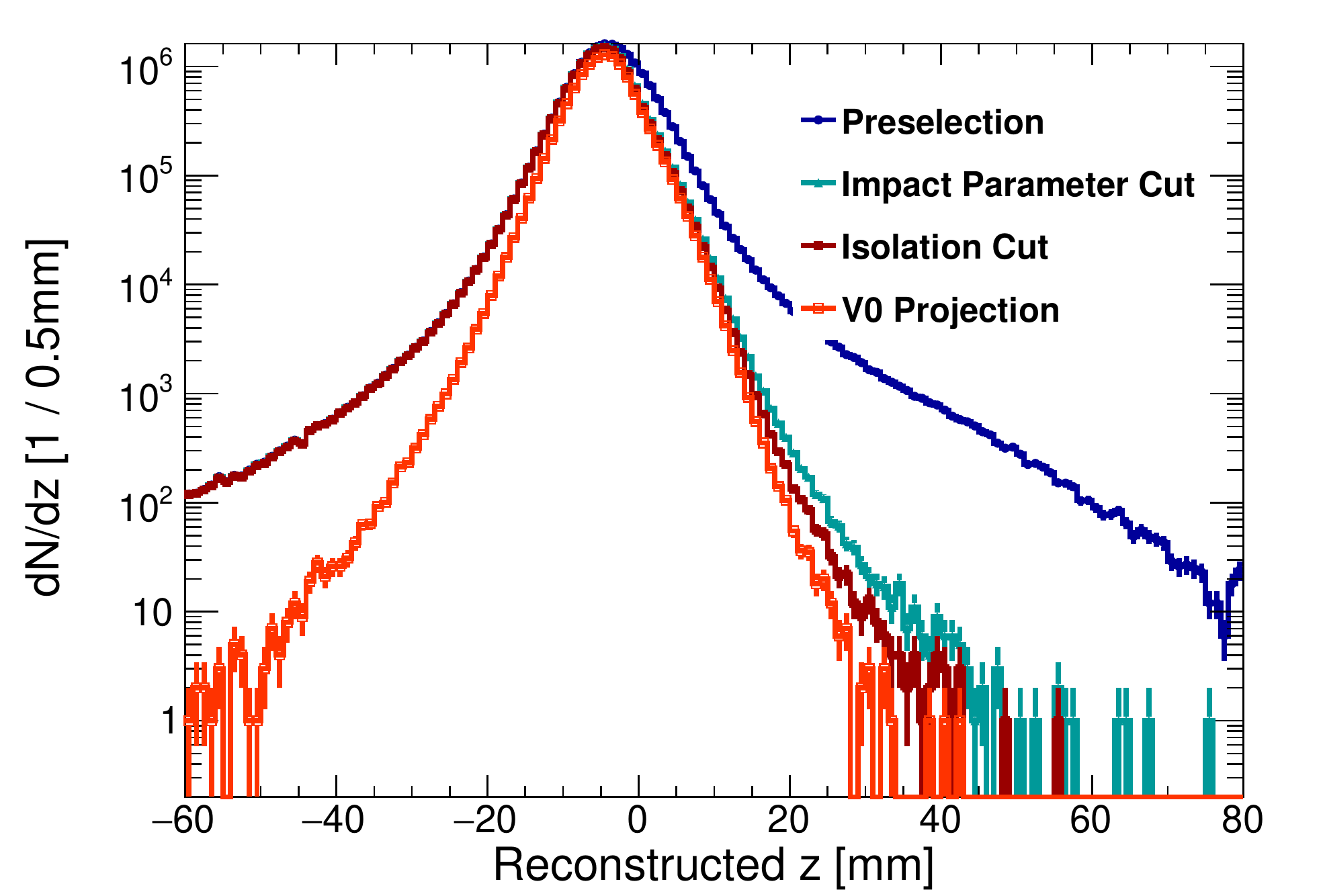}
    \includegraphics[width=.45\textwidth]{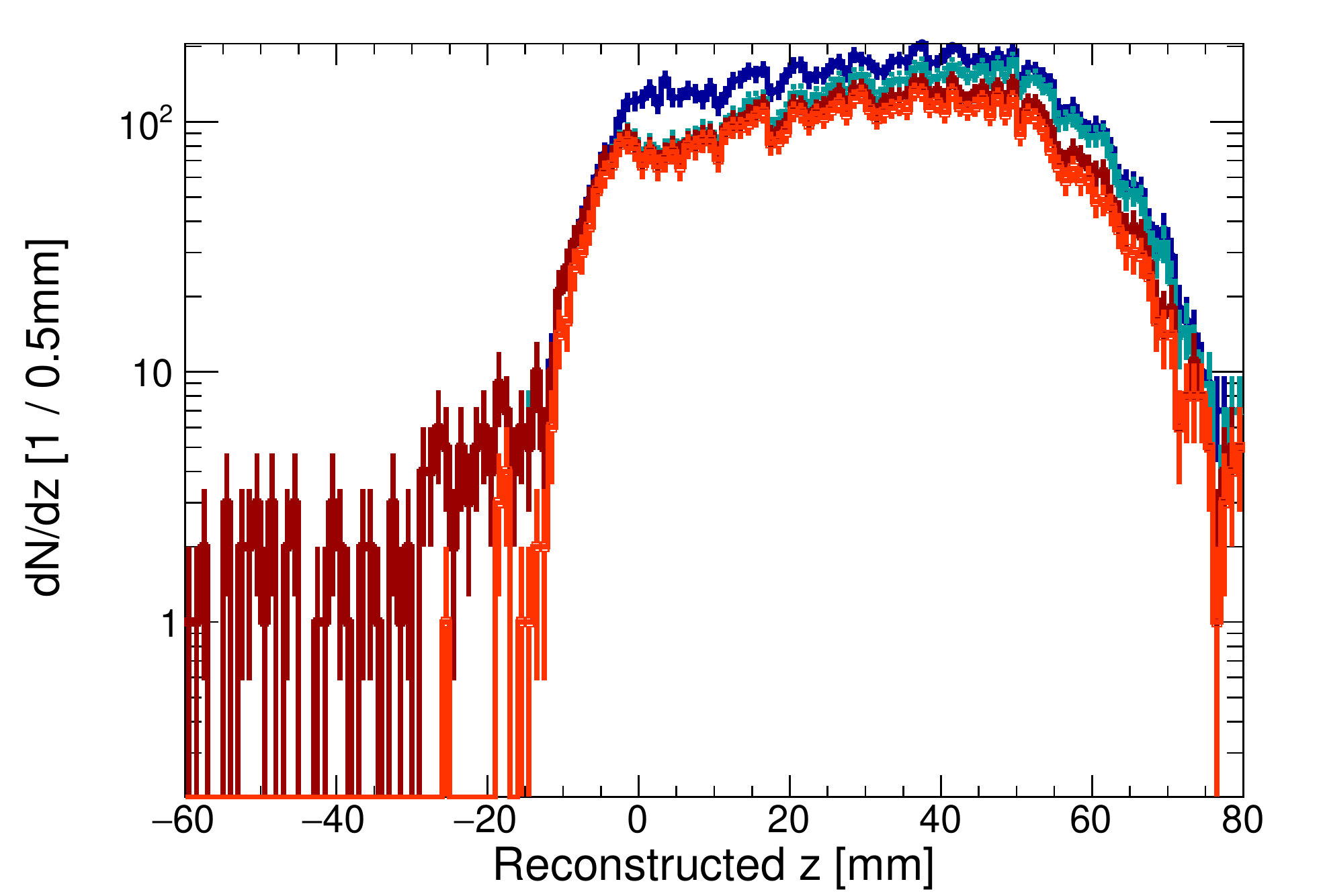}
    \caption{
    	Top: The cut flow  for data events in the L1L2 category. Bottom: The cut flow  for \SI{80}{MeV} displaced $\aprime$ events in the L1L2 category. 
    }
    \label{fig:tightcutflow_L1L2}
\end{figure}

\begin{figure*}
    \centering
    \grinp[width=0.95\tw]{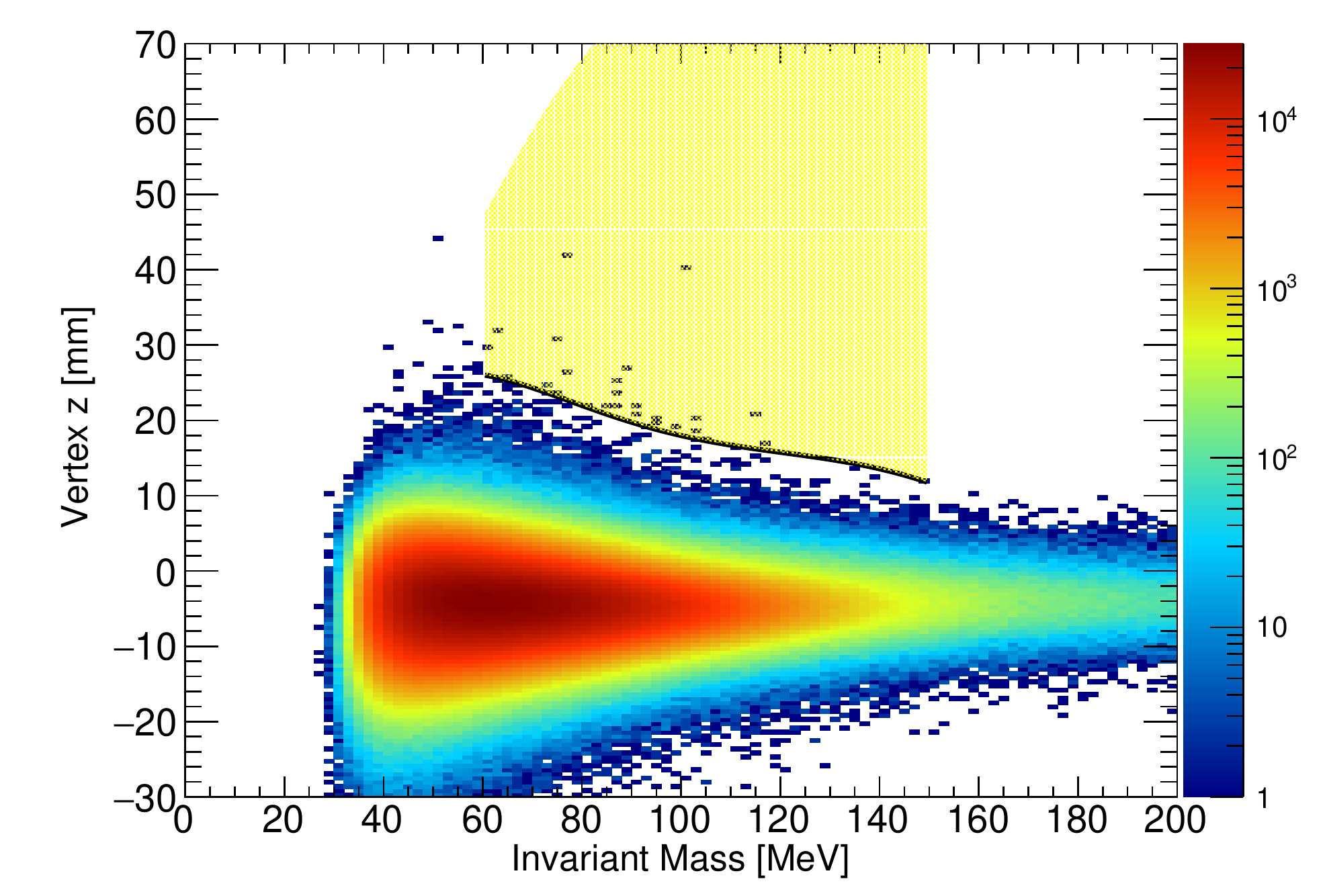}
    \caption{
        The final selection for the L1L2 category is plotted as reconstructed $z$ vs reconstructed $\epem$ mass for the data set.  The black line shows the value of $z_{\rm cut}$ versus mass and the yellow shaded region is, roughly, the region of sensitivity to $\aprime$ events.   
    }\label{fig:singleV0_2D_L1L2}
\end{figure*}

The complete cut flow for all background reduction 
cuts for the L1L1 (L1L2) category is shown in \Cref{fig:tightcutflow_L1L1} (\Cref{fig:tightcutflow_L1L2}).
The resulting reconstructed $z$ vs. mass for events in the L1L1 (L1L2) category is shown in \Cref{fig:singleV0_2D} (\Cref{fig:singleV0_2D_L1L2}). Note that no mass bins in either plot show significant concentrations of events beyond the $z_{\rm cut}$. Further note the nearly complete absence of any events at large decay lengths, beyond 50 mm for L1L1 and out to \SI{75}{mm} for L1L2, although there is acceptance in these regions. 

\subsection{Defining the Signal Region}\label{sec:tailfits}

Because of the low rate of the expected signal, a signal region must be defined such that very little background is expected. Both the background and signal fall exponentially in the $z$-direction; however, the background falls at a much faster rate. Thus a nearly zero background region can be found downstream of a sufficiently large $z$ value. Specifically, this is done as a function of mass since a signal is expected at a specific invariant mass and the vertex position resolution is dependent on the opening angle, and hence mass-dependent. With this in mind, the $z$ vs mass distribution is sliced into overlapping bins of width equal to $\pm 1.9\sigma_m(m)$ for a mass $m$ in the bin center. Each mass slice is fitted in $z$ using the following continuous and differentiable empirical function consisting of Gaussian core and exponential tail.
\begin{equation} \label{eqn:tail_fit}
    F(z) = 
    \begin{cases}
        A e^{-\frac{(z-\mu_z)^2}{2\sigma_z^2}} & \mathrm{if}~ \frac{z-\mu_z}{\sigma_z}<b \\
        A e^{\frac{b^2}{2}-b\frac{z-\mu_z}{\sigma_z}} & \mathrm{if}~ \frac{z-\mu_z}{\sigma_z} \ge b.
    \end{cases}
\end{equation}
The parameter $b$ is the number of standard deviations from the mean that the fit function changes from a Gaussian to an exponential tail. All of $A$, $\mu_z$, $\sigma_z$, and b are determined by the fit for each mass slice. 

Using the results of the fit function, the $z$ value beyond which the background fit function predicts 0.5 background events defines the $z_{\rm cut}$. Or more precisely: 
\begin{equation}
    0.5=\int_{z_{\rm cut}}^{\infty}F(z) \ dz.
    \label{eqn:z_{cut}}
\end{equation}
After this fit is performed in every mass slice and the $z_{\rm cut}$ is found, the final $z_{\rm cut}$ in a given mass slice is found by fitting the $z_{\rm cut}$ distribution as a function of mass without the points in the mass bin of interest in order to be unbiased (i.e. using the mass sidebands). An example background fit to a mass slice in the full data set is shown in \Cref{fig:mass_slice} and the $z_{\rm cut}$ for both the L1L1 and L1L2 categories, along with the signal region in yellow, is displayed on \Cref{fig:singleV0_2D} and \Cref{fig:singleV0_2D_L1L2}, respectively.

\subsection{Computing the Expected Signal Yield}\label{sec:signalyield}

Computing the expected rate of $\aprime$s in the signal region takes  careful consideration of several $z$-dependent factors including decay length distributions (as a function of the model parameters), detector acceptance effects, and efficiency effects. The first step is to compute the truth signal distribution for long-lived $\aprime$s, which is exponential in $z$.  The normalized truth signal shape as a function of $c\tau$ is an exponential given by:
\begin{equation}
    S_{\rm truth}(z,m_{A'},\epsilon)=\frac{e^{-(z_{\rm targ}-z)/\gamma c\tau}}{\gamma c \tau}
    \label{eqn:ap_truth}
\end{equation}
This function is normalized such that the integral from $z_{\rm targ}$, the $z$ position of the target, to infinity is unity so that it gives the expected signal density distribution (i.e. $\int_{z_{\rm targ}}^{\infty} S_{\rm truth}(z,m_{A'},\epsilon) \ dz=1$). In this equation, $\gamma=\frac{E}{m_{A'}}$ is the relativistic constant where the $\aprime$ energy is computed to be $E=0.965 E_{\rm beam}$ (which is the mean of the $x$ distributions across all relevant masses). 

\begin{figure}[!htb]
    \centering
    \includegraphics[width=.45\textwidth]{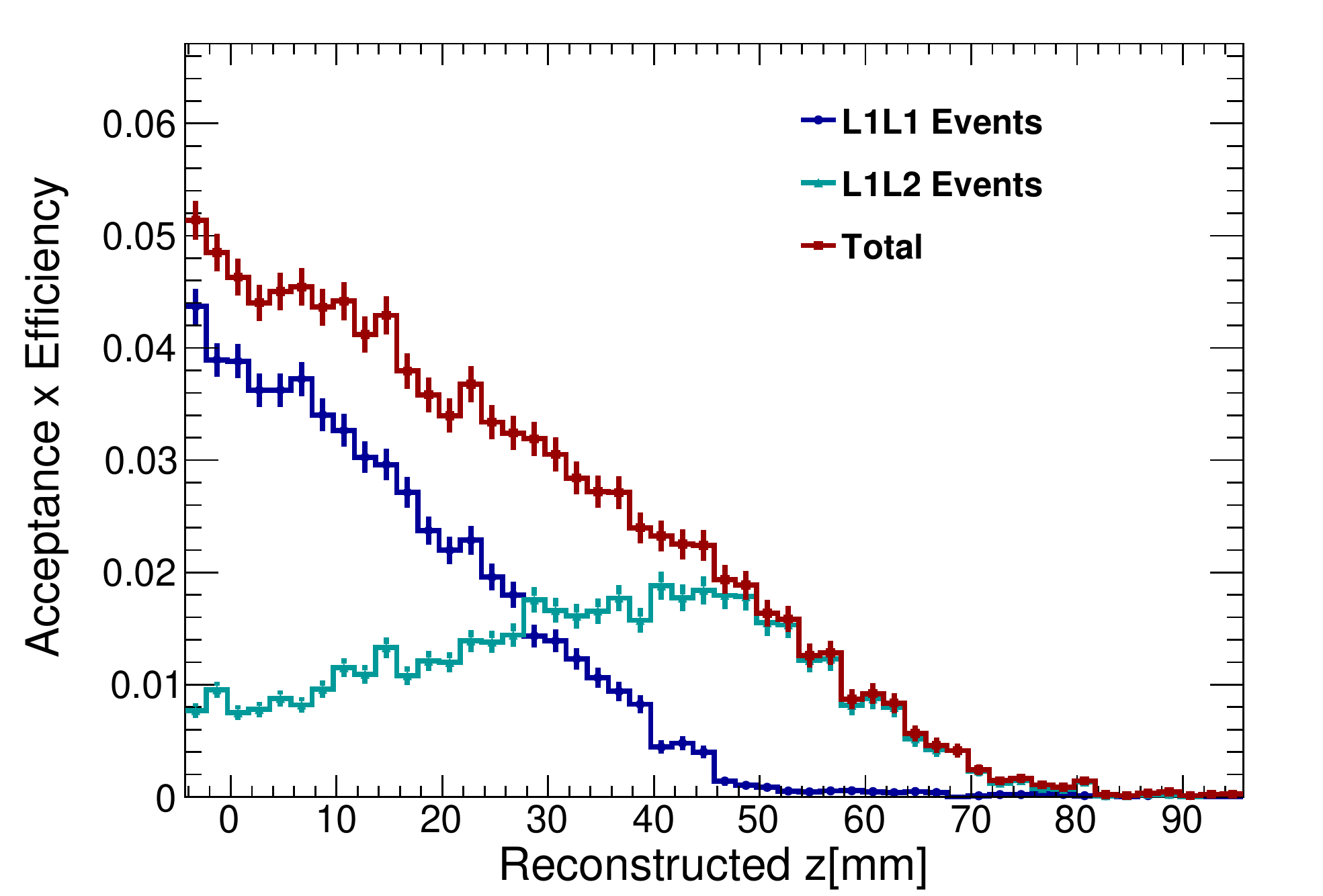}
    \caption{The product of geometrical acceptance and efficiency for displaced $\aprime$ MC for the L1L1 and L1L2 categories as well as their sum for \SI{80}{MeV} displaced $\aprime$s.}
    \label{fig:eff1}
\end{figure}

After computing the truth distributions, detector acceptance must be taken into account. The SVT is designed to have  a $\theta_y$ acceptance beyond \SI{15}{mrad} for prompt decays. However, downstream decays must have a larger opening angle to remain in the acceptance of the SVT, and the farther downstream the decay, the more likely the daughter particles will miss the SVT. The geometrical acceptance drops dramatically with increasing decay length as is shown in \Cref{fig:eff1}. The geometrical acceptance cannot be measured in data and must be derived from simulation. 

\begin{figure}[!htb]
    \centering
    \includegraphics[width=.45\textwidth]{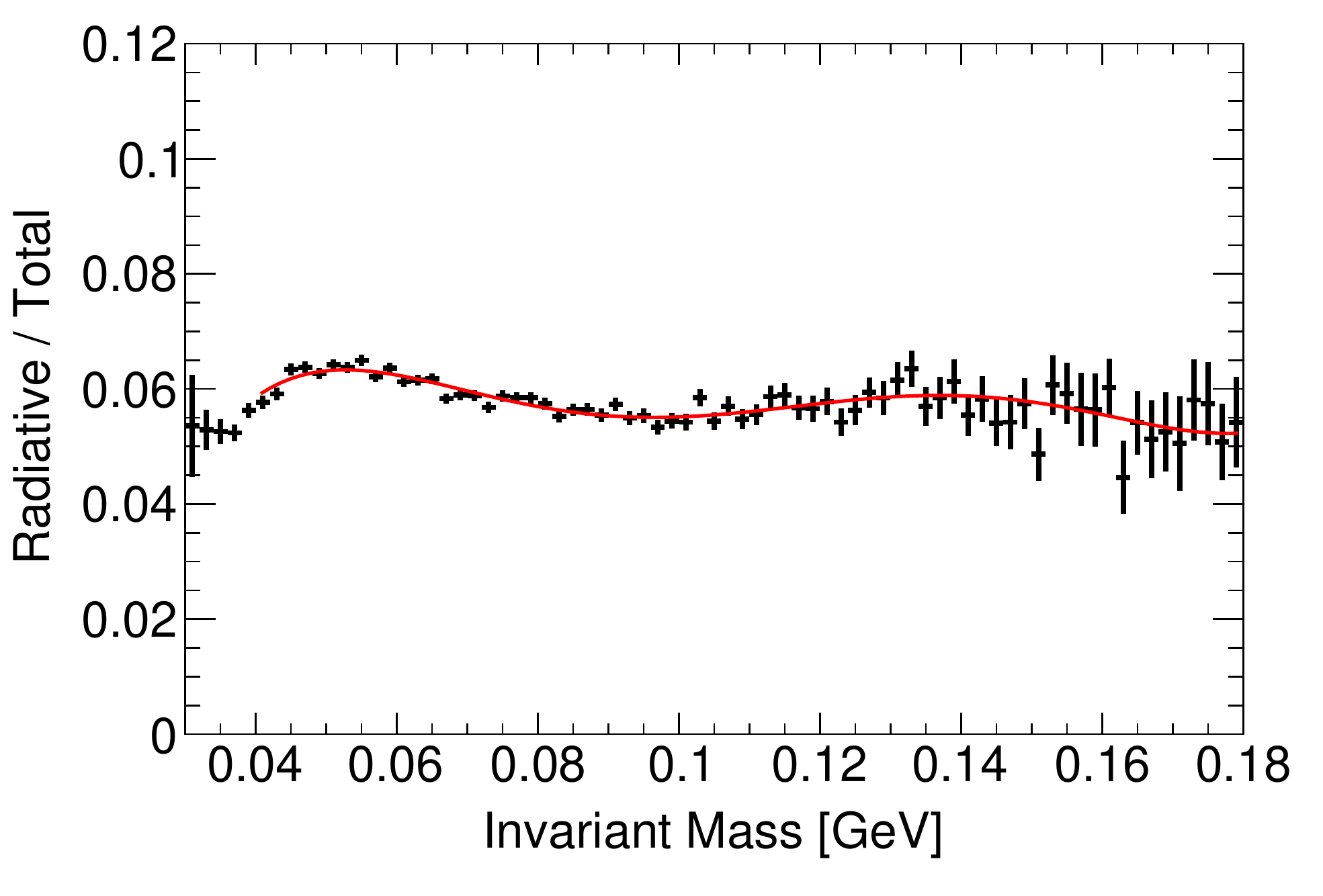}
    \caption{Estimate of the fraction of radiative events as a function of invariant mass. This distribution is obtained from MC.  }
    \label{fig:radfracvert}
\end{figure}

Finally, putting this all together and integrating the signal shape across the $z$ range of interest gives the formula for the expected signal past $z_{\rm cut}$ as a function of mass and $\epsilon$ denoted as $S_{\mathrm{bin},z_{\rm cut}}(m_{A'},\epsilon)$.

\begin{eqnarray}
&&S_{\mathrm{bin},z_{\rm cut}}(m_{A'},\epsilon) = \nonumber \\ &&  S_{\rm bin}(m_{A'},\epsilon) \times 
\int_{z_{\rm targ}}^{z_{\rm max}} S_{\rm truth}(z,m_{A'},\epsilon) \ \epsilon_{\rm vtx}(z,m_{A'}) \ \mathrm{d}z. \nonumber\\
&&
    \label{eqn:signal_yield}
\end{eqnarray}
In this equation, $S_{\rm bin}(m_{A'},\epsilon)$ is the expected signal yield within prompt acceptance and within a finite mass bin computed from the number of $\epem$ pairs and the radiative fraction, shown in  \Cref{fig:radfracvert}, calculated for the specific event selection used in the displaced vertex search (see \Cref{sec:selection}).   Additionally, $\epsilon_{vtx}(z,m_{A'})$ is the normalized efficiency as a function of $z$ including acceptance and all efficiency effects (including analysis cuts and the $z_{\rm cut}$).  The value  $z_{\rm max}$ is the minimum $z$ beyond which signal is not expected.\footnote{Note that the integral is taken starting from the target and not the $z_{\rm cut}$ since the $z_{\rm cut}$ is already applied as an analysis cut at this point. This is done because the $z$ in the integral is a truth value while the $z_{\rm cut}$ is a reconstructed value derived from data.} The expected $\aprime$ rates computed with this equation are used as an input to set the final limit in \Cref{sec:vtxLim}.

\begin{figure}[!htb]
    \centering
    \includegraphics[width=.45\textwidth]{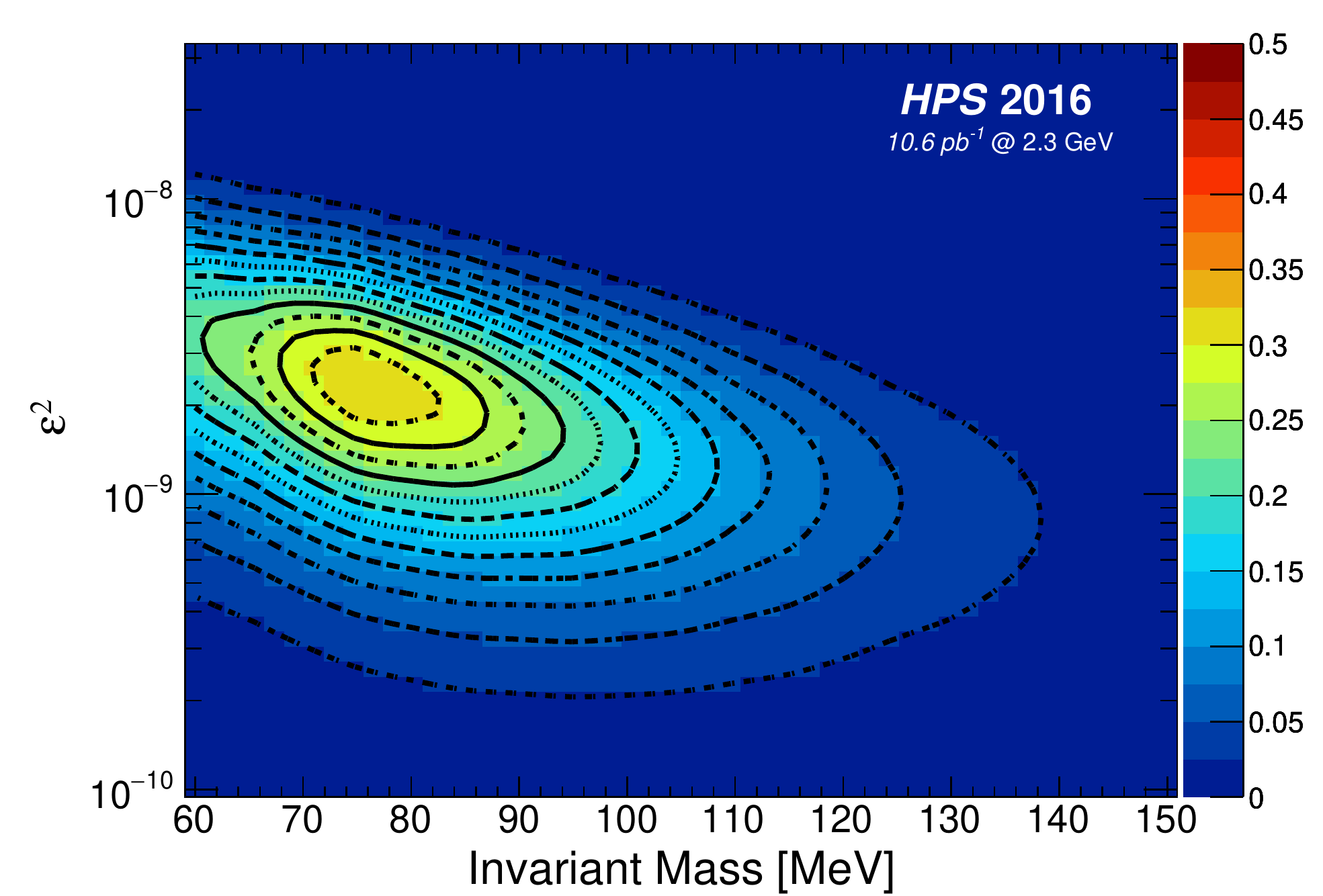}
    \includegraphics[width=.45\textwidth]{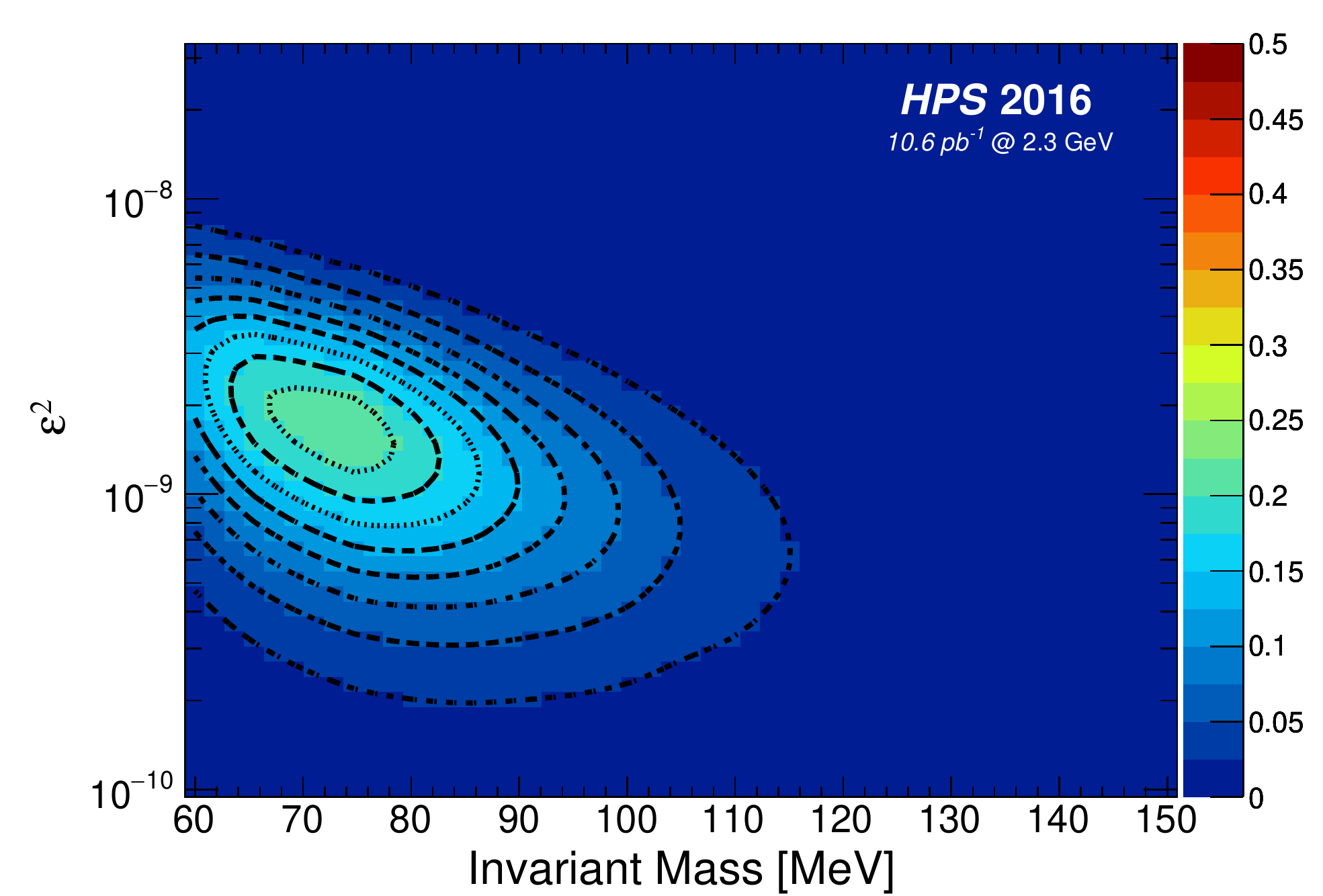}
    \caption{Top: The expected detected $\aprime$ yield in the L1L1 category versus mass and epsilon. Bottom:  The expected detected $\aprime$ yield in the L1L2 category versus mass and epsilon. }
    \label{fig:yield}
\end{figure}

\begin{figure}[!htb]
    \centering
    \includegraphics[width=.45\textwidth]{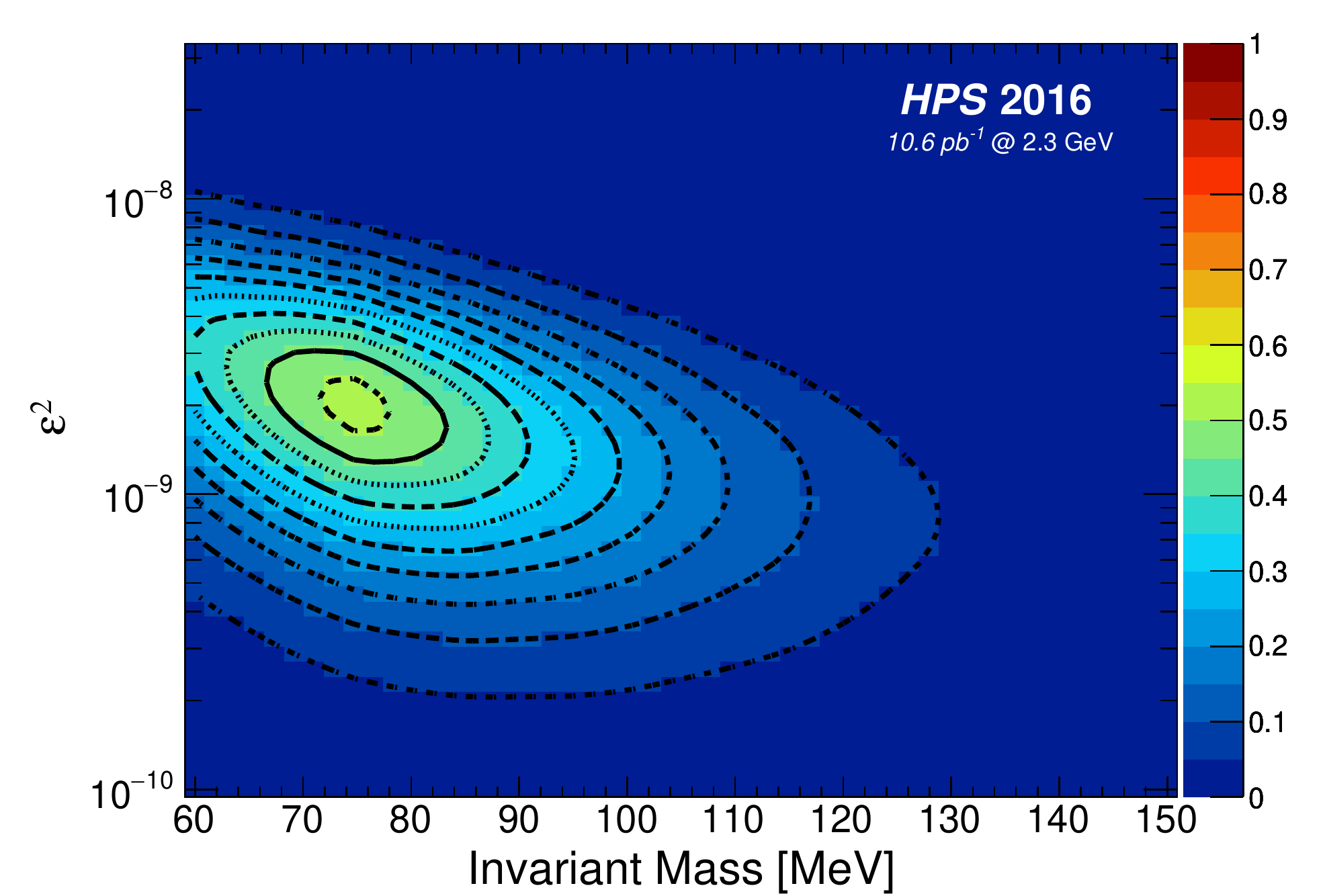}
    \caption{The expected detected $\aprime$ yield  for the combination of both L1L1 and L1L2 categories versus mass and epsilon. }
    \label{fig:yield_comb}
\end{figure}

The expected signal yields for the data set in the L1L1 and L1L2 categories are shown in \Cref{fig:yield}. For the L1L1 category, a peak of 0.32 $\aprime$ events is expected at \SI{75}{MeV} $\aprime$ mass and $\epsilon^2=2.4 \times 10^{-9}$ while for the L1L2 category, a peak of 0.22 $\aprime$ events is expected at \SI{75}{MeV} $\aprime$ and $\epsilon^2=1.7 \times 10^{-9}$.  Adding the two categories, the yield peaks at \SI{75}{MeV} and $\epsilon^2=2.1 \times 10^{-9}$ with 0.52 expected $\aprime$ events as shown in \Cref{fig:yield_comb}.  This shows that for suitable parameters, the HPS sensitivity is closely approaching that needed to exclude some parameter space of canonical $\aprime$ production.  In the following section, we will discuss the systematic uncertainties and the procedure used to set upper limits.

\subsection{Systematic Uncertainties}
\label{sec:systematics}

%%%%%%%%%%%%%%%%%%%%%%%%%%
%   Table: Systematics   %
%%%%%%%%%%%%%%%%%%%%%%%%%%

\begin{table}[!tbh] 
    \centering
    \begin{tabular}{c|cc}
        \toprule
        \textbf{Systematic Description} &\textbf{L1L1 Value} & \textbf{L1L2 Value}\\
        \midrule
        \midrule
            $\epem$ Composition & \multicolumn{2}{c}{$\sim$7\%} \\
            Mass Resolution &  \multicolumn{2}{c}{$\sim$3\%} \\
            Analysis Cuts & $\sim$8\% &$\sim$13\%\\
            $\aprime$ Efficiency &  \multicolumn{2}{c}{$\sim$5\%} \\
             \midrule
             Total in Quadrature  &  ~12\%  & ~16\% \\
             \midrule
            Target position & \multicolumn{2}{c}{$\sim$5-10\% (m/$\epsilon$ dep)} \\
        \midrule
        \midrule
        \bottomrule
    \end{tabular}
    \caption{A summary of systematic uncertainties that impact the final result of the displaced vertex search.  Where there is a single number the systematic effect is the same for L1L1 and L1L2.}
    \label{tab:systematics}
\end{table}

The systematic uncertainties from the experiment and the displaced vertex analysis have been quantified for both the L1L1 and L1L2 samples and are summarized in \Cref{tab:systematics}.  These sources of uncertainties are described below.  

A source of systematic uncertainty that is shared with the resonance search is the uncertainty in the $\epem$ composition that is expressed in the error of the radiative fraction. See \Cref{sec:Uncertainties} for details.  

An underestimate of the mass resolution would result in  signal leaking out of a mass bin. Thus, uncertainty in the mass resolution is a source of systematic uncertainty in the final result.  As described in \Cref{sec:massresolution}, we obtain the mass resolution as a function of $\aprime$ mass using $\aprime$ MC which has the $\mathrm{e}^+$ and $\mathrm{e}^-$ momenta smeared by the data/MC ratio of FEE resolutions.  As a cross-check, we do the same for \Mlr MC and compare that to the resolution seen in \Mlr data.  The \Mlr comparison gives very good agreement between data and MC, with the data having only a 5\% higher mass resolution compared to MC.  We use this 5\% seen in the unconstrained \Mlr samples to estimate a systematic on the number of signal events due to the mass cut and find that it is $\sim 3$\%.  

There are systematic uncertainties associated with the analysis cuts, particularly the cuts to reduce high-$z$ background (see Section \Cref{sec:apvertexcuts}). Recall that we use the radiative fraction to normalize the rates at event selection level while the relative efficiency from going from event selection to the final selection is accounted for using $\aprime$ MC.  There are small differences in the MC and data efficiencies of the final cuts and these have to be accounted for as systematic scaling errors.  

To do this, we calculate the efficiencies of each cut, with all other cuts applied, for data and trident MC events and take the ratio as the relative scaling that must be applied to the final limits.  There are only four categories of cuts to consider:  the V0 projection to the target, the isolation cuts,  the impact parameter cuts, and the shared hits cuts. The results of this study give the product of the efficiency ratios (data/MC) for L1L1 ($\sim 0.92$) and L1L2 ($\sim 0.88$).  The inverse of these ratios is applied to the final limits. 

\begin{figure}[!htb]
    \centering
    \includegraphics[width=.45\textwidth]{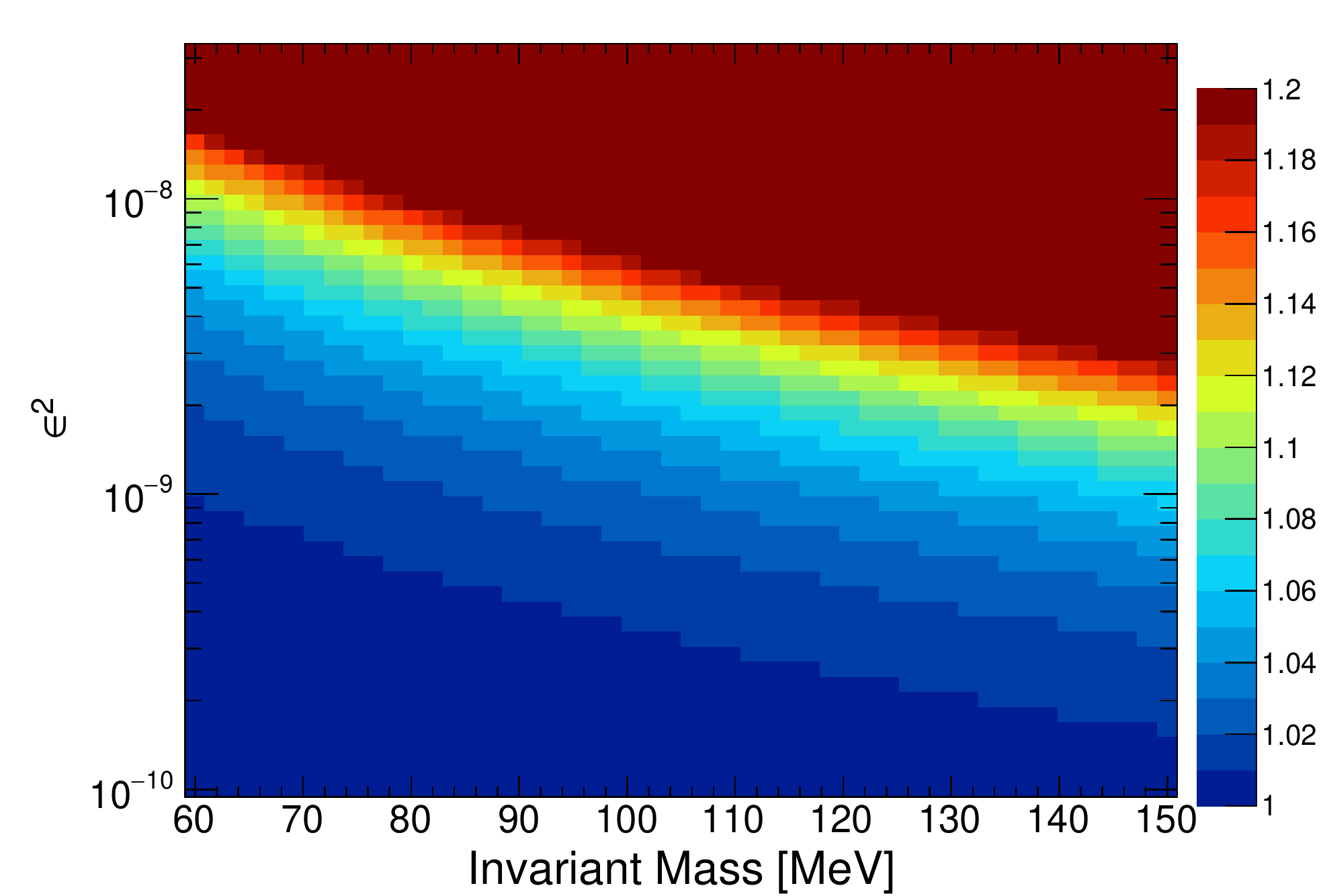}
    \caption{The ratio of the limit for the L1L1 category from the target \SI{0.5}{mm} upstream of the nominal position to the target at the nominal position using the Optimum Interval Method. }
    \label{fig:OIM_syst}
\end{figure}

From mechanical measurements, the target position is estimated to be known within $\pm \SI{0.5}{mm}$ from the nominal position. Any change in the assumed target position will result in  an overall shift in truth $z$ distributions of displaced $\aprime$s, and thus is a source of systematic uncertainty. For example, if the target is 0.5 mm more upstream than assumed, the entire displaced $\aprime$ truth distribution will also shift upstream by \SI{0.5}{mm} (without changing $z_{\rm cut}$) resulting in the actually expected signal yield that is less than the calculated signal yield. For a given $\aprime$ mass, this discrepancy will depend significantly on $\epsilon$ because of varying decay length, and can be calculated by simply recomputing both the signal yield and the limit at a different target position ($\pm \SI{0.5}{mm}$). The ratio of the limit from a target at \SI{0.5}{mm} upstream of the nominal position to the target at the nominal position is shown in \Cref{fig:OIM_syst}. This mass and $\epsilon$ dependence are used in the final estimate of systematic uncertainties. 

We combine the $\epsilon$/mass independent systematic uncertainties in quadrature and then combine those in quadrature at each combination of $\epsilon$ and mass to obtain a map of the uncertainty vs $\epsilon$/mass.  This uncertainty is then used to scale the upper limits we obtain from the data.  

\subsection{Upper Limit on \texorpdfstring{$\aprime$}{} Rate} \label{sec:vtxLim}

The Optimum Interval Method (OIM) \cite{PhysRevD.66.032005} is used to set a limit on the cross-section of the canonical $\aprime$ model. OIM was originally developed for direct detection dark matter experiments in which one expects a small signal where the signal shape in one variable is known and there is a small, but not necessarily understood, background. The OIM is an extension of the Maximum Gap Method, which searches for the largest gap in signal space that has no background events in order to set a limit. The OIM generalizes this method to an arbitrary number of background events between any two events in signal space and sets a limit based on the optimum interval and automatically selects the interval to avoid experimenter bias. In addition, the absolute cross-section of the signal does not need to be known.  Instead the OIM finds the optimum interval and sets a limit at the smallest cross-section at a specified confidence interval $C_0$, 90\% for this analysis.

\begin{figure}[!htb]
    \centering
    \includegraphics[width=.45\textwidth]{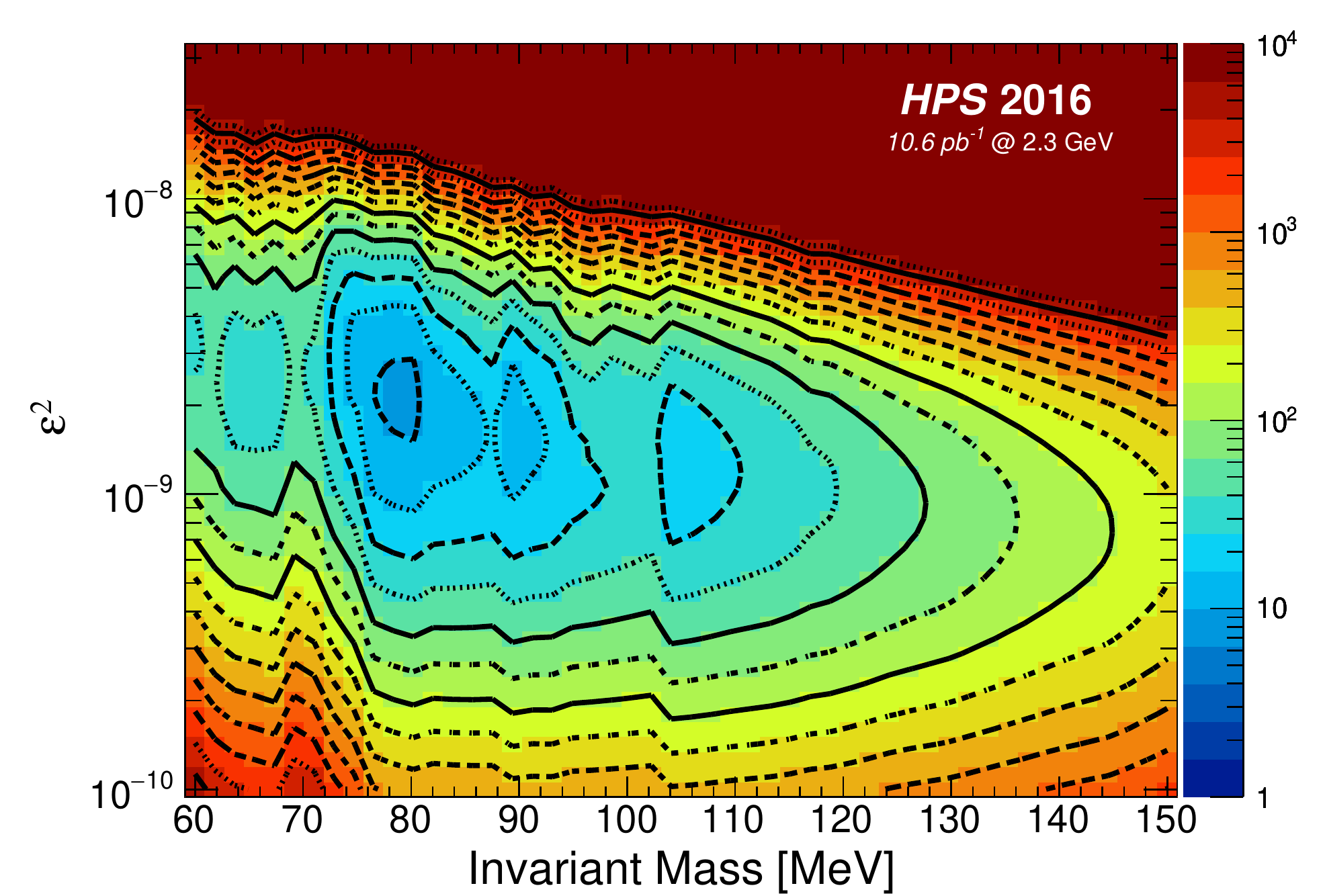}
    \includegraphics[width=.45\textwidth]{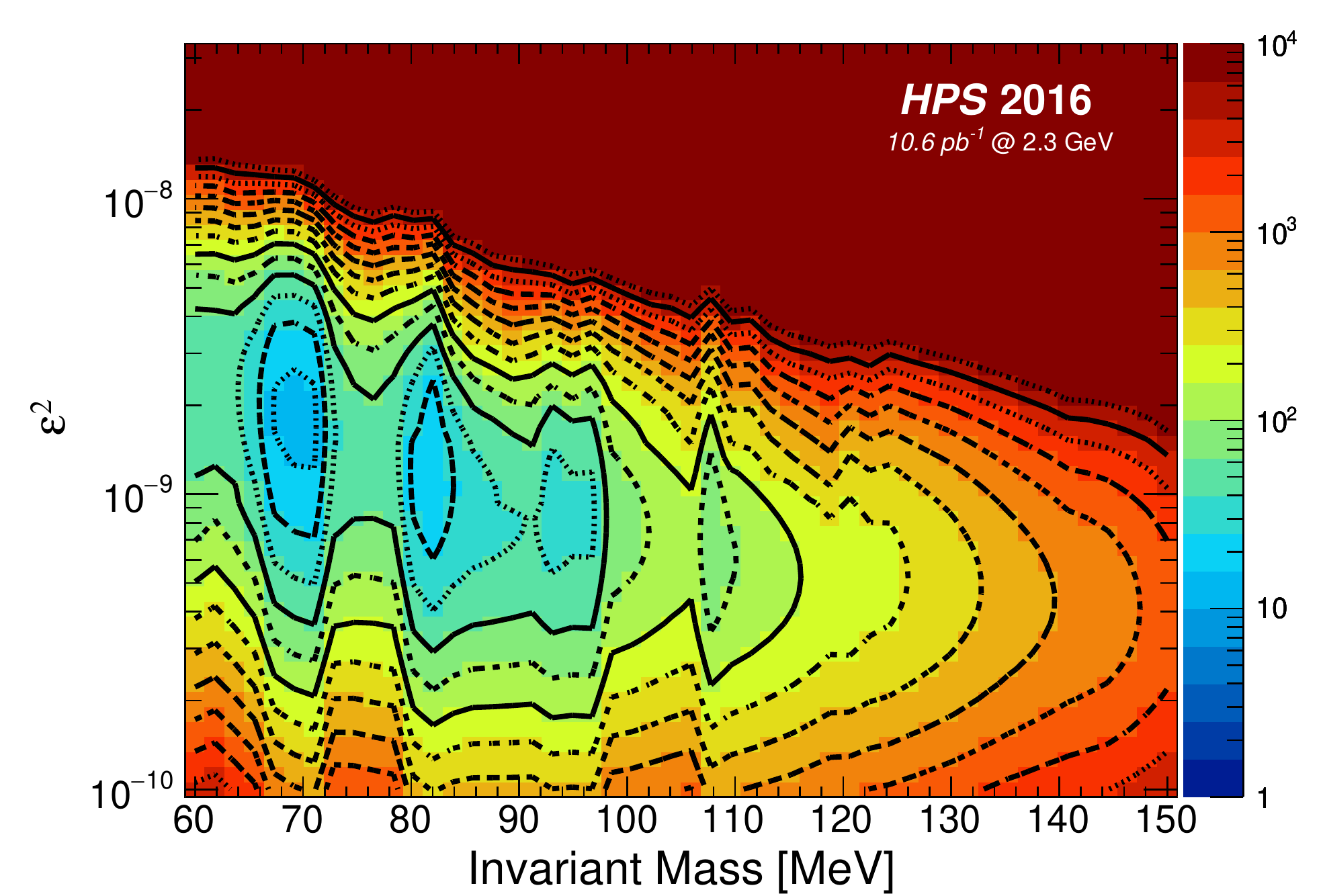}
    \caption{Top: The limit from the Optimum Interval Method for the L1L1 category. Bottom: The limit from the Optimum Interval Method for the L1L2 category. }
    \label{fig:OIM}
\end{figure}

The results for the OIM for the L1L1 and L1L2 categories on the full dataset are shown in \Cref{fig:OIM}. For the full dataset in the L1L1 category, the best limit is set at $m_{\aprime}=\SI{80.2}{MeV}$ and $\epsilon^2=2.1 \times 10^{-9}$ with a factor of 9.1 times the canonical $\aprime$ cross-section. The interpretation of this value is for an $\aprime$-like model with 9.1 times the cross-section.  The model is excluded at that mass and $\epsilon^2$ with 90\% confidence.  For the L1L2 category the best limit is at $m_{\aprime}=\SI{69.2}{MeV}$ and  $\epsilon^2=1.9 \times 10^{-9}$ with a factor of 13.9 times the canonical $\aprime$ cross-section.
These results include the systematic uncertainties described in \Cref{sec:systematics}.  

\begin{figure}[!htb]
    \centering
    \includegraphics[width=.45\textwidth]{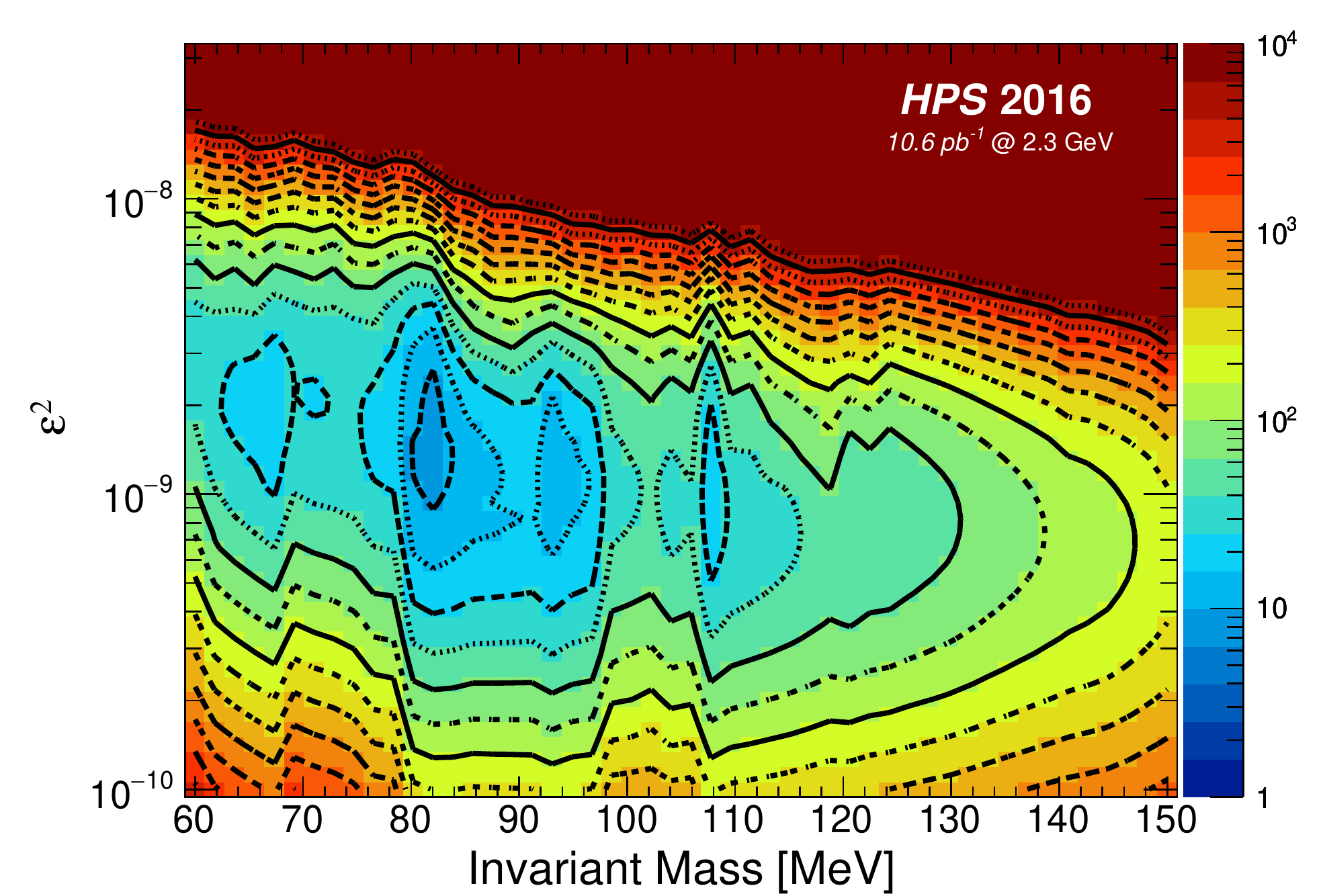}
    \caption{The limit from the Optimum Interval Method for the combination of both L1L1 and L1L2 categories for the full dataset.}
    \label{fig:OIM_comb}
\end{figure}

The limits derived when the L1L1 and L1L2 categories are combined are shown in \Cref{fig:OIM_comb}. Combining the L1L1 and L1L2 categories gives the best limit at  $m_{\aprime}=\SI{82.0}{MeV}$ and  $\epsilon^2=1.7 \times 10^{-9}$  with a factor of 7.9 times the canonical $\aprime$ cross-section.  With the current luminosity
it is not possible to set upper limits on canonical $\aprime$ production in the parameter plane.   

%% file: Summary.tex
\section{Summary}
\label{sec:summary}

\begin{figure}[b]
    \centering
    \grinp[width=0.45\tw]{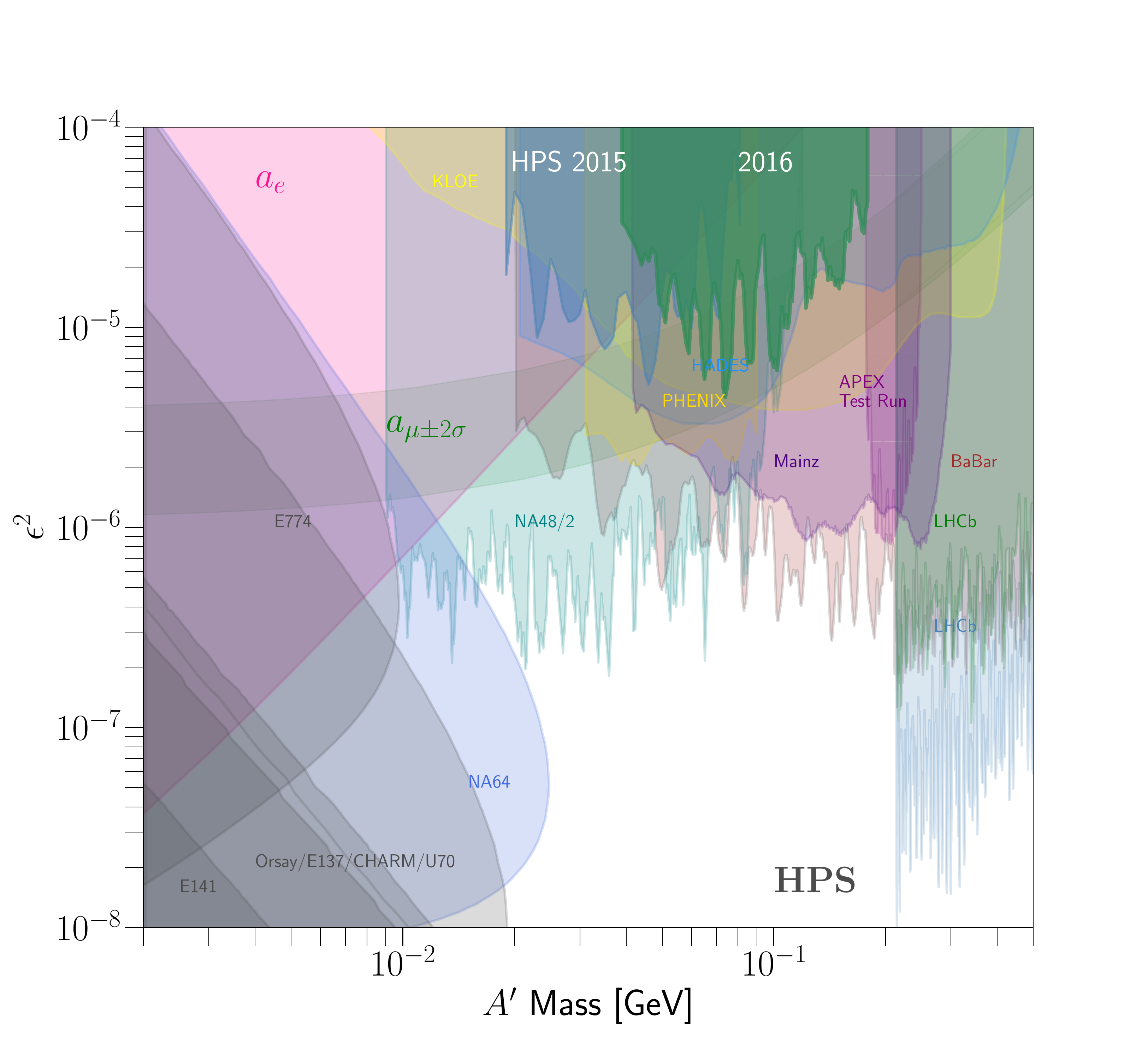}
    \caption{The exclusion of dark photon parameter space by this analysis and statistical recasting of our 2015 result.   Existing 
             limits from beam dump~\cite{Bjorken:1988as, riordan1987, bross1991, konaka1986,
             davier1989, andreas2012, Blumlein:1990ay, Blumlein:1991xh, Banerjee:2019pds, NA64:2019auh, Gninenko:2012eq}, 
             collider~\cite{Aubert:2009cp, Babusci:2012cr, Archilli:2011zc, Aaij:2017rft, PhysRevLett.124.041801, Yamaguchi:2016mbc} 
             and fixed target experiments~\cite{Abrahamyan:2011gv, Merkel:2014avp,
            Agakishiev:2013fwl, Batley:2015lha} are also shown. 
             The region labeled ``$a_e$'' is an exclusion based 
             on the electron $g-2$~\cite{PhysRevLett.106.080801, Aoyama:2012wj, PhysRevLett.100.120801, Davoudiasl:2012ig}
             . The green band labeled ``$a_{\mu} \pm 2\sigma$''
             represents the region that an $A'$ can be used to explain the discrepancy 
             between the measured and calculated muon anomalous magnetic moment~\cite{Pospelov:2008zw, Bennett:2006fi}.   }
    \label{fig:fullResult}
\end{figure}

This paper has presented the HPS results from its \SI{2.3}{GeV} 2016 engineering run. Evidence for heavy photons was searched for with both resonance search and displaced vertex search techniques. Our previous resonance search results, from the \SI{1.06}{GeV} 2015 engineering run, have been updated to use a more modern statistical approach. The 2016 data have extended the coverage in heavy photon mass to \SI{180}{MeV} in the resonance search, and exclude $\aprime$ production over the mass range 40--\SI{180}{MeV} down to the level of $\epsilon^2 = 10^{-5}$ as shown in \Cref{fig:fullResult}. 

The resonance search result confirms the results of previous searches but does not extend their sensitivity. The vertex search, reported here for the first time, explores $\aprime$ masses in the range 60--\SI{150}{MeV/c^2} for $\epsilon^2$ in the region $10^{-8}$ to $10^{-10}$. This is parameter space previously unexplored by other experiments, which is preferred territory for models assuming thermal production of hidden sector dark matter in the early universe. Being statistically limited, the present search does not reach the sensitivity needed to see $\aprime$ production in this region, but it does, at its point of optimal sensitivity, exclude production of long-lived $\epem$ pairs with 7.9 times the expected heavy photon cross-section and has afforded a first sensitive search for $\epem$ secondary vertices in electro-production at low energy. At its peak sensitivity in mass and $\epsilon^2$, the experiment would have expected to see 0.5 $\aprime$ events (on top of the 0.5 expected background), so it is approaching the sensitivity needed for the $\aprime$ search. Over much of the range in $\aprime$ mass, backgrounds were controlled to a level that should allow future vertex searches, with significantly greater luminosity, to explore interesting regions of parameter space. The HPS experiment has taken data runs in 2019 and 2021 and acquired over an order of magnitude more luminosity.   We project sensitivity to $\aprime$ production over a range of mass and $\epsilon^2$ parameters when those data are fully analyzed.

%% file: acknowledgements.tex
\section{Acknowledgements}

The authors are grateful for the outstanding efforts of the Jefferson
Laboratory Accelerator Division, the Hall B engineering group, and Forest
McKinney of UC Santa Cruz in support of HPS. The research reported here is supported by the U.S.
Department of Energy Office of Science, Office of Nuclear Physics, 
Office of High Energy Physics, the French Centre National de la 
Recherche Scientifique, 
United Kingdom's Science and Technology Facilities Council (STFC),
the Sesame project HPS@JLab funded by the French region Ile-de-France 
and the Italian Istituto Nazionale di Fisica Nucleare. Jefferson Science
Associates, LLC, operates the Thomas Jefferson National Accelerator
Facility for the United States Department of Energy under Contract
 No. DE-AC05-060R23177. 